\documentclass[oneside,12pt]{CUEDthesisPSnPDF}

\ifpdf
    \pdfcatalog { /PageMode (/UseOutlines)
                  /OpenAction (fitbh)
                 }
\fi
\title{\Large{An Efficient Quantum Algorithm and Circuit to Generate Eigenstates of $SU(2)$ and $SU(3)$ Representations\\}} 
\ifpdf

 \author{U.Satya Sainadh}
 \collegeordept{Centre for High Energy Physics}
 \university{Indian Institute of Science}
 \crest{\includegraphics[width=2.8cm]{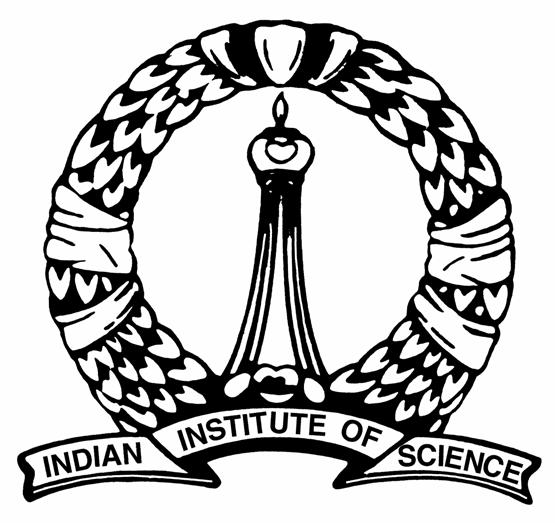}}
\fi
\degree{Master of Science}
\degreedate{August-2013}

\usepackage[hmarginratio=1:1]{geometry} %used for margin left:right
\usepackage[square,sort,numbers]{natbib}
\usepackage[T1]{fontenc}
\usepackage[utf8]{inputenc}
\usepackage{babel}
\usepackage[font=small,labelfont=bf]{caption}
\usepackage{alphalph}
\usepackage{amsmath,amssymb,verbatim}
\usepackage{epsfig}
\usepackage{graphicx, subfigure}
\usepackage{rotating}
\usepackage{bm}
\usepackage{times}
\usepackage[T1]{fontenc}
\usepackage{epstopdf}
\usepackage{youngtab}
\usepackage{young}
\usepackage[toc,page]{appendix}

\usepackage{float}
\begin{document}

%\language{english}

% A page with the abstract on including title and author etc may be
% required to be handed in separately. If this is not so, then comment
% the below 3 lines (between '\begin{abstractseparte}' and 
% 'end{abstractseparate}'), normally like a declaration ... needs some more
% work, mind as environment abstracts creates a new page!
% \begin{abstractseparate}
%   \input{Abstract/abstract}
% \end{abstractseparate}

% Using the watermark package which is in StyleFiles/
% and to remove DRAFT COPY ONLY appearing on the top of all pages comment out below line
%\watermark{DRAFT COPY ONLY}

%-------------------------------------------------------Info to be put on your cover page ends here--------------------------------------
\maketitle
%\include{Abstract/abstract}
%set the number of sectioning levels that get number and appear in the contents

%\include{thessamp}
\frontmatter % book mode only

\setcounter{secnumdepth}{3}
\setcounter{tocdepth}{3}
\pagenumbering{roman}
%\include{Dedication/dedication}
%\include{Acknowledgement/acknowledgement}
% Thesis Acknowledgements ------------------------------------------------

%\begin{acknowledgementslong} %uncommenting this line, gives a different acknowledgements heading
\null
\begin{acknowledgements}      %this creates the heading for the acknowlegments

\hspace*{0.5cm} I would like to express my deep and sincere gratitude to my supervisor Prof. Apoorva D. Patel. It has been a privilege to work with him . He has been a constant source
of inspiration for me. Despite my shortcomings and slow progress in research, he
has been very cooperative and helpful to me. Every discussion with
him has always been productive and has enriched my understanding of the subject.

\hspace*{0.5cm}I would also like to thank my faculty advisor and Int. PhD convenor, Dr. V. Venkataraman and departmental Chairman Prof. B. Ananthanarayan. They have taken all sorts of trouble to make my life easier and joyful. It was due to them that we could enjoy various institute facilities and never felt isolated. I am thankful to all those teachers who taught me during the course work.

\hspace*{0.5cm}It gives me great pleasure to present my gratitude to all my Int. PhD classmates:
G.M. Nagendra, Vaishak, Johnson, Shouvik, Kallol, Anindita, Sundeep, Sudeep Ghosh, Aamir, Moupiya, Maheshwar Maji, for their constant help, advice and support during my stay in IISc. Also, I would like to especially thank my friends Guru Raj, Nagendra, Surya, Sagar, Sandeep, Sri Vallabha, Raghu, Samarat, Vamsi, Ramesh, Sonti Rajesh, Nilesh, Sai Kiran, Sharat, Ramakanth, Krish, and my childhood friends Sagar, Rajesh, Pavan, Harish, for always being there for any help that I needed at any point of time. We really spent an awesome time discussing with each other various issues that helped me become a better person.

\hspace*{0.5cm}Finally, I thank the almighty, and take this opportunity to express the profound gratitude from the deepest of my heart to my beloved parents and my brother for their love and support.  

\end{acknowledgements}
%\end{acknowledgmentslong}

% ------------------------------------------------------------------------

%%% Local Variables: 
%%% mode: latex
%%% TeX-master: "../thesis"
%%% End: 

\tableofcontents
\listoffigures
\listoftables
%\printnomenclature  %% Print the nomenclature
%\addcontentsline{toc}{chapter}{Nomenclature}
%--------------------------------------------------The main body of your thesis------------------------------------------------------------------
%\mainmatter % book mode only
%\include{Introduction/introduction}
%\include{su2/su2}
%\include{copysu3/copysu3}
%\include{sud/sud}
%\appendixpage
%\include{Appendix1/appendix1}
%\include{Appendix2/appendix2}
%\include{Appendix3/appendix3}
\mainmatter % book mode only
%%% Thesis Introduction --------------------------------------------------
\chapter{Introduction}
%\ifpdf
 %   \graphicspath{{Introduction/IntroductionFigs/PNG/}{Introduction/IntroductionFigs/PDF/}{Introduction/IntroductionFigs/}}
%\else
 %   \graphicspath{{Introduction/IntroductionFigs/EPS/}{Introduction/IntroductionFigs/}}
%\fi

Quantum mechanics is one of the greatest achievements of $20^{th}$ century physics. It is an exceptionally successful theory, explaining the natural phenomena at atomic scales to a very high degree of  accuracy. Though quantum mechanics is routinely used to explain the behaviour of individual atoms and particles, there also exist larger systems where quantum effects show up. Exact solutions rarely exist for many-body quantum systems, and many problems are not yet understood completely. To tackle such problems, computer simulations are frequently used. Simulations of quantum processes are inefficient on a classical computer. In 1982, Richard P. Feynman pointed out that simulations of quantum processes become efficient on a controlled quantum system \citep{feynman}. Later in 1985, David Deutsch showed that a controlled quantum system can be a universal Turing machine and called it a \textit{quantum computer} \citep{deutsch}. Since then the mathematical framework for quantum computation has arrived at a well-accepted standard form \citep{elementary-gates}. 

A digital quantum computer uses qudits as its computational basis.  A qudit is the unit of quantum information represented by a unit vector in a  $d$-dimensional complex Hilbert space $\mathbb{C}^d$, \textit{i.e.},   
\begin{eqnarray}
\vert x\rangle = \left(\begin{array}{c}  x_1\\x_2\\. \\. \\x_d 
\end{array}\right)\hspace*{.6cm}\text{with} \hspace*{.2cm}\sum_{k=1}^d |x_k|^2=1  .  \label{a1}
\end{eqnarray}  
Standard computational basis states $\vert i \rangle$, are the \textit{standard basis vectors} of $\mathbb{C}^d$.   The generic qudit state of \eqref{a1} can be obtained from a computational basis state using a unitary transformation $U_d$. Quantum logic gates, which form the  building block of quantum circuits are such unitary transformations. They all have their inverses, and quantum circuits are reversible.

A register of $n$-qudits spans an N-dimensional Hilbert space $(\mathbb{C}^d)^{\otimes n}$,  with  $N=d^n$. The standard computational basis $\vert \textbf{i}\rangle$, of the N-dimensional Hilbert space is given by the tensor product of standard computational basis states of the individual Hilbert spaces, \textit{i.e.}, \begin{equation}
\lbrace\vert \textbf{i}\rangle:\textbf{i}=0,1,2,...,(d^n-1)\rbrace = \lbrace \vert i_1\rangle \otimes \vert i_2\rangle\otimes ... \otimes\vert i_n\rangle :i_1,i_2,...,i_n=0...(d-1)\rbrace .
\end{equation} 
For example, the analogue of a classical bit (with values 0 and 1) in quantum computation is a qudit with $d=2$, known as a qubit (with basis $\vert 0 \rangle$ and $\vert 1 \rangle$). Quantum gates operating on qubits are 2$\times$2 unitary matrices, frequently expressed in terms of the Pauli matrices. The standard computational states are eigenstates of $\sigma_z$, and the quantum $NOT$ gate is $\sigma_x$.
\begin{eqnarray}
\sigma_z = \begin{pmatrix}
1 & 0\\ 0&-1
\end{pmatrix} \hspace*{.6cm} \sigma_x = \begin{pmatrix}
0&1\\1&0
\end{pmatrix}
\end{eqnarray} 
More general unitary transformations can be parametrised in terms of rotation angles, and can also be made part of controlled unitary operations. For example, a 2-qubit control-U operation has the form
\begin{eqnarray}
U = \begin{pmatrix}
\alpha_\theta & \beta_\theta \\ \alpha'_\theta &\beta'_\theta
\end{pmatrix}, \hspace*{.8cm} C_U =\begin{pmatrix}
1&0&0&0\\ 0&1&0&0 \\ 0&0&\alpha_\theta & \beta_\theta \\ 0&0&\alpha'_\theta &\beta'_\theta
\end{pmatrix}.
\end{eqnarray}   
This $C_U$ operation rotates the target qubit by $U$, when the control qubit is in the state $\vert 1 \rangle$. In particular, the $C_{NOT}$ gate with the first qubit as the control qubit and the second one as the target qubit, implements
 \begin{eqnarray}
 \begin{aligned}
C_{NOT}: \begin{array}{c}
\vert 0\rangle\vert 0\rangle \rightarrow \vert 0\rangle\vert 0\rangle \\ \vert 0\rangle\vert 1\rangle \rightarrow \vert 0\rangle\vert 1\rangle \\\vert 1\rangle\vert 0\rangle \rightarrow \vert 1\rangle\vert 1\rangle \\ \vert 1\rangle\vert 1\rangle \rightarrow \vert 1\rangle\vert 0\rangle\end{array}.
\end{aligned}
\end{eqnarray} We will denote a control-U operation, that acts when the control qubit is in the state $\vert 0 \rangle$, by $\tilde{C}_U$. 

Quantum algorithms proposed with the preceding definitions of qudits and quantum gates have turned out to be more efficient than their classical counterparts \citep{grover,shor,deutsch-jozsa,simon,element-distinctness1}. But several quantum algorithms (e.g.  phase estimation and order-finding \citep{shor}), and techniques like measurement based quantum computing \citep{measurement-based-QC,oneway-QC,nielsen-quantummeasurement}, require the initial state of the quantum computer to be an eigenstate of a specific unitary operator, and one needs efficient methods to prepare such states. A generic superposition state in $(\mathbb{C}^d)^{\otimes n}$ can be obtained with $\mathcal{O}(\sqrt{N})$ resources using Grover's algorithm \citep{grover}, while efficiently obtainable states are those that require resources polynomial in $n$. Quantum register states that factorise in terms of component qudits are certainly easy to prepare. But there are also highly entangled $n$-qudit states that can be prepared efficiently. In this thesis, we show that eigenstates of multi-particle systems with unitary representations, labelled by global generalised angular momentum parameters, can be efficiently prepared. Our demonstration uses the Schur transform defined in the next section.

\section{Schur transform}
The Schur transform relies on the mathematical theorem known as `Schur-Weyl duality'~\citep{schur-weyl}. Consider a a system of $n$-qudits, each with a standard local computational basis. The theorem states that there exists a decomposition such that  the joint action of unitary ($U_d$) and permutation ($S_n$) groups  on the total Hilbert space reduces to that on the subspaces  corresponding to the irreducible representations of $U_d$ and $S_n$:  \begin{equation}
 (\mathbb{C}^d )^{\otimes n} \cong \bigoplus_{\lambda\in Part[n,d]} \gamma_{_{\lambda}} \otimes S_\lambda \hspace*{.2cm}.  \label{a}
 \end{equation} Here, $\lambda$'s are the partitions of $n$ into $d$ parts such that  \begin{eqnarray}
  \lambda_1 \geq \lambda_2\geq . . .  \geq \lambda_d \geq 0 \hspace*{2cm} \text{and}  \hspace*{2cm} \sum_{i=1}^d\lambda_i=n  , 
\end{eqnarray}
The key feature of the theorem is that the same set $\{\lambda_i\}$ is used to simultaneously label the irreducible representations of both $U_d$ (\textit{i.e.} $\gamma_{_{\lambda}}$) and the irreducible representation of the permutation group $S_n$ (\textit{i.e.} $S_{\lambda}$). Each subspace $\gamma_{_{\lambda}}\otimes S_{\lambda}$ is orthogonally spanned by eigenstates of irreducible representations of $U_d$ and $S_n$. In other words, there exists a basis, known as the Schur basis and denoted by $\vert [\lambda], q_{_{\lambda}}, p_{_{\lambda}}\rangle_{sch}$ ($q_{_{\lambda}} \in \gamma_{_{\lambda}}$ and $p_{_{\lambda}} \in S_\lambda$), which is a simultaneous eigenstate of unitary  ($Q_\lambda$) and permutation ($P_\lambda$) operators 
\begin{equation}
Q_\lambda \vert [\lambda], q_{_{\lambda}}, p_{_{\lambda}}\rangle_{sch} = \vert [\lambda]\rangle Q_\lambda\vert q_{_{\lambda}}\rangle \vert p_{_{\lambda}}\rangle 
\end{equation}
\begin{equation}
P_\lambda  \vert [\lambda], q_{_{\lambda}}, p_{_{\lambda}}\rangle_{sch} = \vert [\lambda]\rangle \vert q_{_{\lambda}}\rangle P_\lambda\vert p_{_{\lambda}}\rangle 
\end{equation}
The subspaces $\gamma_{_{\lambda}}$ and $S_{\lambda}$ can be conveniently represented by Young diagrams,
with $\lambda_k$ consecutive boxes in $k^{th}$ row [Appendix A]. Each Schur basis state belonging to the subspace can be represented by a unique Young tableau, filling up the Young diagram with $[\lambda]$ labels.\\ 
\indent The computational basis is defined in terms of independent local states, while the Schur basis is defined in terms of specific global properties under $U_d$ and $S_n$. The unitary transformation connecting these two bases is the Schur transformation $U_{sch}$. Specifically, we decompose a generic Schur basis state as a superposition of computational basis states :  
\begin{equation}
\vert \lambda,q_{_{\lambda}},p_{_{\lambda}}\rangle = \sum_{i_1,i_2...i_n=1}^d  [U_{sch}]\vert i_1 i_2 ...i_n\rangle  .
\end{equation}
A quantum circuit performing the Schur transform rotates the input computational basis states to the output Schur basis states. This quantum circuit was efficiently constructed in \citep{schur-2}, by expressing the Schur transform as a recursive function of the Clebsch-Gordan transform $U_{CG}$. $U_{CG}$ is the unitary transform that describes the change in irreducible representations when a single qudit is added to an existing register of qudits. Starting with one qudit, and adding ($n-1$) qudits to it one-by-one, we obtain an $n$-qudit register. At each step, the index $(p_{\lambda})_k$ defines the permutation symmetry property of the $k^{th}$ qudit relative to the existing register of $(k-1)$ qudits, which facilitates the iterative construction. 

\section{Central idea}
The above outlined algorithm for performing the Schur transform can be converted to a quantum logic circuit with resources polynomial in $n$ and $d$. We want to construct the eigenstates $\vert \lambda,q_{_{\lambda}},p_{_{\lambda}} \rangle$ with global properties, as a linear combination of the basis states $\vert i_1,i_2,...,i_n\rangle$ with local properties. Using the fact that quantum logic is reversible \citep{feynman-QC}, we accomplish that using the inverse Schur transform 
\begin{equation}
\vert i_1,i_2,...,i_n\rangle= \sum_{q_{_{\lambda}},p_{_{\lambda}}} [U_{sch}]^{-1}\vert \lambda,q_{_{\lambda}},p_{_{\lambda}} \rangle
\end{equation}
For given $\vert \lambda,q_{_{\lambda}},p_{_{\lambda}} \rangle$, $U_{sch}^{-1}$ takes the desired Schur basis state as input and expresses it as a superposition of computational basis states (which are tensor products of $n$-qudits). 
\begin{figure}[H]
\centering
\framebox{\includegraphics[scale=.7]{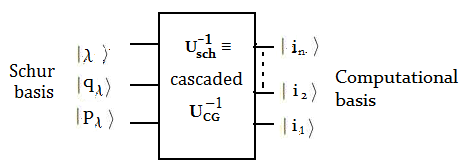}}
\caption{ The inverse Schur transform converts $\vert \lambda,q_{_{\lambda}},p_{_{\lambda}} \rangle$ to a superposition of $\vert i_1,i_2,...,i_n\rangle$ states.}
\end{figure}
The algorithm implementing  $U_{sch}^{-1}$ can be efficiently decomposed as a cascade of the inverse Clebsch-Gordan transform  $U_{CG}^{-1}$. At each step, there are three inputs to $U_{CG}^{-1}$, and change in one of them (i.e. $(p_\lambda)_k$) builds up the iterative process, as shown in Figure 1.2.
\begin{figure}[!]
\centering
\framebox{\includegraphics[scale=.7]{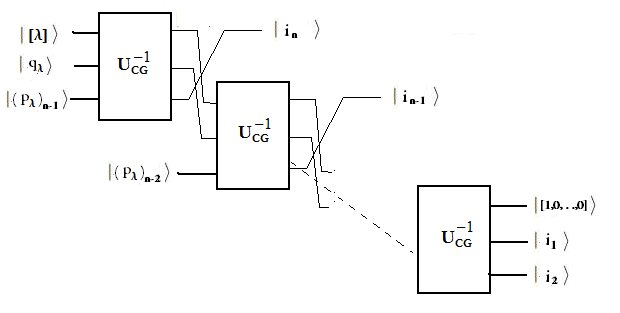}}
\caption{Schematic representation of  $U_{sch}^{-1}$ as a cascaded $U_{CG}^{-1}$}
\end{figure}

We explicitly demonstrate the implementation of the inverse Schur transform for $SU(2)$ and $SU(3)$ groups in the form of quantum circuits in Chapters 2 and 3 respectively. In Chapter 2, we first describe the mapping of a Schur basis state to an $SU(2)$ representation eigenstate and then construct the $U_{CG}$ for the $SU(2)$ group. Subsequently, we present the an efficient quantum circuit with polynomial computational complexity that implements $U_{CG}^{-1}$ and eventually $U_{sch}^{-1}$. A similar analysis is given in Chapter 3 for the case of $SU(3)$. Our method uses $U_{CG}$ for $SU(2)$ as a subroutine to obtain $U_{CG}$ for $SU(3)$. This approach generalises, by induction, to construction of $U_{CG}$ for $SU(d)$ \citep{schur-2}, yielding an algorithm with polynomial complexity. We outline it in the last Chapter. 

The mathematical tools and details of calculations used in constructing the quantum circuit are given in the Appendices. Appendix A gives an introduction to Young diagrams and Young tableaux, and describes how they are used to map the eigenstates of $SU(d)$ representations. Appendix B describes an explicit construction of all the necessary quantum logic gates used in the construction of $U_{CG}^{-1}$ in Chapters 2 and 3, and evaluates their computational complexity. Appendix C contains the derivation of isoscalar factors used to obtain $U_{CG}^{-1}$ for $SU(3)$.

%%% ----------------------------------------------------------------------

%%% Local Variables: 
%%% mode: latex
%%% TeX-master: "../thesis"
%%% End: 

\chapter{$SU(2)$}
%\ifpdf
 %   \graphicspath{{su2/su2Figs/PNG/}{su2/su2Figs/PDF/}{su2/su2Figs/}}
%\else
 %   \graphicspath{{su2/su2Figs/EPS/}{su2/su2Figs/}}
%\fi

  \section{Introduction}
The set of all $d\times d$ unitary matrices of determinant +1 form a unitary group $SU(d)$ under matrix multiplication.  Any element of this group ($U$) can be expressed as an exponential of a traceless hermitian matrix ($H$), $$U=e^{iH} .$$  $H$ can be decomposed using an orthogonal set of basis matrices called the generators. The $SU(2)$ group has 3 generators, and so it is the 3-dimensional special unitary group. The fundamental representation of $SU(2)$ is formed by $2 \times 2$ unitary matrices, and it is convenient to choose its generators as the Pauli matrices: $$ \sigma_1 =  \begin{bmatrix}
0 & 1 \\ 1& 0 \end{bmatrix}, \sigma_2 =  \begin{bmatrix} 0& -i \\i & 0 \end{bmatrix}   , \sigma_3 = \begin{bmatrix}1&0\\0&-1\end{bmatrix}.$$ Every element of $SU(2)$ can be expressed as 
\begin{equation}
U = \exp{(i \frac{\hat{n}.\bar{\sigma}}{2}\theta)} 
\end{equation}   
with the unit vector $\hat{n}$ as the rotation axis, and the rotation angle $\theta \in [0,4\pi]$.
The Pauli matrices obey the commutation and anticommutation relations: \begin{equation}
\begin{bmatrix}
\sigma_i,\sigma_j\end{bmatrix} = 2i\epsilon_{ijk}\sigma_k ,\quad \begin{Bmatrix}
\sigma_i,\sigma_j \end{Bmatrix}
= 2\delta_{ij}\textbf{\textit{I}} .\end{equation}  Here $i$, $j$, $k$ go from 1 to 3,  
and $\epsilon_{ijk}$ is the fully antisymmetric tensor with $\epsilon_{123}=1$. The Lie algebra followed by the Pauli matrices  is defined by the anticommutation relations and the structure constants $\epsilon_{ijk}$, and is denoted by $su(2)$.

The  $2\times 2$ unitary matrices act as norm preserving transformations on 2-component complex vectors. By convention, the eigenvectors of $\sigma_3$ are chosen as the basis for this vector space.  They are $\begin{pmatrix}
1\\0
\end{pmatrix} \text{and} \begin{pmatrix}
0\\1
\end{pmatrix}$, with eigenvalues +1 and -1 respectively.

\section{$SU(2)$ in physics}
The $SU(2)$ algebra is extensively used in quantum mechanics, in particular as the angular momentum algebra. The Pauli matrices $\sigma_1 , \sigma_2 , \sigma_3$ are the operators associated with rotations of a spin-1/2 system about the x,y,z axes. The angular momentum algebra describes the behaviour of an individual spin under rotations, and also that of a composite system of multiple spins. The angular momentum Lie algebra is specified by the commutation rules of its generators$$ [J_i,J_j]=i \epsilon_{ijk} J_k.$$ Its irreducible representations can be obtained as $l \times l$ matrices of any non-negative integer dimension, $$[T_i,T_j]= i \epsilon_{ijk} T_k.$$ $\{T_i\}$ are the generators of the $l$-dimensional representation with $l=2j+1$. In particular, $\{\sigma_i\}$ generate the $j=\frac{1}{2}$ fundamental representation.

The representation theory of $SU(2)$ allows construction of finite dimensional vector spaces, which would describe the behaviour of composite spin systems. To fully specify the state of such a quantum system, we need the complete set of mutually commuting operators.

First, consider a single particle with spin $j$. We  choose one of the quantum labels as the eigenvalue of $J_3$ (the z-component of the angular momentum). From the commutation relations of the generators of $SU(2)$, it is clear that neither $J_1$ nor $J_2$ commute with $J_3$. But the quadratic Casimir operator, 
 \begin{equation} J^2  = J_1^2+J_2^2+J_3^2  , \end{equation} commutes with all $J_i$, and its eigenvalue can be used as another label for the quantum state. In what follows, we replace $J_1,J_2,J_3$ by $J_x,J_y,J_z$ respectively. 
 %----------------------------------------------------------------------------------------
 
Now consider a normalized eigenbasis $\vert a,m \rangle$ with $a,m$ as eigenvalues of $J^2,J_z$, \textit{i.e.} \begin{equation}
    J^2\vert a,m \rangle = a\vert a,m \rangle \hspace*{2cm} J_z\vert a,m \rangle= m\vert a,m\rangle .
\end{equation}  
  We now, define the two ladder operators, $J_+=J_x+ iJ_y $  and $J_-=J_x-iJ_y$. They obey the properties  \begin{equation}
  J_+^\dagger=J_-   \hspace*{0.31cm},\hspace*{0.7cm} [J_z,J_{\pm}]=\pm J_z  \hspace*{.3cm},   \hspace*{1cm} [J^2,J_{\pm}]=0  .
\end{equation}     
These commutators show that action of $J_\pm$ on an eigenstate changes $m$ by $\pm 1$, while leaving $a$ the same. 
So $J_\pm\vert a,m\rangle$ is proportional to $\vert a,m \pm 1\rangle$, \textit{i.e.}
 \begin{equation}
 J_\pm\vert a,m\rangle = C_\pm(a,m)\vert a,m \pm 1\rangle
 \end{equation}
 By taking the adjoint of the above equation we have, 
 \begin{equation}
 \langle a,m\vert J_\mp = C^*_\pm(a,m)\langle a,m \pm 1\vert
 \end{equation}
  Combining the above two equations, we have
   \begin{eqnarray}
  \begin{aligned}
  \langle a,m\vert J_\mp J_\pm \vert a,m\rangle =&|C_\pm(a,m)|^2\langle a,m \pm 1\vert a,m \pm 1 \rangle \\ =& \langle a,m \vert J^2-J_z^2 \mp J_z\vert a,m\rangle \\=& (a-m^2 \mp m) \langle a,m \vert a,m \rangle
      \end{aligned}
  \end{eqnarray}
  The RHS of the above equation is non-negative, because it is a product of the absolute value of $C_\pm$ and the norm of an eigenvector. So $a-m^2 \mp m\geq 0$, even when $m$ can be changed by the action of the ladder operators. That can happen only if the state changes produced by the ladder operators terminate at some stage. So for representations with finite $a$, there must exist values $m_{min},m_{max}$ such that,
          \begin{subequations}\begin{eqnarray} 
     J_+ \vert a,m_{max}\rangle =0 \hspace*{1.5cm}
     J_-\vert a,m_{min}\rangle = 0 \\ a=m_{max}(m_{max}+1)=m_{min}(m_{min}-1).\end{eqnarray}\end{subequations}
    With $m_{max} \geq m_{min}$, the only solution of the above equation is $m_{max}=-m_{min}$. Furthermore, since ladder operator can connect $\vert a,m_{max}\rangle$ and $\vert a,m_{min}\rangle$, $m_{max}-m_{min}$ has to be an integer. Consequently, all allowed values of $m$ are integer or half-integer, and they are symmetrically located around zero. Conventionally, the largest eigenvalue of $J_z$, \textit{i.e.} $m_{max}$ is labelled $j$. Then $a=j(j+1)$. The standard notation represents the eigenstate $\vert a,b \rangle$ as $\vert j,m \rangle$, whereby\\\vspace*{-1cm}   
    \begin{subequations}\begin{eqnarray} 
  &J^2 \vert j,m\rangle = j(j+1)\vert j,m\rangle \hspace*{2cm} j=0,\dfrac{1}{2},1,\frac{3}{2}.... \\
  &J_z \vert j,m\rangle = m\vert j,m\rangle  \hspace*{2cm} m=j,j-1,.....,-j+1,-j   .
  \end{eqnarray}\end{subequations} 
  Thus a spin-$j$ system  has 2$j$+1 eigenstates corresponding to different $m$ values. 
   Since, $ J^2 ,J_z $ are Hermitian operators, their eigenstates for distinct eigenvalues are orthogonal, and one can choose normalisations such that
  \begin{equation}
  \langle j_1 m_1\vert j_2 m_2\rangle = \delta_{j_1,j_2}\delta_{m_1,m_2} . 
  \end{equation}
Given any eigenstate of a spin-$j$ system, one can obtain all the others by applying the ladder operators. 

Now we can calculate the proportionality constants $C_\pm (j,m)$ in (2.6).
From (2.8), (2.10) and the orthogonality relation (2.11) we have
  \begin{eqnarray}
      |C_\pm (j,m)|^2 = (j \mp m)(j\pm m+1) .                         
  \end{eqnarray}  
 It is conventional to choose the relative phases of the states $\vert j,m \rangle$ such that all $C_\pm (j,m)$ are real and positive.
%-------------------------------------------------------------------------------------------------
\section{Addition of angular momenta} 
This section is about answering the question, \textit{what happens when we take two systems with different angular momenta and combine them?} 

Let us consider two systems with angular momenta $j_1$ and $j_2$. The allowed values of their $J_z$ are, $ -j_1 \leq m_1  \leq +j_1$ with $ 2j_1 + 1$ states and $ -j_2 \leq m_2  \leq +j_2$ with $2j_2 + 1$ states, respectively. Therefore, when we combine the two angular momenta as the tensor product $j_1 \otimes j_2$, we have in total $ (2j_1 +1)(2j_2 +1) $ states consisting of all possible combinations with $ m=m_1+m_2$ . 

In working with a system of combined angular momenta, it is useful to construct a change of basis. The tensor product basis is $\vert j_1,m_1\rangle \otimes\vert j_2,m_2\rangle $ with  eigenoperators $ J_1^2,J_{1z},J_2^2,J_{2z}$. Each individual system needs two quantum numbers to specify a state, and we need 4 quantum numbers to specify the combined system. We can also describe the whole system in its total angular momentum basis, \textit{i.e.}  the basis specified by the operators $ J^2= (J_1+J_2)^2,J_{z}=J_{1z}+J_{2z}$. In the tensor product notation,
\begin{eqnarray}
\begin{aligned}
&J_1 = J \otimes \textit{\textbf{I}}  \hspace*{2cm} J_2 = \textit{\textbf{I}} \otimes J  \\
 &J_{1\pm} = J_\pm \otimes \textit{\textbf{I}}  \hspace*{1.6cm} J_{2\pm} = \textit{\textbf{I}} \otimes J_\pm \ \\  
&J^2 = J_1^2 + J_2 ^2 + J_{1+}J_{2-} + J_{1-}J_{2+} + 2 J_{1z}J_{2z} 
\end{aligned} 
\end{eqnarray}
  We can see that the operators $J^2,J_{z},J_1^2,J_2^2 $ commute with each other. Therefore we specify this total angular momentum basis by $\vert  j,m; j_1, j_2 \rangle$. The eigenvalues of $J_{z}$ \textit{i.e.} $m$ are obvious in both the bases:
  \begin{eqnarray}
  \begin{aligned}
  J_{z}\vert  j,m; j_1, j_2 \rangle =&  m\vert  j,m, j_1, j_2 \rangle ,\\ (J_{1z} +J_{2z})  \left( \vert j_1,m_1\rangle\otimes\vert j_2, m_2 \rangle\right)  =& (m_1+m_2) \left( \vert  j_1,m_1\rangle\otimes\vert j_2, m_2 \rangle\right). \end{aligned}
\end{eqnarray}    
  The non-trivial question is \textit{what are the possible values for $j$ ?}
  
Without loss of generality, let us consider the case $j_1\geq j_2$. Using (2.10) and (2.14) it is clear that the values of $m$ range from max($m_1+m_2$) to min($m_1+m_2$), \textit{i.e.}, $j_1+j_2$ to $-(j_1+j_2)$.  With the values of $j_1$, $j_2$ fixed, the degeneracy $g_{j_{1},j_{2}}(m)$ of  $m$ is given by the number of ways one can obtain the value $m$ from algebraic addition of $m_1$ and $m_2$. It is quite straightforward to see that 
\begin{eqnarray*}
\begin{aligned}
&g_{j_{1},j_{2}}(m=j_1+j_2)=1 \hspace*{1cm} \text{ $m_1=j_1,m_2=j_2$},\\
&g_{j_{1},j_{2}}(m=j_1+j_2-1)=2 \hspace*{.3cm} \text{ $m_1=j_1-1,m_2=j_2$ or $m_1=j_1,m_2=j_2-1 $},\end{aligned}
\end{eqnarray*}\begin{eqnarray}
\begin{aligned}&g_{j_{1},j_{2}}(m=j_1+j_2-2) =3 \hspace*{.3cm} \text{ $m_1=j_1-2,m_2=j_2$}\\
 &\hspace*{5.6cm}\text{or $m_1=j_1,m_2=j_2-2 $}\\ &\hspace*{5.6cm}\text{or $m_1=j_1-1,m_2=j_2-1$}\\
& \hspace*{5cm} \vdots \\ 
&g_{j_{1},j_{2}}(0 \leq m \leq j_1-j_2)=2j_2+1.\end{aligned}
\end{eqnarray} 
 The degeneracy cannot increase beyond $2j_2+1$, because $m_2< -j_2$ is not allowed. Furthermore, by symmetry, $g_{j_{1}j{_{2}}}(-m)=g_{j_{1}j{_{2}}}(m)$.

The invariant subspace $\xi(j)$ of angular momentum $j$ contains ($2j+1$) states with $|m|\leq j$. Since the tensor product contains no state with $m>j_1+j_2$, and the state $m=j_1+j_2$ only once, we can infer that there exists no invariant subspace of $\xi(j_1) \otimes \xi(j_2)$ with value $j>j_1+j_2$, and only one invariant subspace with $j=j_1+j_2$. Now let $k_{j_{1},j_{2}}(j)$ be the degeneracy of the invariant subspaces $j \in \xi(j_1) \otimes \xi(j_2) $ (with $j_1,j_2$ fixed). Since for all $m \in \xi(j)$ each $m$ occurs only once, we have the simple relation
\begin{eqnarray}
g_{j_{1},j_{2}}(m)=k_{j_{1},j_{2}}(j=|m|)+k_{j_{1},j_{2}}(j=|m|+1)+k_{j_{1},j_{2}}(j=|m|+2)+...
\end{eqnarray}
 By writing the above relation for all possible values of $m$ and inverting, we get
 \vspace*{-.4cm}\begin{eqnarray}
 \begin{aligned}
 k_{j_{1},j_{2}}(j) &=& g_{j_{1},j_{2}}(m=j)- g_{j_{1},j_{2}}(m=j+1).
  \end{aligned}
 \end{eqnarray}
 Iterative solution gives, using (2.15),
  \begin{eqnarray}
 \begin{aligned}
 &k_{j_{1},j_{2}}(j>j_1+j_2)=0  ,\\
 &k_{j_{1},j_{2}}(j=j_1+j_2)=g_{j_{1},j_{2}}(m=j_1+j_2)=1 ,\\
 &k_{j_{1},j_{2}}(j=j_1+j_2-1)=g_{j_{1},j_{2}}(m=j_1+j_2-1)-g_{j_{1},j_{2}}(j=j_1+j_2)=1,\\
 & \hspace*{5cm} \vdots \\
 &k_{j_1,j_2}(j<j_1-j_2)=0.
 \end{aligned}
 \end{eqnarray}
We thus find that the degeneracies of all the invariant subspaces of $\xi(j_1) \otimes \xi(j_2)$ is 1, with the allowed values of $j$ being $j_1+j_2, j_1+j_2-1,....,j_1-j_2$. Overall, we have the result
\vspace*{-.4cm} \begin{eqnarray}
   \xi(j_1) \otimes \xi(j_2) = \xi(j_1+j_2) \oplus \xi(j_1+j_2-1) \oplus......\oplus \xi(|j_1-j_2|).  
\end{eqnarray} 

\vspace*{-.4cm} It is easily verified that the sum of the dimensions of the vector spaces on RHS equals the dimension of the vector space on LHS. 
\begin{eqnarray}\begin{aligned}
\sum_{j=j_1-j_2}^{j_1+j_2} Dim(\xi(j))=\sum_{j=j_1-j_2}^{j_1+j_2} (2j+1)=& \sum_{j=j_1-j_2}^{j_1+j_2}[(j+1)^2-j^2]\\ =&(j_1+j_2+1)^2-(j_1-j_2)^2\\ =&(2j_1+1)(2j_2+1) \\ =& Dim(\xi(j_1) \otimes \xi(j_2))\end{aligned}
\end{eqnarray}
Now, let us see how to obtain the unitary transformation that relates the tensor product basis to the total angular momentum basis and vice versa. 

 \subsection{Clebsch-Gordan coefficients and recurrence relations }
  When we express the total angular momentum basis  $\vert jm;j_1j_2\rangle$ in terms of the tensor product basis $\vert j_1m_1\rangle \vert j_2 m_2\rangle$, the coefficients that appear in the expansion are called the \textit{Clebsch-Gordan (CG)} coefficients, denoted by $C _{m m_1 m_2}^{j j_1 j_2} $:
 \begin{eqnarray}
 \vert j m;j_1 j_2\rangle = \sum_{j_1,j_2,m_1,m_2} C _{m m_1 m_2}^{j j_1 j_2} \vert j_1 m_1 \rangle \vert j_2 m_2 \rangle  
 \end{eqnarray} These coefficients are non-zero only for $ m=m_1+m_2 $ and $ |j_1-j_2| \leq j \leq j_1+j_2$. Here both the LHS and the RHS individually span orthonormal Hilbert spaces. Hence the CG-coefficients are defined as\\
\vspace*{-.4cm} \begin{equation}
 C _{m m_1 m_2}^{j j_1 j_2}= \langle j_1m_1;j_2,m_2 \vert j m;j_1j_2 \rangle.
 \end{equation}
  It is possible to choose a phase convention \citep{condon-shortley} such that all the CG-coefficients are real. Then $\langle j_1,m_1,j_2,m_2 | j, m, j_1,j_2\rangle = \langle j, m, j_1,j_2| j_1,m_1,j_2,m_2 \rangle$,  and we also have the inverse relation,
  \vspace*{-.4cm}\begin{eqnarray}
 \vert j_1 m_1 \rangle \vert j_2 m_2 \rangle = \sum_{j,m} C _{m m_1 m_2}^{j j_1 j_2} \vert j m;j_1 j_2\rangle. 
 \end{eqnarray}\vspace*{-.4cm}Furthermore, orthonormality  of the states (2.11) gives the constraints
  \begin{subequations}\begin{eqnarray}
 & \sum_{m_1 = -j_1}^{j_1} \sum_{m_2=-j_2}^{j_2} C _{m m_1 m_2}^{j j_1 j_2} C _{m' m_1 m_2}^{j' j_1 j_2} =\delta_{j j'} \delta_{m m'},  \\ 
  &\sum_{j=|j_1-j_2|}^{j_1+j_2} \sum_{m=-j}^{j} C _{m m_1 m_2}^{j j_1 j_2} C _{m m'_1 m'_2}^{j j_1 j_2} = \delta_{m_1 m'_1} \delta_{m_2 m'_2}  .
 \end{eqnarray}\end{subequations}

\vspace*{-.4cm}We can calculate the CG-coefficients using the ladder operators. Let us consider the action of $J_-= J_{1-}+J_{2-}$ on the total angular momentum basis state $\vert j,m;j_1 j_2\rangle$, 
   \vspace*{-.4cm}\begin{eqnarray}
   J_- \vert j,m;j_1 j_2\rangle = C_-(j,m)\vert j,m-1;j_1 j_2\rangle.
   \end{eqnarray}
  \vspace*{-.4cm} Now, using the relations (2.13) and (2.21), we have
   \begin{eqnarray}\begin{aligned}
   C_-(j,m)\vert j,m-1;j_1 j_2\rangle =& \sum_{j_1,j_2,m_1,m_2} C _{m m_1 m_2}^{j j_1 j_2} \big( C_-(j_1,m_1)\vert j_1,m_1-1\rangle \vert j_2,m_2\rangle \\ +& C_-(j_2,m_2)\vert j_1,m_1\rangle \vert j_2,m_2-1\rangle \big) .
\end{aligned}\end{eqnarray}  
Multiplying this with $\langle j_1 m_1'\vert \langle j_2 m_2'\vert $,  with $m=m_1'+m_2'+1$: \footnotesize
\begin{eqnarray}
\begin{aligned}
   C_-(j,m)\langle j_1 m_1';j_2 m_2'\vert j,m-1;j_1j_2\rangle = C_-(j_1,m_1'+1)\langle j_1,m_1'+1; j_2,m_2'\vert j,m;j_1j_2\rangle \\ + C_-(j_2,m_2'+1)\langle j_1,m_1'; j_2,m_2'+1\vert j,m;j_1j_2\rangle.
\end{aligned}  
\end{eqnarray} \normalsize
 Similarly we have another recurrence relation, following from the action of $ J_+= J_{1+}+ J_{2+}$ on $\vert j,m;j_1,j_2\rangle$, \footnotesize
 \begin{eqnarray}
\begin{aligned}
  C_+(j,m)\langle j_1 m''_1;j_2 m_2''\vert j,m+1;j_1j_2\rangle = C_+(j_1,m''_1-1)\langle j_1,m''_1-1; j_2,m''_2\vert j,m;j_1j_2\rangle \\ + C_+(j_2,m''_2-1)\langle j_1,m''_1; j_2,m''_2-1\vert j,m;j_1j_2\rangle ,
\end{aligned}  
\end{eqnarray} \normalsize 
 with $ m=m''_1+m''_2-1 $. We already know $C_\pm(j,m)$ from (2.12). Therefore, starting with a particular known CG-coefficient, the others can be obtained from these recurrence relations. The standard phase convention \citep{condon-shortley} is that $ C _{j m_1 m_2}^{j j_1 j_2} \geq 0 $ and real. Thereafter, application of the recurrence relations ensures that all the other CG-coefficients are real.
\subsection{\textbf{\textit{J+S}} problem}
Let us now solve the simple case of adding spin-$\frac{1}{2}$ to spin-$j$. This problem is sufficient for constructing our desired algorithm to generate eigenstates of any $SU(2)$ representation.

From the rules of combining angular momenta, the only allowed combinations are the states with $j'=j+1/2$ and $j'=j-1/2$ (when $j>0$). For a particular combination with $J_z$-eigenvalue $m$ , we have 
  \begin{subequations}\begin{eqnarray}
  \vert j+\dfrac{1}{2},m; j,\dfrac{1}{2}\rangle = \alpha \vert j,m-\dfrac{1}{2}\rangle\vert \dfrac{1}{2},\dfrac{1}{2}\rangle + \beta\vert j,m+\dfrac{1}{2}\rangle\vert \dfrac{1}{2},-\dfrac{1}{2}\rangle,  \\ \nonumber\\
  \vert j-\dfrac{1}{2},m; j,\dfrac{1}{2}\rangle = \alpha' \vert j,m-\dfrac{1}{2}\rangle\vert \dfrac{1}{2},\dfrac{1}{2}\rangle + \beta'\vert j,m+\dfrac{1}{2}\rangle\vert \dfrac{1}{2},-\dfrac{1}{2}\rangle . 
  \end{eqnarray}\end{subequations}
 To determine the four real coefficients ($\alpha, \alpha', \beta, \beta'$) we need four equations. Three of them are the orthonoramilty conditions, as per (2.24):
 \vspace*{-.4cm}\begin{subequations}\begin{eqnarray}
 \alpha^2+\beta^2 = 1,\\   
 \alpha'^2+\beta'^2 =1, \\
 \alpha\alpha'+\beta\beta' =0.
 \end{eqnarray} 
  \end{subequations}
 The fourth equation is the equality that results from (2.29), when the two sides are acted upon by \begin{equation}
 J'^2=(J + S)^2= J^2+S^2+ 2J_zS_z+ J_+S_- + J_-S_+ , 
\end{equation}  in the total angular momentum basis and the tensor product basis respectively. Explicitly, (2.29a) gives
\footnotesize
\begin{eqnarray}
\begin{aligned}
\left[(j+\frac{1}{2})(j+\frac{3}{2})\right] \vert j+\dfrac{1}{2},m; j,\dfrac{1}{2}\rangle =& \left( \alpha \left[ j(j+1) +\frac{1}{4} +m)\right] + \beta \left[\sqrt{(j+m+\frac{1}{2})(j-m+\frac{1}{2})}\right] \right) \\ &\times\vert j,m-\dfrac{1}{2}\rangle\vert \dfrac{1}{2},\dfrac{1}{2}\rangle\\ +&\left( \beta \left[ j(j+1) +\frac{1}{4} -m)\right] + \alpha \left[\sqrt{(j+m+\frac{1}{2})(j-m+\frac{1}{2})}\right] \right) \\ &\times\vert j,m+\dfrac{1}{2}\rangle\vert \dfrac{1}{2},-\dfrac{1}{2}\rangle
\end{aligned}
\end{eqnarray}
\normalsize
Again using (2.29a) in the LHS of (2.32), and then equating the coefficients of the kets, we get the relation.\vspace*{-.4cm}
\begin{eqnarray}
\left(\frac{\alpha}{\beta}\right)=\left(\dfrac{j+m+1/2}{j-m+1/2}\right)^{\frac{1}{2}}
\end{eqnarray}   
A similar analysis based on (2.29b) gives,\vspace*{-.4cm}
\begin{eqnarray}
\left(\frac{\alpha'}{\beta'}\right)=-\left(\dfrac{j-m+1/2}{j+m+1/2}\right)^{\frac{1}{2}}
\end{eqnarray}  
The orthonormality conditions (2.30), together with (2.33) and (2.34), thereafter uniquely determine the coefficients: \vspace*{-.2cm}
 \begin{eqnarray} \begin{aligned}
 \alpha = \sqrt{\dfrac{j+m+1/2}{2j+1}}=\beta', \hspace*{2cm} \beta = \sqrt{\dfrac{j-m+1/2}{2j+1}}=-\alpha' . \end{aligned}
\end{eqnarray}  
 These are the Clebsch-Gordan coefficients connecting  the total angular momentum basis to the corresponding tensor product basis.
 \section{Computational basis for $SU(2)$}
In Chapter 1, we defined the standard computational basis.When specialized to the $SU(2)$ group, each building block becomes a spin-$\frac{1}{2}$ system, which is called a qubit. The eigenstates of a qubit are easily mapped to classical bits. The total angular momentum of a collection of spin-$\frac{1}{2}$ systems is a half integer or an integer. The whole state  $\vert j,m\rangle$ is obtained as the tensor product of 2$j$ qubits,
   \begin{equation}
   \vert j,m\rangle = \sum_{k=1}^n C_k[\bigotimes_{i=1}^{2j} \vert \dfrac{1}{2},m_i\rangle_i]_k , 
   \end{equation}
where $m_i$ is the $J_z$-component of $i^{th}$  spin-$\frac{1}{2}$ system, and $C_k$ is the CG-coefficient of the $k^{th}$ term in the tensor product basis. The coefficients $C_k$ can be calculated using (2.35) in a recursive manner. The label $k$ goes from 1 to $n$, where $n$ is the possible number of ways of obtaining $m$ from $2j$ spin-$\frac{1}{2}$ systems. Some values of $ C_k$ might vanish when evaluated using the recurrence relations. 

Now, instead of writing the individual spin-$\frac{1}{2}$ systems  as $\vert \frac{1}{2},\pm\frac{1}{2}\rangle$,  we choose the computational basis notation, mapping $ m=+1/2$ to $\vert 1\rangle$ and $m=-1/2$ to $\vert 0\rangle$. This becomes a simple convention whereby the  2$j$+1 states of a spin-$j$ system are denoted by $\vert j+m \rangle$ instead of $\vert m \rangle$. Then (2.36) transforms to 
 \begin{equation}
   \vert j,m\rangle = \sum_{k=1}^n C_k[\bigotimes_{i=1}^{2j} \vert  d \rangle_i]_k  \equiv \sum_{k=1}^n C_k \vert d_1d_2...d_{2j} \rangle_k
   \end{equation}
   where each $d_k$ is either a 0 or 1. For example,  using (2.35) and (2.29a) for the state $j=1,m=0$, we have $$\vert 1,0\rangle =\sqrt{\dfrac{1}{2}}\hspace*{.2cm} [\vert 1 \rangle\vert0\rangle+\vert 0 \rangle\vert 1 \rangle] .$$ 
\section{Young diagrams and Schur basis}
The Schur-Weyl duality ~\citep{schur-weyl}, introduced  in Chapter 1, assures that there exists a decomposition of the  N dimensional Hilbert space ($N = 2^n$ for a Hilbert space of $n$ qubits), such that the joint action of both unitary ($U_N$) and permutation ($S_n$) transformations is reduced to transformations on the subspaces corresponding to their
irreducible representations, 
 \vspace*{-.4cm} \begin{eqnarray}
  (\mathbb{C}^2 )^{\otimes n} \cong \bigoplus_{\lambda\in Part[n,2]} \gamma_{_{\lambda}} \otimes S_\lambda .
\end{eqnarray} 
Here $ \gamma_{_{\lambda}}$ is an irreducible representation of $SU(2)$ (\textit{i.e}, a spin-$j$ representation), and $ S_\lambda$ is an irreducible representation of $ S_n$  (\textit{i.e}, a Young diagram [Appendix A]). The index $\lambda$ is specified by the  partition $\lambda_1,\lambda_2$, satisfying the conditions 
  \begin{eqnarray}
  \lambda_1 \geq \lambda_2\geq0 \hspace*{2cm} \text{and}  \hspace*{2cm} \lambda_1+\lambda_2=n .  
\end{eqnarray} 
 
\vspace*{-.4cm}The subspaces on the RHS of equation (2.38) are spanned by the Schur basis states $\vert \lambda_1,\lambda_2, q_\lambda, p_\lambda \rangle_{sch}$, as discussed in  Chapter 1. Also, every Schur basis state is associated with a unique Young tableau [Appendix A].  
The value $j$ for the unitary representation $\gamma_\lambda$ is given by  
\begin{equation}
j =\frac{\lambda_1 -\lambda_2}{2} .  
\end{equation} $ \vert q_\lambda \rangle \in \gamma_{_{\lambda}} $
 represents the $m $ value of spin-$j$, and $\vert p_\lambda \rangle\in S_\lambda$ represents the way the Young diagram is built up in a stepwise process adding one box at a time.  

For example, for the Hilbert space $(\mathbb{C}^2 )^{\otimes 2}$  of 2 qubits, we have $$\vert i_1,i_2\rangle \Longleftrightarrow\vert \frac{1}{2},m_1\rangle \otimes \vert \frac{1}{2},m_2 \rangle $$ where, $ i_1, i_2$ are either 0 or 1 in the computational basis, and $m_1, m_2$ are $\pm\frac{1}{2}$.  There are four possible states. From the angular momentum algebra of (2.31) and (2.21), we can say that they are the spin-1 triplet and the spin-0 singlet states. The singlet and triplet states are also the basis states for the  irreducible representations of $S_2$, \textit{i.e.} symmetric or anti-symmetric under the exchange of two identical particles. The symmetric and antisymmetric representations are the Young diagrams (2,0) and (1,1) respectively.\footnotesize $$
\Yvcentermath1 \young(1) \otimes \young(2)=\young(12) \oplus \young(1,2) 
$$  
\normalsize
As per the rules of Young tableau [Appendix A], the symmetric representation has 3 valid Young tableaux and the antisymmetric one has 1 valid Young tableau. These tableaux represent the component states with each box representing a qubit.\\\footnotesize
 \begin{eqnarray}\begin{aligned}
 \Yvcentermath1 \young(12)&  \leftrightarrow
\begin{array}{rr}
\tiny\Yvcentermath1 \young(++)\hspace*{0.4cm} \vert 1,+1\rangle  =& \tiny{  \vert 11 \rangle}   \\
\tiny\Yvcentermath1 \young(+-)\hspace*{0.4cm}\vert 1,\hspace*{.3cm} 0\rangle  =& \tiny{ \sqrt{\frac{1}{2}} (\vert 10 \rangle + \vert 01 \rangle)} \\
\tiny\Yvcentermath1 \young(--)\hspace*{0.4cm}\vert 1,-1\rangle  =& \tiny{\vert 00 \rangle}  
\end{array}\\\\ 
\Yvcentermath1 \young(1,2)&  \longleftrightarrow
\begin{array}{rrr}
\tiny \hspace*{.3cm}\Yvcentermath1 \young(+,-) &\vert 0,0\rangle  =& \tiny{ \sqrt{\frac{1}{2}} \vert 10 \rangle - \vert 01 \rangle} 
\end{array}\end{aligned}
\end{eqnarray}\normalsize

Next,  consider a 3-qubit Hilbert space with  8 possible eigenstates . 
\begin{equation}
(\mathbb{C}^2 )^{\otimes 3} \cong  \big(\gamma_{_{\frac{3}{2}}} \otimes S_{(3,0)}\big) 
 \oplus   2\big(\gamma_{_{\frac{1}{2}}} \otimes S_{(2,1)} \big)
\end{equation} 
The quadruplet spin-$\frac{3}{2} $ appears only once here. But the spin-$\frac{1}{2}$ doublet appears twice because $ S_{(2,1)}$ has two possibilities, one symmetric and the other antisymmetric, under the exchange of the first two identical particles.\footnotesize 
$$ \Yvcentermath1 \young(1) \otimes \young(2) \otimes \young(3) = \left( \young(12) \oplus \young(1,2)\right)\otimes \young(3)= \left( \young(123) \oplus \young(12,3)\right)  \oplus \young(13,2)  $$\normalsize 
Using  recurrence relations of the  CG-coefficients, we have
\begin{subequations}
 \begin{eqnarray}
\Yvcentermath1 \young(123)  \leftrightarrow 
 & \begin{array}{rr}
\tiny\Yvcentermath1 \young(+++) \hspace*{0.2cm} \vert \frac{3}{2},+\frac{3}{2}\rangle  =& \tiny{  \vert 111 \rangle } \\
\tiny\Yvcentermath1 \young(++-)\hspace*{0.2cm}\vert \frac{3}{2},+\frac{1}{2}\rangle = &\sqrt{\frac{1}{3}} ( \vert 110 \rangle + \vert 011\rangle  +  \vert 101 \rangle)  \\
\tiny\Yvcentermath1 \young(+--)\hspace*{0.2cm}\vert \frac{3}{2},-\frac{1}{2}\rangle = &\sqrt{\frac{1}{3}}( \vert 001\rangle + \vert 100\rangle + \vert 010 \rangle)\\
\tiny\Yvcentermath1 \young(---)\hspace*{0.2cm}\vert \frac{3}{2},-\frac{3}{2}\rangle  =& \tiny{  \vert 000 \rangle }
\end{array} &\\\nonumber\\\nonumber\\
\Yvcentermath1 \young(12,3) \leftrightarrow 
&\begin{array}{rr}
\hspace*{0.2cm}\tiny\Yvcentermath1 \young(++,-)\hspace*{0.2cm}\vert \frac{1}{2},+\frac{1}{2}\rangle \hspace*{0.6cm}  =& \tiny{\sqrt{\frac{2}{3}}  \vert 110 \rangle -\sqrt{\frac{1}{6}} (\vert 101\rangle  + \vert 011\rangle)} \\
\tiny\Yvcentermath1 \young(+-,-)\hspace*{0.2cm}\vert \frac{1}{2},-\frac{1}{2}\rangle \hspace*{0.6cm}= &\sqrt{\frac{2}{3}}  \vert 001 \rangle -\sqrt{\frac{1}{6}}  (\vert 010\rangle  +  \vert 100   \rangle)  
\end{array} &
%\end{aligned}
\end{eqnarray} 
\begin{eqnarray}
%\begin{aligned} 
 \Yvcentermath1 \young(13,2)  \leftrightarrow
&\begin{array}{cc}
\tiny\Yvcentermath1 \young(++,-)\hspace*{0.2cm}\vert \frac{1}{2},+\frac{1}{2}\rangle  =& \tiny{\sqrt{\frac{1}{2}}  (\vert 101 \rangle -\vert 011\rangle) }  \\
\tiny\Yvcentermath1 \young(+-,-)\hspace*{0.2cm}\vert \frac{1}{2},-\frac{1}{2}\rangle = &\tiny{\sqrt{\frac{1}{2}}  (\vert 010 \rangle -\vert 100\rangle) }
\end{array}
%\end{aligned}
\end{eqnarray} 
\end{subequations}
In the above equation 1,2,3 represent the order of addition of boxes. For example, in the Young diagram \scriptsize$\young(12,3)$ \normalsize,  the $1^{st}$ and $2^{nd}$ spins appearing in the state are symmetric under exchange. That is not specific to a particular state, but it is true for all the states of that irreducible representation. Symmetry or antisymmetry is the property of the irreducible representation (Young diagram ), and not just the states (Young tableaux). Going further, the $1^{st}$ and $3^{rd}$ spins in \scriptsize$\young(12,3)$ \normalsize are not antisymmetric though they appear in a column. Such representations are said to have mixed symmetry.

Every bit of the label $ \vert p_\lambda \rangle$,  either 1 or 0, tells us whether a box is  added to the Young diagram in the first row or the second row respectively. For an $n$-box Young diagram, we sequentially add n-1 boxes to the first box,  and so we have a register of ($n$-1) bits as  $ \vert p_\lambda \rangle$. The Schur basis for an $n$-qubit system is then specified by the convention, $$ \vert (\lambda_1,\lambda_2);q=j+m;p_1,....,p_{n-2},p_{n-1}\rangle.$$ 

For the examples considered above: \footnotesize
\begin{eqnarray}\begin{aligned}
 \Yvcentermath1 \young(12)  &\longleftrightarrow & \vert (2,0);q;1\rangle_{sch}  \\  \\
 \Yvcentermath1 \young(1,2)  &\longleftrightarrow & \vert (1,1);q;0\rangle_{sch} \\ \\
\Yvcentermath1 \young(123)  &\longleftrightarrow & \vert (3,0);q;1,1\rangle_{sch} \\ \\
\Yvcentermath1 \young(12,3)  &\longleftrightarrow & \vert (2,1);q;1,0\rangle_{sch} \\ \\
\Yvcentermath1 \young(13,2)  &\longleftrightarrow & \vert (2,1);q;0,1\rangle_{sch}\end{aligned}
\end{eqnarray}
 \normalsize
\section{Algorithm and circuit to generate eigenstates of $SU(2)$ representations}
 In the previous section,  the relationship between the computational basis and the Schur basis was established with the help of Young tableaux. In this section, we see how that helps us in constructing angular momentum eigenstates in the computational basis, using the inverse Schur transform $U_{sch}^{-1}$.
 \subsection{CG-transform ($U_{CG}$) and  Schur transform ($U_{sch}$)}
 $U_{sch}$ is the unitary operation that transforms the computational basis to the Schur basis. Its basic unit is the unitary  Clebsch-Gordan transform $U_{CG}$.  $U_{CG}$ provides the change in the irreducible representation states as qubits are added to an existing state, one at a time. When $U_{CG}$ is recursively operated on a computational basis state, formed by a register of $n$-qubits, we obtain a superposition of Schur basis states.

  The matrix elements of $U_{CG}$ are the CG-coefficients. These CG-coefficients relate the way states of $k$-qubits get combined with one more qubit, giving rise to ($k+1$) qubit states. In terms of angular momentum algebra it is  the \textit{\textbf{J+S}} problem. In terms of Young tableaux, it  describes the way one more box is added to a Young tableau of $k$ boxes, to give another valid Young tableau of ($k+1$) boxes. This addition specifies whether the new box is symmetrised or antisymmetrised with the earlier boxes, \textit{i.e.} whether $p_k$ is 1 or 0.
  
From (2.29) and (2.35), the CG-transform is given by
 \begin{eqnarray}
 \left[ \begin{array}{c}
 \vert j+1/2,m,p=1\rangle  \\
 \vert j-1/2,m,p=0\rangle 
 \end{array}
 \right]
 = \begin{bmatrix}
 \cos\theta & \sin\theta \\ -\sin\theta & \cos\theta
 \end{bmatrix}  \left[ \begin{array}{cc}
 \vert j,m-1/2\rangle\vert1/2,+1/2\rangle \\
 \vert j,m+1/2\rangle\vert1/2,-1/2\rangle  
 \end{array}
 \right]  ,
\end{eqnarray}
 $ \mbox{with} \hspace*{.5cm} \cos\theta = \sqrt{\dfrac{j+m+1/2}{2j+1}}.$\\  In this form, the spin-$j$ system combines with a qubit, with the information contained in  $\vert p_\lambda \rangle$, to give LHS. In the process, the computational basis states on RHS are transformed to the Schur basis on LHS. The complete Schur basis state needs the information of ($n-1$) bits in $\vert p_\lambda \rangle$, and we have to cascade  $U_{CG}$ ($n$-1) times in the complete Schur transform.
 
  The complete information to specify the Schur basis eigenstate can be put together as follows. Let the desired $n$-qubit state be $ \vert j,m\rangle$. Then
 \begin{itemize}\vspace*{-.3cm}
 \item $\lambda_1,\lambda_2$ are determined uniquely by the equations (2.39), (2.40).
 \vspace*{-.3cm}\item $q = j+m$, which is easy to encode into classical bits. With known $j$, there is no loss of generality. 
 \vspace*{-.3cm}\item The bits of $\vert p_{\lambda} \rangle$ are determined by knowing how the spin-$j$ is built from spin-$\frac{1}{2}$ components. The following diagram illustrates the values for $p_k$, for the tensor product of 5 qubits.
 \begin{figure}[H]
\centering
\includegraphics[scale=.53]{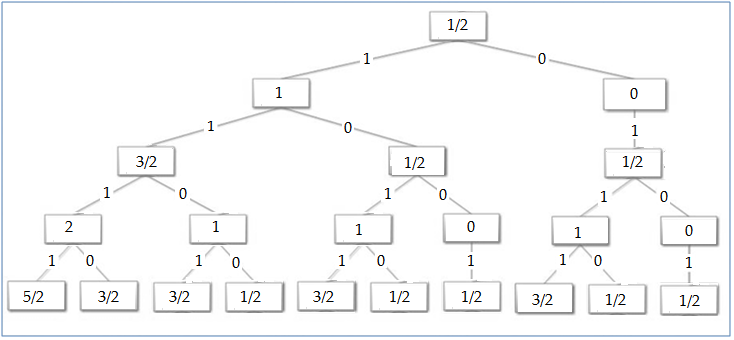}
\caption{\small{Construction of tensor product states for 5 qubits. The numbers in the boxes are $j$'s. $p_k$'s are the numbers in the lines connecting the boxes.}}
\end{figure}
 \vspace*{-.3cm}\item Note that the same $\vert j,m \rangle$ may be obtained in many ways for $n$-qubits, using different sequences of $\vert p_\lambda \rangle$. A particular $\vert p_{\lambda} \rangle$  picks  a unique construction route for the desired state. 
\end{itemize}
 
\subsection{Algorithm for $U_{sch}^{-1}$}
Our aim  is to express the total angular momentum eigenstates in terms of the computational basis states. We achieve that by implementing $U_{sch}^{-1}$. From preceding discussion, $U_{sch}^{-1}$  can be explained as the transformation which outputs a superposition of computational basis states, given any  Schur basis state as the input. The operation again consists of ($n-1$) recursive steps, with one  $p_k$ and $U_{CG}^{-1}$ used at each step. In terms of the flowchart in Figure \ref{e1}, this gets translated  as follows:
\begin{figure}[H]
\centering
\includegraphics[scale=.56]{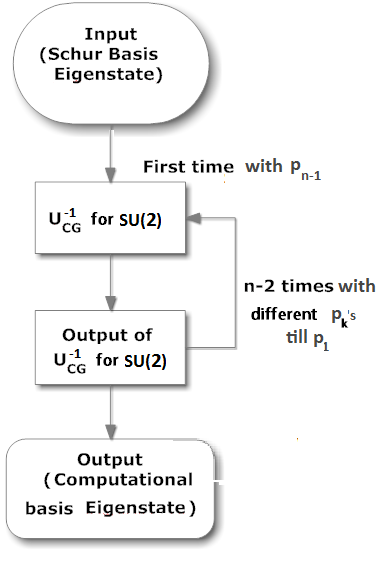}
\caption{Flow chart for implementing $U_{sch}^{-1}$ \label{e1}}
\end{figure}
 \begin{itemize}\vspace*{-.4cm}
\item \textbf{Input:} Schur basis description of the desired eigenstate.\vspace*{-.4cm} $$ \vert \lambda_1,\lambda_2;q;p_1,p_2,...p_{n-1} \rangle$$
\item \textbf{Processing:} 
\begin{itemize}
\item In the first iteration, we transform  $\lambda_1, \lambda_2, q, p_{n-1}$ using the CG-transform, leaving the other   $p_k$  unchanged. \end{itemize}
   \small   
\begin{eqnarray}
\begin{aligned}
&\vert \lambda; q_\lambda;p_\lambda\rangle_{sch} \hspace{0.2cm} \xrightarrow{\hspace{0.1cm}U_{CG}^{-1}}  \hspace{0.2cm} \vert \lambda';q_\lambda'; p'_\lambda \rangle_{sch} \\
& = \vert \lambda_1- p_{n-1},\lambda_2 - 1+p_{n-1};q+1-p_{n-1}-p_{n-1}';p_1...., p_{n-2}, p'_{n-1} = R_Y(\theta)p_{n-1}\rangle
\end{aligned}
\end{eqnarray}  \normalsize 
\begin{itemize}
\item We run the subsequent iterations of CG-transform with $\lambda_1', \lambda_2', q'$ (which are the outputs of previous iterations) and a new $p_k$.
 \item We run totally ($n-1$) iterations, \textit{i.e.} till $p_1$ is used. 
 %$\vert p_k' \rangle = \vert i_k \rangle$ and $q'=\vert i_0\rangle$ are combined into the output.
   \end{itemize}
\item \textbf{Output:} Desired eigenstate is obtained in the computational basis as the $n$-qubit register formed by $\vert p_k' \rangle = \vert i_k \rangle$ and $q'=\vert i_0\rangle.$ At the end of the execution, $ \vert \lambda_1,\lambda_2 \rangle$ gets reduced to $\vert 1,0 \rangle$.
% and the $n$ qubits replace ($n-1$) $\vert p_\lambda \rangle$'s and $\vert q \rangle$.
\end{itemize}
\subsection{Efficient quantum circuit for $U_{sch}^{-1}$}
To explicitly demonstrate that $U_{sch}^{-1}$ can be implemented efficiently, we now convert the above algorithmic description into a logic circuit. First we break up $U_{sch}^{-1}$ into ($n-1$) $U_{CG}^{-1}$ blocks, as schematically shown below. 
\begin{figure}[H]
 \centering
 \includegraphics[scale=.7]{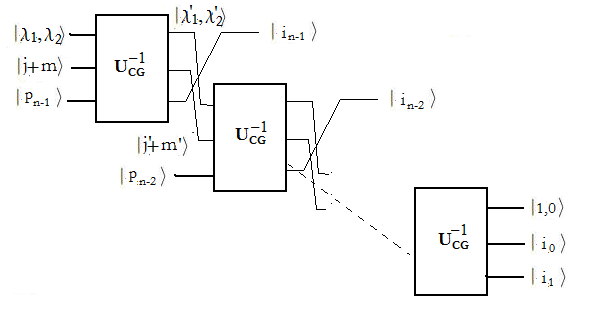}
 \caption{\footnotesize Decomposition  of $U_{sch}^{-1}$   into $U_{CG}^{-1}$ blocks for $SU(2)$. \normalsize}
\end{figure} The algorithm ends  with $\vert \lambda_{1}, \lambda_{2} \rangle = \vert 1, 0 \rangle$, 
 implying that the final state is a spin-$\frac{1}{2}$ representation. 
 The corresponding value of $q=i_{0}$ is  either 1 or 0, which stands for a $+\dfrac{1}{2}$ or $-\dfrac{1}{2}$  spin component. Together with all the $\vert p_k \rangle$ that get converted to the $\vert i_k \rangle$, we have the complete computational basis state formed by $\vert i_{n-1}\rangle...\vert i_0 \rangle$. \\

The essential  task is to construct an efficient circuit for $U_{CG}^{-1}$, in terms of logic gates. Since $U_{CG}$ is a unitary matrix, from (2.45) we have 
 \footnotesize  \begin{eqnarray}
  \left[ \begin{array}{c}
  \vert j,m-1/2\rangle \otimes \vert 1\rangle \\
 \vert j,m+1/2\rangle \otimes\vert 0 \rangle 
 \end{array}
 \right]=R_Y(\theta)
 \left[ \begin{array}{c}
 \vert j+1/2,m,p=1\rangle \\
 \vert j-1/2,m,p=0\rangle 
 \end{array}
 \right] , 
\end{eqnarray}\normalsize 
 $$ \mbox{where} \hspace*{.5cm} R_Y(\theta)=\begin{bmatrix}
 \cos\theta & -\sin\theta \\ \sin\theta & \cos\theta
 \end{bmatrix}, \hspace{.2cm} \mbox{with}\hspace{.2cm} \theta = \cos^{-1}\left(\sqrt{\dfrac{j+m+1/2}{2j+1}}\right) \in [0,\frac{\pi}{2}].$$\\
    This transformation  can be implemented using binary adders, subtractors [see Appendix B], and a unitary rotation gate $R_Y(\theta)$   for a qubit, as illustrated below.
     \begin{figure}[H]
    \centering
   \includegraphics[scale=.77]{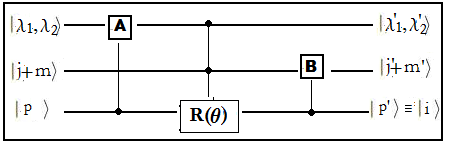}
   \caption{\label{c2}Schematic circuit for $U_{CG}^{-1}$. $\lambda_1, \lambda_2$ and $j+m$ are registers of at most $\log_2 n+1$ bits when there are $n$ boxes in the Young diagram. The qubit $\vert p \rangle$ is rotated to $\vert p' \rangle \equiv \vert i \rangle .$ }
    \end{figure} 
\noindent Here the registers for $\vert \lambda_1,\lambda_2\rangle$ and $\vert j+m\rangle$ need sufficient carry bits to allow addition/subtraction. The bitwise addition/subtraction operations that go into  \textbf{A} and \textbf{B} are  explained below. $\vert p \rangle$ is a single qubit that undergoes a rotation.
\subsubsection*{Detailed circuit implementation of $U_{CG}^{-1}$} 
The diagram in Figure \ref{c2} contains  the following transformations mentioned in (2.46). 
%\begin{itemize}
  \begin{subequations}    
    \begin{eqnarray}
   \vert \lambda_1,\lambda_2 \rangle &\longrightarrow & \vert \lambda_1 -p,\lambda_2 +p-1 \rangle \end{eqnarray}\begin{eqnarray}
   \vert j \rangle &\longrightarrow & \vert j+\frac{1}{2}-p \rangle, \hspace*{1cm} \vert m \rangle \longrightarrow \vert m+\frac{1}{2}-p' \rangle \\
    \vert q \rangle &\longrightarrow & \vert q+1-p-p' \rangle
     \end{eqnarray}
     \end{subequations}
   Using an ancilla bit to store the value of $1-p$, we can construct the following circuit for $U_{CG}^{-1}$ in terms of elementary quantum logic gates \citep{elementary-gates}.
  % \end{itemize}
   \begin{figure}[H]
    \centering
    \framebox{\includegraphics[scale=.5]{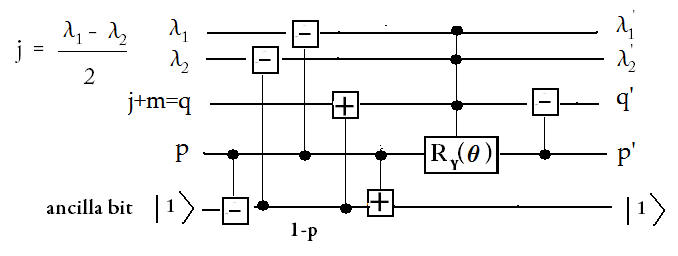}}
    \caption{\label{c3}\small{Detailed Circuit for $U_{CG}^{-1}$}}
    \end{figure}
 \vspace*{-.2cm} \noindent Here $\boxplus$ is a binary adder and  $\boxminus$ is a binary subtractor described in [Appendix B]. $R_Y(\theta)$ is the single qubit rotation gate defined in (2.47). 
    
Explicitly, the controlled additions/subtractions in \textbf{A} are:
  \begin{subequations}   
   \begin{eqnarray}
   C_-\vert p\rangle\vert 1\rangle &=& \vert p\rangle\vert 1-p\rangle \\
   C_-\vert 1-p\rangle\vert \lambda_2\rangle &=& \vert 1-p\rangle\vert \lambda_2-1+p\rangle \equiv \vert  1-p \rangle \vert \lambda_2' \rangle\\
   C_-\vert p\rangle\vert \lambda_1\rangle &=& \vert p\rangle\vert \lambda_1-p\rangle \equiv \vert p \rangle \vert \lambda_1' \rangle \\
   C_+\vert 1-p\rangle \vert q \rangle &=& \vert 1-p\rangle\vert q+1-p \rangle\\
   C_+\vert p\rangle\vert 1-p\rangle &=&\vert p\rangle\vert 1\rangle
   \end{eqnarray}
   \end{subequations}
   with the first operation modifying the ancilla bit and the last one resetting it.
   These operations take $\lambda_1, \lambda_2,q$ values to the ones required to implement the single qubit rotation as a controlled operation.
   % with the first operation modifying the ancilla bit and the last one resetting it. 

The controlled rotation then gives,
   \begin{equation}
 R_Y(\theta)\vert p\rangle = (\cos \theta+ i \sigma_Y\sin \theta )\vert p\rangle =\vert p'\rangle ,
\end{equation}  with cos$\theta $  given by $ \left( \frac{q+(1-p)}{(\lambda_1-\lambda_2)+2(1-p)}\right)^{\frac{1}{2}}= \left( \frac{q'+p'}{\lambda_1'-\lambda_2'+1} \right)^{\frac{1}{2}}$.

The last controlled subtraction in \textbf{B} takes the modified q value to its final value.
 %The final controlled operation takes the modified $\vert j+m \rangle$ register to its correct value.
 \vspace*{-.4cm} \begin{eqnarray}
 C_-\vert p'\rangle\vert q+1-p\rangle &=& \vert p'\rangle \vert q+1-p-p'\rangle = \vert p'\rangle \vert q'\rangle
\end{eqnarray}

\subsection{Examples}
 We give some examples to illustrate how the algorithm works. For convenience, we pad together $q$ and $p_k$ in the state notation, and represent in bold the $q$ and $p_i$ undergoing the $U_{CG}^{-1}$ transformation.
 \small
 \begin{enumerate}
 \item ${n=2, j=0,m=0, p_1=0}$ \\
 \textit{$1^{st}$ iteration} 
 \begin{eqnarray}
 \begin{aligned}
   & \cos\theta=  \sqrt{\frac{q+1-p}{\lambda_1'-\lambda_2'+1}} = \sqrt{\frac{1}{2}} \\
  &\vert 1,1;\textbf{0,0}\rangle \longrightarrow \sqrt{\frac{1}{2}}(\vert 1,0;\textbf{1,0}\rangle - \vert 1,0;\textbf{0,1}\rangle )
 \end{aligned}
 \end{eqnarray}
  \item $ {n=3 ,j=1/2 ,m=1/2 ,p_2=1,p_1=0 }$\\
  \textit{$1^{st}$ iteration} 
\begin{subequations}
 \begin{eqnarray}
 \begin{aligned}
    & \cos\theta= 1 \\
  &\vert 2,1;\textbf{1},0,\textbf{1}\rangle \longrightarrow \vert 1,1;\textbf{0},0,\textbf{1}\rangle  
   \end{aligned}
 \end{eqnarray}
  \textit{$2^{nd}$ iteration} 
  \begin{eqnarray}
 \begin{aligned}
    &\cos\theta=  \sqrt{\frac{1}{2}} \\
  &\vert 1,1;\textbf{0,0},1\rangle \longrightarrow \sqrt{\frac{1}{2}}(\vert 1,0;\textbf{1,0},1\rangle - \vert 1,0;\textbf{0,1},1\rangle)  
  \end{aligned}
 \end{eqnarray} \end{subequations}
  \item $n=3,j=1/2,m=1/2 , p_2=0,p_1=1$ \\\\
  \textit{$1^{st}$ iteration} 
  \begin{subequations}\begin{eqnarray}
 \begin{aligned}
   &\cos\theta =  \sqrt{\frac{2}{3}} \\
& \vert 2,1;\textbf{1},1,\textbf{0}\rangle \longrightarrow \sqrt{\frac{2}{3}}\vert 2,0;\textbf{2},1,\textbf{0}\rangle - \sqrt{\frac{1}{3}}\vert 2,0;\textbf{1},1,\textbf{1}\rangle  
    \end{aligned}
 \end{eqnarray} \\
   \textit{$2^{nd}$ iteration} \\
   \begin{eqnarray}
 \begin{aligned}
&\cos\theta=   1  \\
   &\sqrt{\frac{2}{3}}\vert 2,0;\textbf{2,1},0\rangle \longrightarrow \sqrt{\frac{2}{3}}\vert 1,0;\textbf{1,1},0\rangle  \\   
   &\cos\theta=   \sqrt{\frac{1}{2}}  \\
&\sqrt{\frac{1}{3}}\vert 2,0;\textbf{1,1},1\rangle \longrightarrow   \sqrt{\frac{1}{3}}(\sqrt{\frac{1}{2}} (\vert 1,0;\textbf{0,1},1\rangle + \vert 1,0;\textbf{1,0},1\rangle)) \\   \\  
    \mbox{Together, }
    & \vert 2,1;\textbf{1,1,0}\rangle \longrightarrow \sqrt{\frac{2}{3}}\vert 1,0;\textbf{1,1,0}\rangle - \sqrt{\frac{1}{6}}(\vert 1,0;\textbf{0,1,1}\rangle + \vert 1,0;\textbf{1,0,1}\rangle).\nonumber\\\nonumber\\
    \end{aligned}
 \end{eqnarray} \end{subequations}
\end{enumerate} \normalsize

\section{Complexity analysis}
From the structure of algorithm, we observe the following:
\begin{itemize}
 \item The number of iterations increase linearly with the number of qubits, $n$. Each iteration  implements one $U_{CG}^{-1}$ containing 6 controlled addition/subtraction gates and 1 controlled rotation gate. 
 \item  The resources needed by the algorithm for generating an $n$-qubit state are
 \begin{itemize}
 \item \textit{\textbf{Space:}} Each of $\lambda_1, \lambda_2, q$ are represented by $\log_2 n +1$ qubits. Furthermore, $n-1$ $\vert p_k \rangle$'s and a single ancilla qubit are required to run the algorithm. 

 Therefore, we need $3 (log_{_{2}}n+1)+n $ qubits, which scales as $ \mathcal{O} \left[ log_{_{2}} \left(N\left(log_{_{2}}N\right)^3\right)\right]$ for large $N$.
\item  \textit{\textbf{Time:}} Each addition/subtraction operation involving $\left( log_{_{2}}n+1\right) $ qubit registers uses $\left( log_{_{2}}n+1\right) $ $C_{NOT}$ gates and $\left( log_{_{2}}n\right)$ $C^2_{NOT}$ gates [Appendix B]. 

 We need 3 subtractors, 1 adder for registers, and 2 $C_{NOT}$ gates for the ancilla qubit  in each iteration. Therefore for $n-1$ iterations, the total  gates are: $$
(n-1)\left( 4log_{_{2}}n+6\right)\text{$C_{NOT}$} +(n-1)\left( 4log_{_{2}}n\right)\text{$C^2_{NOT}$ } ,
$$ which scales as
$ \mathcal{O} \left[ log_{_{2}}N\cdot log_{_{2}}\left( log_{_{2}}N\right)  \right]$ for large $N$.
\item \textbf{\textit{Controlled Rotation:}} These are $(n-1)$ in number. The resources needed to implement each one depend on the available hardware, but are independent of $n$. 
 \end{itemize}
 \item Clearly, the space and time resources are polynomial in $n$, and so the algorithm belongs to the class BQP \citep{elementary-gates}.  
 \item  Since the Schur transform and the Schur-Weyl duality are fundamental properties of the representation theory of unitary and permutation groups, the algorithm works for any eigenstate of $SU(2)$ and with any number of qubits.
 
 \end{itemize}

\chapter{$SU(3)$}
%\ifpdf   \graphicspath{{copysu3/copysu3Figs/PNG/}{copysu3/copysu3Figs/PDF/}{copysu3/copysu3Figs/}}
%\else   \graphicspath{{copysu3/copysu3Figs/EPS/}{copysu3/copysu3Figs/}}
%\fi

\section{Introduction}
 $SU(3)$ is the multiplication  group of  $3 \times 3$ unitary matrices of determinant +1. Its generators, traceless hermitian matrices $\hat{X}_i$ ($i=1$ to 8), are often chosen in the fundamental representation as half the Gell-Mann matrices \citep{georgii}, $ X_i=\dfrac{1}{2}g_i$.
 $$ g_1=\begin{pmatrix}
 0&1&0\\1&0&0\\0&0&0 \end{pmatrix}, g_2=\begin{pmatrix}0&-i&0\\+i&0&0\\0&0&0 \end{pmatrix},g_3=\begin{pmatrix}
 1&0&0\\0&-1&0\\0&0&0 \end{pmatrix},g_4=\begin{pmatrix}0&0&1\\0&0&0\\1&0&0 \end{pmatrix}   $$ \\
  \small $$ g_5=\begin{pmatrix}0&0&-i\\0&0&0\\+i&0&0 \end{pmatrix}, g_6=\begin{pmatrix}0&0&0\\0&0&1\\0&1&0 \end{pmatrix},g_7=\begin{pmatrix}0&0&0\\0&0&-i\\0&+i&0 \end{pmatrix},g_8=\sqrt{\dfrac{1}{3}}\begin{pmatrix}1&0&0 \\0&1&0\\0&0&-2 \end{pmatrix}   $$ \\
\normalsize 

The generators satisfy the  commutation and anticommutation relations,
\begin{eqnarray}
[\hat{X}_k,\hat{X}_l]=i\sum_{m=1}^8f_{_{klm}}\hat{X}_m , \hspace*{1.7cm} \lbrace \hat{X}_k,\hat{X}_l \rbrace= \dfrac{1}{3} \delta_{kl}\textbf{I}+ \sum_{m=1}^8d_{_{klm}}\hat{X}_m    ,
\end{eqnarray}   
where the structure constants $f_{_{klm}}$ are real and fully antisymmetric, and $d_{_{klm}}$ are real and fully symmetric. The  non-vanishing constants are
\begin{eqnarray}
 f_{_{123}}=1,f_{_{458}}=f_{_{678}}=\dfrac{\sqrt{3}}{2} , \nonumber\\
 f_{_{147}}=f_{_{246}}=f_{_{345}}=f_{_{257}}=-f_{_{156}}=-f_{_{367}}=\dfrac{1}{2} , 
\end{eqnarray}
and \vspace*{-1cm}
\begin{eqnarray}
d_{_{118}}=d_{_{228}}=d_{_{338}}=-d_{_{888}}=\dfrac{1}{\sqrt{3}} , \nonumber\\
 d_{_{448}}=d_{_{558}}=d_{_{668}}=d_{_{778}}=-\dfrac{1}{2\sqrt{3}}  ,\nonumber\\
 d_{_{146}}=d_{_{157}}=d_{_{256}}=d_{_{344}}=d_{_{355}}=-d_{_{247}}=-d_{_{366}}=-d_{_{377}}=\dfrac{1}{2} .
\end{eqnarray} 
In the fundamental representation, $X_1, X_2,X_3$ are half the Pauli matrices with an extra null row and null column. Hence, they generate an $SU(2)$ subgroup. 

\begin{figure}[h]
 \centering
 \includegraphics[scale=.6]{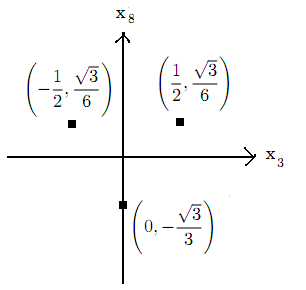}
 \caption{\label{y1}Weights of the fundamental representation of $SU(3)$}
 \end{figure}

\vspace{-0.4cm} To characterise $SU(3)$ representations, we look for a complete set of commuting operators ($CSCO$). Relations (3.1) and (3.2) show that $\hat{X}_1,\hat{X}_2,\hat{X}_3$ commute with $\hat{X}_8$. Therefore we start with choosing the diagonal generators $\hat{X}_3$ and $\hat{X}_8$ as part of  the $CSCO$. In the fundamental representation, the eigenvectors and eigenvalues ($x_{_{3}},x_{_{8}}$) corresponding to these two generators are:
\small 
 \begin{align}
 \begin{pmatrix}
 1\\0\\0 \end{pmatrix} \leftrightarrow \left( \dfrac{1}{2},\dfrac{\sqrt{3}}{6}\right) 
  \hspace*{1cm}
  \begin{pmatrix}
 0\\1\\0 \end{pmatrix} \leftrightarrow \left( -\dfrac{1}{2},\dfrac{\sqrt{3}}{6}\right) 
\hspace*{1cm}
  \begin{pmatrix}
 0\\0\\1 \end{pmatrix} \leftrightarrow \left( 0,-\dfrac{\sqrt{3}}{3}\right) . 
 \end{align} 
 \normalsize When these eigenvalues, also called  weights, are plotted in the  $x_{_{3}}-x_{_{8}}$ plane, they form the vertices of an equilateral triangle centred at the origin, as shown in Figure \ref{y1}.
 
From the $SU(2)$ subgroup, we have $\hat{X}^2=\hat{X}_1^2+\hat{X}_2^2+\hat{X}_3^2$. Also $\hat{X}_1 \pm i\hat{X}_2$ act as  ladder operators for the eigenvalues of $\hat{X}_3$, leaving the eigenvalues of $\hat{X}_8$ and $\hat{X}^2$ untouched. By the action of $\hat{X}_4 \pm i\hat{X}_5$  and $\hat{X}_6 \pm i\hat{X}_7$ on the weights, we observe that even these matrices act as ladder operators changing  the eigenvalues of $\hat{X}_3$ and $\hat{X}_8$. 

The structure of the algebra described above  happens to explain the flavour symmetry of the strong interactions, in the limit where Isospin($T$) and Hypercharge ($Y$) are conserved. In a hadron multiplet, each particle's state is specified as  $\vert T,T_3,Y\rangle$. The convention used is \citep{georgii}:  \begin{equation}
\hat T_3=\hat{X}_3  , \hspace*{1cm} \hat Y=\dfrac{2}{\sqrt{3}}\hat{X}_8 ,
\end{equation}. \vspace*{-1.5cm}
 \begin{subequations}\begin{eqnarray}
\hat T^2= \hat T_1^2+\hat T_2^2+\hat T_3^2= \hat{X}_1^2+\hat{X}_2^2+\hat{X}_3^2 ,  \\
\hat T_\pm = \hat{X}_1 \pm i\hat{X}_2, \hspace*{1cm} \hat V_\pm = \hat{X}_4\pm i\hat{X}_5, \hspace*{1cm}\hat U_\pm = \hat{X}_6\pm i\hat{X}_7 .
\end{eqnarray}\end{subequations} From (3.1) and (3.2), we have the following commutation relations:
\begin{align}
&[\hat T_3,\hat T_\pm]=\pm \hat T_\pm, \hspace*{1cm} [\hat T_3,\hat V_\pm]=\pm \dfrac{1}{2}\hat V_\pm \hspace*{1cm} [\hat T_3,\hat U_\pm]=\mp \dfrac{1}{2}\hat U_\pm \nonumber\\
&[\hat Y,\hat T_\pm]=0, \hspace*{1.7cm} [\hat Y,\hat V_\pm]=\pm \hat V_\pm \hspace*{1.5cm} [\hat Y,\hat U_\pm]=\pm \hat U_\pm
\end{align}

In this notation, the states in (3.4) forming the fundamental triangle are the three flavours of quarks,
\begin{equation}
\vert u\rangle=\vert \dfrac{1}{2},\dfrac{1}{2},\dfrac{1}{3}\rangle \hspace*{1cm} \vert d\rangle=\vert \dfrac{1}{2},-\dfrac{1}{2},\dfrac{1}{3}\rangle \hspace*{1cm} \vert s\rangle=\vert 0,0,-\dfrac{2}{3}\rangle 
\end{equation} for $up$, $down$ and $strange$ quarks respectively  \cite{georgii}. The action of the 6 ladder operators translates weights of states from one vertex of the triangle to another as shown below.\vspace*{-0.4cm}
\begin{figure}[H]
\centering
\includegraphics[scale=.5]{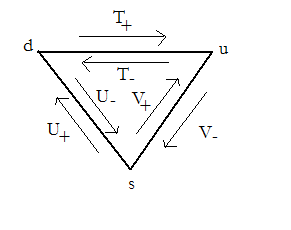}
\caption{\label{y2}Action of ladder operators on the weights of $SU(3)$ representation}
\end{figure} \vspace*{-0.4cm}\noindent The actions of $\hat U_-,\hat V_+,\hat T_+$ annihilate the state $\vert u\rangle$, which is the highest weight state of the fundamental representation.

An irreducible representation of a group is always specified by the eigenvalues of the Casimir invariants. In case of the $SU(3)$ group, they are the quadratic and the cubic invariants  \cite{biedenharn}. \vspace*{-0.4cm}
\begin{subequations}
\begin{eqnarray}
\hat F=\sum_{i,k=1}^3 \hat A_k^i\hat A_i^k=\sum_{i=1}^8 \hat{X}_i^2 \\
\hat H=\dfrac{1}{2}\sum_{i,k,l=1}^3 (\hat A_l^i\hat A_i^k\hat A_k^l+\hat A_i^l\hat A_k^i\hat A_l^k)
 \end{eqnarray}
 \end{subequations}
 with eigenvalues $f,h$ respectively. Here $\hat A$'s are the traceless matrices given by,
 \footnotesize \begin{eqnarray}
 \hat A_1^1=\dfrac{\hat Y}{2} + \hat T_3, \hspace*{1cm} \hat A_2^2=\dfrac{\hat Y}{2} - \hat T_3,\hspace*{1cm} \hat A_3^3=-\hat Y \nonumber\\
\hat A_2^1=\hat T_+, \hspace*{1.5cm} \hat A_3^1=\hat V_+,\hspace*{1.5cm} \hat A_3^2=\hat U_+ \nonumber \\
 \hat A_1^2=\hat T_-, \hspace*{1.5cm} \hat A_1^3=\hat V_-,\hspace*{1.5cm} \hat A_2^3=\hat U_-  .
 \end{eqnarray}\normalsize
 The eigenstates of all $SU(3)$ representations are fully specified as $ \vert f,h;T,T_3,Y\rangle$, with
 \vspace*{-0.4cm} \begin{subequations}
  \begin{eqnarray}
   \hat F  \vert f,h;T,T_3,Y\rangle &=& f  \vert f,h;T,T_3,Y\rangle \\
  \hat H  \vert f,h;T,T_3,Y\rangle &=& h  \vert f,h;T,T_3,Y\rangle\\
  \hat T^2\vert f,h;T,T_3,Y\rangle &=& T(T+1) \vert f,h;T,T_3,Y\rangle\\
  \hat T_3 \vert f,h;T,T_3,Y\rangle &=&T_3\vert f,h;T,T_3,Y\rangle \\
  \hat Y\vert f,h;T,T_3,Y\rangle &=& Y\vert f,h;T,T_3,Y\rangle 
  \end{eqnarray}
 \end{subequations}
 \vspace*{-1.8cm}\section{Specification of general representations and eigenstates}
A general quantum state can be constructed from tensor products of its components. In particular,  hadron states are often described as tensor products of individual quark flavours. In this section, we try to answer the question: \textit{What will be the final state $\vert f,h;T,T_3,Y \rangle$,  when the two states $\vert f_1,h_1;T_1,T_{1_{3}},Y_1\rangle$ and $\vert f_2,h_2;T_2,T_{ 2_{3}},Y_2\rangle$ are combined?} Similar to the case of $SU(2)$, the weights simply get added in a tensor product, while the Casimir invariants combine in a more complicated fashion.  Using the relation $SU(2)\subset SU(3)$, we have 
 \vspace*{-0.4cm}  \begin{subequations}
\begin{eqnarray}
 &&Y=Y_1+Y_2 \hspace*{1cm} T_3=T_{1_{3}}+T_{2_{3}}\\
  &&|T_1-T_2|\leq T \leq T_1+T_2  .
\end{eqnarray}   
   \end{subequations}     
 The possible values of $f$ and $h$ can be deduced  using tensor analysis, as shown in the following subsection.
\subsection{Young diagrams and irreducible representations of $SU(3)$}
The Schur-Weyl duality  \citep{schur-weyl} introduced in Chapter 1 allows every irreducible representation of $SU(3)$  to be mapped  to an irreducible representation of $S_n$. So the representation  can be denoted as a valid Young diagram of $n$ boxes ($n$ quarks), with three partitions ($\lambda_1,\lambda_2,\lambda_3$),  such that
 \vspace*{-0.4cm} \begin{eqnarray}
&&\lambda_1 \geq \lambda_2 \geq \lambda_3 \geq 0, \hspace*{1cm} \lambda_1 +\lambda_2 +\lambda_3 = n,\nonumber\\ 
   &&\lambda_1 - \lambda_2 =P, \hspace*{2cm} \lambda_2 - \lambda_3 = Q      . 
\end{eqnarray}  
Since a column of three boxes represents an $SU(3)$ singlet, every irreducible representation is identified by the numbers  $(P,Q)$. The Casimir invariants ($f,h$) are functions of ($P,Q$), as obtained in  \citep{su3casimir}.
\begin{subequations}
\begin{eqnarray}
f=\dfrac{P^2+PQ+Q^2}{3}+P+Q \\ h=\dfrac{1}{9}(P-Q)(2P+Q+3)(P+2Q+3) 
\end{eqnarray}\end{subequations}
These relations let us replace $(f,h)\leftrightarrow (P,Q)$, and label the $SU(3)$ eigenstates as $ \vert P,Q;T,T_3,Y\rangle$. With two $SU(3)$ irreducible representations ($f_1,h_1$) and ($f_2,h_2$) represented as valid Young diagrams $(P_1,Q_1)$ and $(P_2,Q_2)$, it is possible to determine the irreducible representations occurring in the tensor product of the two, using  systematic rules  \citep{georgii}.

Consider the simplest case of the tensor product of the representation $(1,0)$ with the irreducible representation ($P,Q$). This is the case needed to  construct the algorithm that generates arbitrary $SU(3)$ eigenstates. In terms of Young diagrams, the single box corresponding to (1,0) can be added to any of the three rows of boxes corresponding to $(P,Q)$. For example,
\small  
\begin{figure}[H]
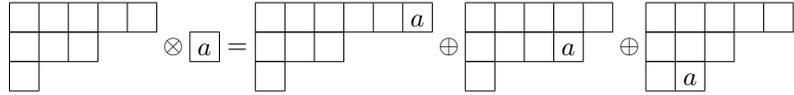
\footnotesize
\begin{equation}
  \Yvcentermath1 \yng(5,3,1)\otimes \young(a) =\Yvcentermath1 \young(\ \ \ \ \ a,\ \ \ ,\ ) \oplus \young(\ \ \ \ \ ,\ \ \ a,\ )\oplus\young(\ \ \ \ \ ,\ \ \ ,\ a)\nonumber .
  \end{equation}\normalsize
  \caption{\label{y4}Young diagram representation of the $SU(3)$ tensor product $(P,Q)\otimes(1,0)$}
\end{figure} 
\normalsize 
\noindent As per the restrictions (3.13), the second and the third diagrams on the r.h.s are allowed only if $P>0$ and $Q>0$ respectively. As a result, we have
  \begin{eqnarray}
(P,Q)  \otimes (1,0) = \left\lbrace  \begin{array}{cc}
  (P+1,Q),\\\oplus \\ (P-1,Q+1), & \text{if} (P>0)\\\oplus \\ (P,Q-1), & \text{if} (Q>0) 
 \end{array} \right. .  
  \end{eqnarray}
  \subsubsection*{Eigenstates and Young tableaux}
Every irreducible representation of $SU(N)$ can be represented as a Young diagram, and a state belonging to it can be  represented as a valid Young tableau [Appendix A]. For $SU(3)$ representations, we substitute the quarks $u,d,s$ for numbers $1,2,3$ respectively, and form  valid Young tableau. For example, the Young diagram (2,1,0) has the possible Young tableaux given by \footnotesize \begin{figure}[H]
$$ \Yvcentermath1 \young(\ \ ,\ )\sim \Yvcentermath1 \young(uu,d)+\young(uu,s)+\young(ud,d)+\young(ud,s)+\young(us,d)+\young(us,s)+\young(dd,s)+\young(ds,s) .$$ 
\caption{\label{y3}Possible Young tableaux for Young diagram (2,1,0)}
\end{figure} \normalsize
 
\vspace*{-0.4cm}A valid Young tableau of representation $(P,Q)$ with $n$ boxes can be represented by the set of integers  $n_u, n_d, n_{d1}$. Here $n_u(n_d)$ represents the number of $u(d)$ quarks in the Young tableau, and $n_{d1}$ is the number of $d$ quarks in the first row of Young tableau. We have, of course,
\vspace*{-0.4cm}\begin{eqnarray}
  n = n_u +n_d+ n_s = \lambda_1 + \lambda_2 + \lambda_3 .
\end{eqnarray} 
\indent Young tableaux for $SU(3)$ can not have more than 3 rows, and a fully filled column implies a trivial singlet (completely antisymmetric state). Fully filled columns do not contribute to the  specification of Young tableaux (provided $n$ is specified). For instance, the two Young diagrams below are equivalent.\\\vspace*{-0.4cm}
$$\Yvcentermath1 \young(\ \ \ ,\ \ ,\ ) \approx \young(\ \ ,\ ) $$
Therefore we concentrate only on the Young tableaux which have two rows also called reduced Young tableaux. They have unique specification ($P,Q,n_u',n_d',n_{d1}'$), with \begin{subequations}\begin{eqnarray}
 &&n'_i=n_i-\lambda_3 \hspace*{1cm}  ( i=u,d,s)\\
 &&n_{d1}'=n_{d1} .
\end{eqnarray}\end{subequations}
With $n$ qutrits, we have
\vspace*{-0.4cm}\begin{eqnarray}
%&&n =n_u+n_d+n_s = \lambda_1+ \lambda_2+ \lambda_3 \\
 &&n-3\lambda_3 = P+2Q = n'_u+n'_d+ n'_s .
 \end{eqnarray}
Since each of the individual quarks is associated with additive  quantum numbers, $T_3$ and $Y$, the net $T_3$ and $Y$ of the eigenstate represented by the Young tableau are simply the algebraic sum of the individual quark values given in (3.8).  Therefore, 
 \begin{subequations}\begin{eqnarray}
 &&T_3 = \dfrac{n'_u-n'_d}{2}  ,\\
&&Y= \dfrac{n'_u+n'_d-2n'_s}{3} .
  \end{eqnarray}\end{subequations}
It follows that
\begin{equation}
Y = n'_u+n'_d -\dfrac{2}{3}( P+2Q) .
\end{equation}
\indent The action of ladder operators on these states (Young tableaux) produces new valid Young tableaux according to Figure \ref{y2}. The state is annihilated by a ladder operator, when the action cannot produce another  valid Young tableau.

 The highest weight state (HWS) for a Young diagram is the state with the highest $T_3$ possible (for that reason, $\hat T_+,\hat U_-,\hat V_+$ annihilate this state). Using the rules of Young tableaux and ladder operators, this state has the first row boxes filled with $u$ quarks and the second row boxes filled with $s$ quarks. Therefore for the HWS,\\\vspace*{-0.6cm}\begin{subequations}\begin{eqnarray}
&&T_3 = \dfrac{\lambda_1-\lambda_3}{2} = \dfrac{P+Q}{2} = \dfrac{n'_u}{2}, \\
 &&Y = \dfrac{\lambda_1-\lambda_3-2(\lambda_2-\lambda_3)}{3} = \dfrac{P-Q}{3} = \dfrac{n'_u-2n'_s}{3}.
\end{eqnarray}\end{subequations}

\subsubsection*{Isospin}
  Isospin is the $SU(2)$ subgroup of $SU(3)$ with $\hat T_\pm$ as the ladder operators. A given Young diagram may contain  many different isospin multiplets. In particular, the set of integers $(P,Q, n,n_u',n_d')$ specifies $T_3$ and $Y$ of a state, but does not specify the isospin $T$ of the state. To uniquely specify an eigenstate of $SU(3)$ (Young tableau), we use the additional integer $n_{d1}'=n_{d1}$ (introduced in the previous subsection). That helps in assigning a unique $T$ to the state as follows.\\
\indent Every action of $\hat T_+$ converts a $d$ quark to a $u$ quark without changing $T$. So the value of isospin for a multiplet is the value of $T_3$ of the state when further action of $\hat{T}_+$ annihilates it. Since a $u$ quark can appear only in the first row, it is clear that the number of times one can apply $\hat T_+$  on a state, without annihilating it, is the number of $d$ quarks in the first row. Therefore for a given Young tableau, we have
 \vspace*{-0.4cm} \begin{eqnarray}
  T &=& \dfrac{n'_u-n'_d}{2}+n_{d1}.
   \end{eqnarray}
  \subsection{Convenient notation for quantum numbers}
The values of hypercharge and isospin are often fractional numbers. It is useful to define a set of integer parameters from which all the above quantum numbers can be obtained. They are  suitable for use in a digital algorithm. A convenient choice is to label each Young tableau by $k, l, m$  \citep{williams} :\vspace*{-0.4cm} 
  \begin{subequations}\begin{eqnarray}
  k &=& n'_u+n_{d1}  \\
  l &=& n'_d-n_{d1} =n_{d2} \hspace*{1cm}\text{(number of $d$ quarks in second row)}\end{eqnarray}\begin{eqnarray}
  m &=& n'_u 
  \end{eqnarray}\end{subequations}
From the rules of Young tableaux, these parameters satisfy the following conditions:
\vspace*{-0.4cm}\begin{subequations}  
  \begin{eqnarray}
  Q \leq &k&\leq P+Q\\
  0 \leq &l&\leq Q \\ 
  l \leq &m&\leq k.
  \end{eqnarray}
  \end{subequations}
Within these restrictions, all integer values are allowed for ($k,l,m$), and they are non-degenerate. The dimension of the irreducible representation $(P,Q)$ is given by
 \begin{eqnarray}
 \sum_{k=Q}^{P+Q}\sum_{l=0}^Q\sum_{m=l}^k 1= \dfrac{(P+1)(Q+1)(P+Q+2)}{2} .
 \end{eqnarray}
The conventional quantum numbers, and the number of quarks in the (reduced) Young tableaux, are related to ($k,l,m$) by 
\vspace*{-0.4cm}\begin{subequations}
\begin{eqnarray}
T &=& \dfrac{k-l}{2} =\dfrac{n'_u-n'_d}{2}+n_{d1} ,\\
T_3 &=& m-\dfrac{k+l}{2} =\dfrac{n'_u-n'_d}{2} ,\\
Y &=& k+l -\dfrac{2}{3}( P+2Q) =n'_u+n'_d -\dfrac{2}{3}( P+2Q) .
\end{eqnarray}
\end{subequations}
The inverse relations are  
\vspace*{-0.4cm}\begin{subequations}
\begin{eqnarray}
k &=& \dfrac{P+2Q}{3}+\dfrac{Y}{2}+T , \\
l &=& \dfrac{P+2Q}{3}+\dfrac{Y}{2}-T  ,\\
m &=& \dfrac{P+2Q}{3}+\dfrac{Y}{2}+T_3 . 
\end{eqnarray}
\end{subequations}
These relations show that given the irreducible representation $(P, Q)$, the choice of $(k,l,m)$ or $(T,T_3,Y)$ or $(n_u',n_d',n_{d1}')$ is equivalent; they are sets of quantum numbers that are interconvertible by linear transformations.\\
\indent Clearly, $k=m$  corresponds to the highest weight state in any isomultiplet within the representation. Furthermore $l=0$ corresponds to the largest isospin multiplet for a given $Y$, and $k=m=P+Q,$ $l=0$ describes the highest weight state (HWS) of the $SU(3)$ irreducible representation.\\
\indent For example, in Figure \ref{y3}, the first state is the highest weight state of  isospin 1/2 with $k=m=2,l=1$. But, the second state with $k=m=2,l=0$ and isospin 1 is the HWS of the irreducible representation $P=Q=1$.\\
\indent From now on, we denote an arbitrary eigenstate by $ \vert P,Q;k,l,m\rangle$ and the highest weight state by  $\vert P,Q;HWS\rangle$. 

 \subsection{Matrix elements of $SU(3)$ ladder operators}
 We have described the action of the ladder operators on quarks in Figure 3.2. The action of these operators on arbitrary states produce the same changes in quantum numbers $T_3,Y$. The action of $\hat{T}_\pm$ does not change $T$, and so results in a unique state. But the actions of $\hat{U}_\pm,\hat{V}_\pm$ change $T$ by 1/2, and so result in a linear combination of two states with $T \pm 1/2$. The proportionality constants accompanying each state are the $SU(3)$ versions  of the normalisation constants, and have to be evaluated using the algebra of the generators  \citep{biedenharn}.
\newcommand{\ket}[1]{|#1\rangle}
\vspace*{-0.4cm}\begin{subequations}
\begin{eqnarray} 
\hat{T}_{+}\ket{ P,Q;k,l,m } &=& \sqrt{(k-m)(m-l+1)} \ket{ P,Q;k,l,m+1 } \\ 
\hat{T}_{-}\ket{ P,Q;k,l,m } &=& \sqrt{(k-m+1)(m-l)} \ket{ P,Q;k,l,m-1 } \\ \nonumber\\
\hat{V}_{+}\ket{ P,Q;k,l,m } &=& \sqrt{\frac{(k+2)(m-l+1)(k-Q+1)(P+Q-k)}{(k-l+1)(k-l+2)}} \ket{ P,Q;k+1,l,m+1 }\nonumber \\ \nonumber &+& \sqrt{\frac{(l+1)(k-m)(Q-l)(P+Q-l+1)}{(k-l)(k-l+1)}} \ket{ P,Q;k,l+1,m+1 }  \\\end{eqnarray}\begin{eqnarray}
 \hat{V}_{-}\ket{ P,Q;k,l,m }  &=& \sqrt{\frac{(k+1)(m-l)(k-Q)(P+Q-k+1)}{(k-l)(k-l+1)}} \ket{ P,Q;k-1,l,m-1 } \nonumber\\  &+& \sqrt{\frac{l(k-m+1)(Q-l+1)(P+Q-l+2)}{(k-l+1)(k-l+2)}} \ket{ P,Q;k,l-1,m-1 } \nonumber\\  \\\nonumber 
\hat{U}_{+}\ket{ P,Q;k,l,m } &=& \sqrt{\frac{(k+2)(k-m+1)(k-Q+1)(P+Q-k)}{(k-l+1)(k-l+2)}} \ket{ P,Q;k+1,l,m } \nonumber\\ &-& \sqrt{\frac{(m-l)(l+1)(Q-l)(P+Q-l+1)}{(k-l)(k-l+1)}} \ket{ P,Q;k,l+1,m } \nonumber\\\\ 
 \hat{U}_{-}\ket{ P,Q;k,l,m }  &=& \sqrt{\frac{(k+1)(k-m)(k-Q)(P+Q-k+1)}{(k-l)(k-l+1)}} \ket{ P,Q;k-1,l,m } \nonumber\\ &-& \sqrt{\frac{l(m-l+1)(Q-l+1)(P+Q-l+2)}{(k-l+1)(k-l+2)}} \ket{ P,Q;k,l-1,m }  \nonumber\\
 \end{eqnarray}
 \end{subequations}

\section{$SU(3)$ tensor products}
 A general multi-qutrit register state is obtained as a tensor product of individual qutrit states. To construct such states, we need the rules for forming tensor products of $SU(3)$ representations. Once again the algebra is a unitary change of orthonormal basis, and the corresponding CG-coefficients are defined by \footnotesize 
\begin{eqnarray}
 \vert P,Q;k,l,m\rangle =\sum_{\substack{P_1,Q_1,P_2,Q_2\\k_1,l_1,m_1\\k_2,l_2,m_2}}\left( \begin{array}{ccc}
 P,Q & P_1,Q_1 & P_2,Q_2 \\
 k,l,m & k_1,l_1,m_1 & k_2,l_2,m_2
 \end{array}\right) \vert P_1,Q_1;k_1,l_1,m_1\rangle\vert P_2,Q_2;k_2,l_2,m_2\rangle \nonumber\\
 \end{eqnarray} 
 \normalsize
 The summation is over the terms corresponding to the states $(P_1,Q_1,T_1,T_{1_{3}},Y_1)$ and $(P_2,Q_2,T_2,T_{2_{3}},Y_2)$ that result in the irreducible representation $(P,Q,T,T_3,Y)$, when added box by box as in (3.15), with each combination satisfying (3.12).
 %\begin{subequations}\begin{eqnarray}&&|T_1-T_2|\leq T \leq T_1+T_2\\&&T_3=T_{1_{3}}+T_{2_{3}}, \hspace*{1cm}Y=Y_1+Y_2 .\end{eqnarray}\end{subequations}
 By the relations (3.24) and (3.26), the constraints  can be rewritten in $k,l,m$ notation. Similarly,  we can also write the inverse relation of (3.29),  which will be used in the algorithm,
 \footnotesize 
\begin{eqnarray}
 \vert P_1,Q_1;k_1,l_1,m_1\rangle\vert P_2,Q_2;k_2,l_2,m_2\rangle = 
 \sum_{\substack{P,Q \\ k,l,m}} \left( \begin{array}{ccc}
  P_1,Q_1 & P_2,Q_2 & P,Q  \\
  k_1,l_1,m_1 & k_2,l_2,m_2 & k,l,m 
 \end{array}\right) \vert P,Q;k,l,m\rangle. \nonumber\\
 \end{eqnarray} 
 \normalsize
 
As a consequence of the Wigner-Eckart theorem \citep{wigner}, the $SU(3)$ CG-coefficients can be factored as products of $SU(2)$ CG-coefficients and isoscalar factors \citep{deswart}.
 \footnotesize
 \begin{equation}
 \left( \begin{array}{ccc}
 P,Q & P_1,Q_1 & P_2,Q_2 \\
 T,T_3,Y & T_1,T_{1_{3}},Y_1 & T_2,T_{2_{3}},Y_2
 \end{array}\right)  = 
\left( \begin{array}{ccc}
 T & T_1 & T_2\\
 T_3 & T_{1_{3}} & T_{2_{3}}
 \end{array}\right)  
 \left( \begin{array}{ccc}
 P,Q & P_1,Q_1 & P_2,Q_2 \\
 T,Y & T_1,Y_1 & T_2,Y_2
 \end{array}\right)  
 \end{equation}\normalsize
\noindent On the RHS, the first parentheses contain the $SU(2)$ CG-coefficient and the second one contains the isoscalar factor. Isoscalar factors depend on $P,Q,T$ and $Y$ but not on  $T_3$. Multiplying the isoscalar factor with an appropriate $SU(2)$ CG-coefficient gives the $SU(3)$ CG-coefficient. 

In the ($k,l,m$) notation, the role of $T_3$ is played by $m$, and
 \footnotesize
 \begin{equation}
 \left( \begin{array}{ccc}
 P,Q & P_1,Q_1 & P_2,Q_2 \\
 k,l,m & k_1,l_1,m_1 & k_2,l_2,m_2
 \end{array}\right)  = 
\left( \begin{array}{ccc}
 k,l & k_1,l_1 & k_2,l_2\\
 m & m_1 & m_2
 \end{array}\right)  
 \left( \begin{array}{ccc}
 P,Q & P_1,Q_1 & P_2,Q_2 \\
 k,l & k_1,l_1 & k_2,l_2
 \end{array}\right)     .
 \end{equation}\normalsize
Also, the $SU(3)$ CG-coefficients, the $SU(2)$ CG-coefficients and the isoscalar factors satisfy the orthonormality relations individually \citep{deswart}. 

Note that in all the derivations that follow, the conditions on $k,l,m$ listed in  (3.24) should be satisfied in each of the Hilbert spaces separately. Otherwise that term is equated to zero.
 
\subsection{$(P_1,Q_1)\otimes(1,0)$ problem}
In our algorithm, we only add a single qutrit to the state register at a time. That becomes the simple problem of constructing the tensor product $(P_1,Q_1) \otimes (1,0)$ in $SU(3)$, analogous to the $J+S$ problem for $SU(2)$. The function of $U_{CG}$ is thus to add a single quark to the state $\vert P_1,Q_1;T_1,T_{1_{3}},Y_1\rangle$, and the corresponding $SU(3)$ CG-coefficients describe the resultant superposed state.  We use the ($k,l,m$) notation, and find the appropriate isoscalar factors of this operation using recurrence relations. \\
\indent We abbreviate the isoscalar factors as $F(k_1,l_1:k_2,l_2;k,l)$ leaving the labels $(P,Q)$ implicit. In the $\vert P,Q;k,l,m\rangle$ notation, the single quark states are 
\vspace*{-0.4cm}\begin{eqnarray}
\vert u\rangle = \vert 1,0; 1,0,1 \rangle , \hspace*{1cm}
\vert d\rangle = \vert 1,0; 1,0,0 \rangle , \hspace*{1cm}
\vert s\rangle = \vert 1,0; 0,0,0 \rangle  .
\end{eqnarray}
Therefore the required isoscalar factors are $F(k_1,l_1:1,0;k,l)$ and $F(k_1,l_1:0,0;k,l)$.
The isoscalar factors are independent of $T_3$, and hence are the same for $u$ and $d$. So we simplify the problem by of determining them setting $k_1=m_1$, \textit{i.e.} only consider the states with $T_1=T_{1_{3}}$. \\
\indent We determine the isoscalar factors using recursion relations [Appendix C]. The isoscalar factors for an addition of a $u$ quark and an $s$   quark are (C.55) and (C.40) respectively:
\begin{subequations}\begin{eqnarray}
&&F(k_1,l_1:1,0;k,l)=\delta_{l,l_1+1}(-\sqrt{1-|F(k_1,l_1+1:0,0;k,l)|^2})\nonumber\\&&+c[P_1,Q_1,k_{1},l_{1},l]\hspace*{.2cm} A[P,Q,k,l]\hspace*{.2cm}F(k_{1},l_{1}-l:1,0:k,0),\\
 &&F(k_{1},l_{1}:0,0:k,l) = c[P_1,Q_1,k_{1},l_{1},l]\hspace*{.2cm} A[P,Q,k,l]\hspace*{.2cm}F(k_{1},l_{1}-l:0,0:k,0),\nonumber\\
\end{eqnarray} with $A,c$ as defined in (C.38) and (C.39). The values of $F(k_{1},l_{1}-l:0,0:k,0)$ and $F(k_{1},l_{1}-l:1,0:k,0)$ are determined by the relations
\begin{eqnarray}
F(k_{1},l_{1}:1,0;k,0) =&& B[P,Q,s] \hspace*{.1cm} d[P_1,Q_1,k_{1},l_{1},s] \hspace*{.1cm}F(k_{1}+s,l_{1}:1,0;P+Q,0 ),\nonumber\\
F(k_{1},l_{1}:0,0;k,0) =&& B[P,Q,s]  
\left\lbrace d[P_1,Q_1,k_{1},l_{1},s]F(k_{1}+s,l_{1}:0,0:P+Q,0 ) \right. \nonumber\\ &&\hspace*{-.5cm}\left.+s \times d[P_1,Q_1,k_{1},l_{1},s-1] F(k_{1}+s-1,l_{1}:1,0;P+Q,0 ) \right\rbrace,\nonumber\\ 
\end{eqnarray} with $s=P+Q-k$ and $B, d$ as defined in (C.22) and (C.23). Finally, the isoscalar factors for the HWS are given by: \footnotesize
\begin{eqnarray}
\begin{aligned} 
&F(k_{1},l_{1}-1:1,0;P+Q,0 ) = \delta[k_1,l_1-1]+(1-\delta[k_1,l_1-1])\sqrt{\frac{(u_{2+})^{2}(k_{1}-l_{1}+2)}{G[P_1,Q_1,k_{1},l_{1}]}}, \\ 
&F(k_{1}-1,l_{1};1,0;P+Q,0 ) =\delta[k_1-1,l_1]-(1-\delta[k_1-1,l_1])\sqrt{\frac{(u_{1+})^{2}}{G[P_1,Q_1,k_{1},l_{1}]}}, \\ 
&F(k_{1},l_{1}:0,0:P+Q,0 ) = \delta_{k_1-l_1,P+Q}\sqrt{\frac{(v_{1-}u_{1+}+v_{2-}u_{2+})^2}{G[P_1,Q_1,k_{1},l_{1}]}},\\
& G[P_1,Q_1,k_{1}l_{1}] = (u_{2+})^{2}(k_{1}-l_{1}+2) + (u_{1+})^{2} + (v_{1-}u_{1+}+v_{2-}u_{2+})^2.
\end{aligned}
\end{eqnarray}
\end{subequations}\normalsize The arguments of $u_{i \pm}$ are $(P_1,Q_1,k_1-1,l_1-1)$ and $v_{i \pm}$ are $(P_1,Q_1,k_1,l_1)$. Detailed expressions for the values of $u_{i\pm}$ and $v_{i\pm}$, where $i\in \{1,2\}$, are  in [Appendix C].
\subsubsection*{$SU(3)$ CG-coefficients}
By construction, the $SU(3)$ CG-coefficients are simply the product of the $SU(2)$ CG-coefficients with the respective isoscalar factors. 

In the case of $s$ quark, the $SU(2)$ CG-coefficient is unity and the $SU(3)$ CG-coefficient is the isoscalar factor. When it comes to $u$ and $d$ quarks, the $SU(2)$ CG-coefficient is obtained   as a $J+S$ problem. Then, the $SU(3)$ CG-coefficient is the product of $F(k_1,l_1:1,0;k,l) $ with \footnotesize$ \sqrt{\dfrac{T_1+1/2+T_3}{2T_1+1}}$ \normalsize or \footnotesize $ \pm\sqrt{\dfrac{T_1+1/2-T_3}{2T_1+1}}$,  \normalsize as per the relations (2.35).

\section{Computational basis and Schur basis}
A computational basis is required to define and construct the unitary logic gates of $SU(3)$ group operations. The standard computational basis for qutrits is $\{0,1,2\}$, related to the weights of the fundamental representation. We do not use this basis. Instead, we use the $SU(3)$ computational basis defined as the following padded $SU(2)$ computational basis: 
\vspace*{-0.4cm}\begin{equation}
\vert 0\rangle_{su(3)} =\vert \underline{00}\rangle_{su(2)} \hspace*{1cm}  \vert 1\rangle_{su(3)} =\vert \underline{01}\rangle_{su(2)} \hspace*{1cm}\vert 2\rangle_{su(3)} =\vert \underline{10}\rangle_{su(2)}
\end{equation}
The bars below two qubits indicate that they are padded qubits, which together represent a qutrit. Now we can construct combinations of well-known $SU(2)$ unitary gates, which act on a set of these qutrits (padded qubits), to give desired $SU(3)$ transformations. We call such operations $SU(3)$ unitary gates.

The standard computational basis is mapped to the weights of quark states with the help of the $k,l,m$ notation. Then we have,
\vspace*{-0.4cm}\begin{equation}
\vert u\rangle=\vert 2\rangle =\vert \underline{10}\rangle, \hspace*{1cm}\vert d\rangle =\vert 1\rangle = \vert\underline{01}\rangle,\hspace*{1cm} \vert s\rangle =\vert 0\rangle =\vert \underline{00}\rangle ,
\end{equation} with the quarks satisfying $\vert q\rangle = \vert k+m\rangle$. Noting the appearance of $k+m$ in (3.36), we trade $k,l,m$ for $k+l,l+m,k+m$. This  change of variables makes encoding and construction of logic gates easier.

Every Schur basis state can be represented by a unique Young tableau. In section 3.2.2, we described how    $SU(3)$ eigenstates are labelled as  ($P,Q,k,l,m$), and how they can be inferred from a Young tableau. To construct tensor products, we need to track how the exchange symmetry of particles changes, when one  quark is added to a representation. Including this information, we label each eigenstate of an $SU(3)$ representation  of $n$ quarks as \vspace*{-0.4cm} $$ \vert \lambda_1,\lambda_2,\lambda_3;k+l,l+m,k+m:p_1,p_2....,p_{n-1}\rangle,$$  where $p_i \in \{0,1,2\}$,  based on whether the additional box got added to the $3^{rd}/2^{nd}/1^{st}$ row of the Young diagram respectively. In case of $SU(2)$, as displayed in  Figure 2.3, all the $p_i$ together with $j+m$ yield  the $n$-qubit state in the  computational basis, after $n-1$ iterations of $U_{CG}^{-1}$. Similarly, in case of $SU(3)$, combinations of $k,l,m$ together with all the $p_i$ help us infer the $n$-quark state in the computational basis. 

To illustrate our notation, we write down  states of some baryons which correspond to  representations with 3 quarks \citep{georgii}. The quark notation  can be  converted to the computational basis  using (3.36). The tensor product of 3 quarks can be decomposed into irreducible representations as  
\vspace*{-.4cm}\footnotesize
\begin{align}
 &\Yvcentermath1 \young(1) \otimes \young(2) \otimes \young(3) =  \Yvcentermath1 \left( \young(12) \oplus \young(1,2)\right)\otimes \young(3)=  \Yvcentermath1 \left( \young(123) \oplus \young(12,3)\right)  \oplus \left(\young(13,2)\oplus \young(1,2,3)\right) \nonumber \\\nonumber \\
&\Yvcentermath1 \young(123)  \leftrightarrow  \vert (3,0,0);k+l,m+l,k+m;2,2\rangle_{sch} \hspace*{.3cm}\text{(All quarks are symmetric) }\nonumber \\\nonumber \\
&\Yvcentermath1 \young(12,3)  \leftrightarrow  \vert (2,1,0);k+l,m+l,k+m;2,1\rangle_{sch} \hspace*{.7cm}\text{(1 and 2 quarks are symmetric )}\nonumber  \\\nonumber \\
&\Yvcentermath1 \young(13,2)  \leftrightarrow  \vert (2,1,0);k+l,m+l,k+m;1,2\rangle_{sch} \hspace*{.7cm}\text{(1 and 2 quarks are antisymmetric) }\nonumber \\\nonumber \\
&\Yvcentermath1 \young(1,2,3)  \leftrightarrow  \vert (1,1,1);k+l,m+l,k+m;1,0\rangle_{sch}\hspace*{.8cm}\text{(All quarks are antisymmetric) }
\end{align}\normalsize

In the above decomposition \scriptsize$ \young(12,3)$ \normalsize and  \scriptsize$ \young(13,2)$ \normalsize  are representations of mixed symmetry. The individual  eigenstates of the irreducible representations  can be obtained using the CG expansion. Some of them, in the tensor product basis, are
\footnotesize
\begin{subequations}
\begin{eqnarray}
&\Yvcentermath1 \young(123)&\leftrightarrow
\begin{array}{cc}
\scriptsize\Yvcentermath1 \young(uuu)& (uuu) \\
\scriptsize\Yvcentermath1 \young(uus)& \frac{1}{\sqrt{3}}(uus+usu+suu)\\
\scriptsize\Yvcentermath1 \young(uds)& \frac{1}{\sqrt{6}}(uds+dsu+sud+sdu+dus+usd)
\end{array} \\ \nonumber\\
&\Yvcentermath1 \young(12,3)&\leftrightarrow
\begin{array}{cc}
\scriptsize\Yvcentermath1 \young(uu,s)& \frac{1}{\sqrt{6}}(usu+suu-2uus)\\
\scriptsize\Yvcentermath1 \young(ud,s)& \frac{1}{\sqrt{12}}(dsu+sdu+sud+usd-2uds-2dus)\\
\scriptsize\Yvcentermath1 \young(us,d)& \frac{1}{2}(sud-sdu+usd-dsu)
\end{array} \\ \nonumber\\
&\Yvcentermath1 \young(13,2)&\leftrightarrow
\begin{array}{cc}
\scriptsize\Yvcentermath1 \young(uu,s)& \frac{1}{\sqrt{2}}(usu-suu)\\
\scriptsize\Yvcentermath1 \young(ud,s)& \frac{1}{2}(sud+sdu-usd-dsu)\\
\scriptsize\Yvcentermath1 \young(us,d)& \frac{1}{\sqrt{12}}(2uds-2dus-dsu+sdu-sud+usd)
\end{array}\end{eqnarray}
\begin{eqnarray}
&\Yvcentermath1 \young(1,2,3)&\leftrightarrow
\begin{array}{cc}
\scriptsize\Yvcentermath1 \young(u,d,s)& \frac{1}{\sqrt{6}}(uds+dsu+sud-sdu-dus-usd)
\end{array}
\end{eqnarray}
\end{subequations}\normalsize
 
\section{Algorithm and circuit to generate eigenstates of $SU(3)$ representations}
 To generate the eigenstates of $SU(3)$ representation with $n$ quarks, we use the $U_{CG}^{-1}$ transform recursively $n-1$ times. Schematically, this is the same as in Figure 2.3, with $\vert \lambda_1, \lambda_2 \rangle$ replaced by $\vert \lambda_1, \lambda_2, \lambda_3 \rangle$ and $\vert j+m \rangle$ replaced by $ \vert k+l, l+m, k+m\rangle$.
% \end{comment}
\subsection{Inverse Clebsch-Gordan transform}
We have already obtained the CG-coefficients when $SU(3)$ eigenstates are combined with a single quark state. For every application of $U_{CG}$, we have the relation
  \footnotesize
   \begin{eqnarray}
   \vert P,Q;T,T_3,Y\rangle_{2/1/0} &=& \alpha \vert P',Q';T\pm\frac{1}{2},T_3-\frac{1}{2},Y-\frac{1}{3}\rangle \vert u\rangle + \beta \vert P',Q';T\pm\frac{1}{2},T_3+\frac{1}{2},Y-\frac{1}{3}\rangle \vert d\rangle \nonumber\\&+&\gamma\vert P',Q';T,T_3,Y+\frac{2}{3}\rangle \vert s\rangle .
   \end{eqnarray}\normalsize
  Here $ \vert P,Q;T,T_3,Y\rangle_{2/1/0}$ represents the state formed upon adding a box to the $1^{st}/2^{nd}/3^{rd}$ row of the Young diagram respectively, and $ \alpha,\beta,\gamma$ are the CG-coefficients. For a completely specified Schur basis state, isospin $T'$ takes a value which is either $T+1/2$ or $T-1/2$ on the RHS of the above equation. For these Schur basis states, the  CG-coefficients and the quantum numbers are easily converted to the $k,l,m$ notation. 
  
For the states appearing in (3.39), let us denote the $SU(2)$ CG-coefficients as $cos(\theta _{k,l,m})$, $sin(\theta_{k,l,m})$, and the isoscalar factors as $F^{u,p}_{k,l},F^{s,p}_{k,l}$:
\begin{subequations}  
  \vspace*{-0.4cm}\begin{eqnarray}\begin{aligned}
cos(\theta _{k,l,m})&=cos(\theta_{T',T_{3}})\\&=\sqrt{\dfrac{T'+T_3+1/2}{2T'+1}} =\left\lbrace \begin{array}{c}
\sqrt{\dfrac{m-l+1}{k-l+2}} \quad \text{for $T'=T+1/2$}\\ \sqrt{\dfrac{m-l}{k-l}}\quad\qquad \text{for $T'=T-1/2$}
\end{array} \right., \end{aligned}\end{eqnarray}\begin{eqnarray}\begin{aligned}
sin(\theta _{k,l,m})&=sin(\theta_{T',T_{3}})\\&=\sqrt{\dfrac{T'-T_3+1/2}{2T'+1}}=\left\lbrace \begin{array}{c}
\sqrt{\dfrac{k-m+1}{k-l+2}}\quad \text{for $T'=T+1/2$}\\ \sqrt{\dfrac{k-m}{k-l}}\quad\qquad \text{for $T'=T-1/2$}
\end{array} \right., \end{aligned}\end{eqnarray} \begin{eqnarray}
F^{u,p}_{k',l'}= F(P',Q',k',l':1,0;P,Q,k,l),\hspace*{4.5cm}\\
F^{s,p}_{k',l'}=F(P',Q',k',l':0,0;P,Q,k,l). \hspace*{4.5cm}
  \end{eqnarray}
  \end{subequations}
  Here $p$ expresses the relation between $(P,Q)$ and $(P',Q')$ according to (3.15). Note that (3.40a-b) agrees with (2.29) and (2.35). Using these coefficients in (3.39), we have the complete $U_{CG}$:
  \footnotesize
\begin{eqnarray}
 \left[ 
\begin{array}{c}
\vert P'+1,Q';T,T_3,Y\rangle_2 \\ 
\vert P'-1,Q'+1;T,T_3,Y \rangle_1 \\ 
\vert P',Q'-1;T,T_3,Y \rangle_0
\end{array}
\right] &=& \left[\hat R(\theta,F)\right]\left[ 
\begin{array}{cc}
\vert P',Q';T',T_3-1/2,Y-1/3\rangle & \vert u \rangle \\ 
\vert P',Q';T',T_3+1/2,Y-1/3\rangle & \vert d \rangle   \\ 
\vert P',Q';T,T_3,Y+2/3 \rangle &  \vert s \rangle 
\end{array}
\right]\nonumber\\ \end{eqnarray} \normalsize 
 with $\left[\hat R(\theta,F)\right] =\delta_{T',T+1/2}\left[\hat\rho (\theta,F)\right]+\delta_{T',T-1/2} \left[\hat{\sigma}(\theta,F)\right]$, such that  
 \begin{subequations}\begin{eqnarray}
 \left[\hat \rho(\theta,F)\right]&=&\begin{bmatrix}
-sin(\theta_{k,l,m})F^{u,2}_{k,l-1} & cos(\theta_{k,l,m})F^{u,2}_{k,l-1} & F^{s,2}_{k,l}\\
-sin(\theta_{k,l,m})F^{u,1}_{k,l-1} & cos(\theta_{k,l,m})F^{u,1}_{k,l-1} & F^{s,1}_{k,l}\\
-sin(\theta_{k,l,m})F^{u,0}_{k+1,l} & cos(\theta_{k,l,m})F^{u,0}_{k+1,l} & F^{s,0}_{k+1,l+1} \\
\end{bmatrix}, \\
\left[\hat{\sigma}(\theta,F)\right]&=&\begin{bmatrix}
cos(\theta_{k,l,m})F^{u,2}_{k-1,l} & sin(\theta_{k,l,m})F^{u,2}_{k-1,l} & F^{s,2}_{k,l}\\
cos(\theta_{k,l,m})F^{u,1}_{k-1,l} & sin(\theta_{k,l,m})F^{u,1}_{k-1,l} & F^{s,1}_{k,l}\\
cos(\theta_{k,l,m})F^{u,0}_{k,l+1} & sin(\theta_{k,l,m})F^{u,0}_{k,l+1} & F^{s,0}_{k+1,l+1} \\
\end{bmatrix} .
 \end{eqnarray}\end{subequations}

 According to (3.34), for $(k_1,l_1) \equiv (k',l')=(k,l-1)$, the isoscalar factor $F^{u,p}_{k,l-1}$ alone is non-zero, and for $(k_1,l_1) \equiv (k',l')=(k-1,l)$, the isoscalar factor $F^{u,p}_{k-1,l}$ alone is non-zero. Therefore, for a completely defined Schur basis state, the isoscalar factors are unique, \textit{i.e.} $T'$ takes the value $T+1/2$ if $F^{u,p}_{k,l-1} \neq 0$ and $T-1/2$ if $F^{u,p}_{k-1,l} \neq 0$. We have used this fact to write the transformation, where a single qutrit is added to a state, such that $\hat R(\theta,F)$ reduces either to $\hat{\sigma}$ or to $\hat\rho$. Note that for fixed values of $p,p'$, the orthonormality condition,
$$F^{u,p}_{k',l'}F^{u,p'}_{k',l'}+F^{s,p}_{k,l}F^{s,p'}_{k,l}=\delta_{p,p'},$$ guarantees unitarity of $\hat{R}(\theta,F)$.\\

Since $U_{CG}^{-1}=(U_{CG})^T$, the inverse relationship is given by 
 \footnotesize
 \begin{eqnarray}
 \left[\begin{array}{cc}
\vert P',Q';T',T_3-1/2,Y-1/3\rangle & \vert u \rangle \\ 
\vert P',Q';T',T_3+1/2,Y-1/3\rangle & \vert d \rangle   \\ 
\vert P',Q';T',T_3,Y+2/3 \rangle &  \vert s \rangle
\end{array} \right] = \underbrace{\left[\hat{R}(\theta,F)\right]^T}_{U_{CG}^{-1}} \left[ \begin{array}{c}
\vert P'+1,Q';T,T_3,Y\rangle_2 \\ 
\vert P'-1,Q'+1;T,T_3,Y \rangle_1 \\ 
\vert P',Q'-1;T,T_3,Y \rangle_0
\end{array}\right].
 \end{eqnarray} \normalsize \\In the $k,l,m$ notation, the initial state $\vert k,l,m\rangle$ transforms to $ \vert k',l',m'\rangle$ according to 
 \begin{equation}
\vert k',l',m'\rangle \vert q \rangle= U_{CG}^{-1}\left( \vert k,l,m \rangle \vert p \rangle\right) ,
\end{equation} with the changes in the quantum numbers given by: 
 \begin{center}
 \begin{tabular}{|c|c|c|c|}
 \hline $\vert p \rangle$ & $\vert q \rangle$  & $\hat{\sigma}$& $\hat{\rho}$\\\hline 
&$\vert u \rangle$ & $\vert k-1,l,m-1 \rangle$ &$\vert k,l-1,m-1 \rangle$\\
$\vert 2 \rangle$,$\vert 1 \rangle$&$\vert d \rangle$ &  $\vert k-1,l,m \rangle$ & $\vert k,l-1,m \rangle$ \\
&$\vert s \rangle$&$\vert k,l,m \rangle$&$\vert k,l,m \rangle$\\ \hline
&$\vert u \rangle$ & $\vert k,l+1,m \rangle$ &$\vert k+1,l,m \rangle$\\
$\vert 0 \rangle$&$\vert d \rangle$  & $\vert k,l+1,m+1 \rangle$ & $\vert k+1,l,m+1 \rangle$ \\
&$\vert s \rangle$&$\vert k+1,l+1,m+1 \rangle$&$\vert k+1,l+1,m+1\rangle$\\\hline 
\end{tabular} \captionof{table}{The quantum numbers $(k',l',m')$ after applying $U_{CG}^{-1}$}
 \end{center}  
\subsection{Algorithm and its transformation equations}
 
\begin{figure}[H]
 \centering
 \includegraphics[scale=.7]{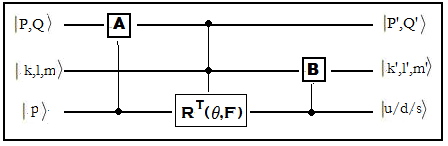}
 \caption{\label{y5} Schematic construction of $U_{CG}^{-1}$ for $SU(3)$. \textbf{A} and \textbf{B} represent combinations of unitary gates and $R^T(\theta,F)$ is a rotation gate.}
 \end{figure}

In our algorithm, we apply $U_{CG}^{-1}$ recursively $n-1$ times to construct an $n$-quark state. The flow chart for it is same as Figure 2.2 for the $SU(2)$ case, and only the transformation equations making up $U_{CG}^{-1}$ are different. The construction of $U_{CG}^{-1}$, defined in (3.43), can   be separated into three parts, depicted as the three boxes in Figure \ref{y5}. We consider them explicitly  in the following.
\begin{itemize}
\item \textbf{Control-A:} This operation produces the Young diagram labels for the new representation according to
\vspace*{-0.4cm}\begin{eqnarray}
\vert \lambda_1,\lambda_2,\lambda_3\rangle \longrightarrow \left\lbrace\begin{array}{cc}
\vert \lambda_1-1,\lambda_2,\lambda_3\rangle & \text{if} \hspace*{.2cm}\vert p\rangle =\vert 2\rangle =\vert \underline{10}\rangle\\
\vert \lambda_1,\lambda_2-1,\lambda_3\rangle & \text{if} \hspace*{.2cm}\vert p\rangle =\vert 1\rangle =\vert \underline{01}\rangle\\
\vert \lambda_1,\lambda_2,\lambda_3-1\rangle & \text{if} \hspace*{.2cm}\vert p\rangle =\vert 0\rangle =\vert \underline{00}\rangle
\end{array} \right.\nonumber
\end{eqnarray} Let the padded qutrit be specified as $\vert p \rangle=\vert \underline{p}(1)\underline{p}(2)\rangle$. Then these transformations can be realised using the following $C_{-}$ gates representing ternary subtractors [Appendix B]:
\vspace*{-0.4cm}\begin{subequations}
\begin{eqnarray}
C_-\vert \underline{p}(1)\rangle\vert \lambda_1\rangle &=& \vert \underline{p}(1)\rangle\vert \lambda_1-\underline{p}(1)\rangle \\
C_-\vert \underline{p}(2)\rangle\vert \lambda_2 \rangle &=& \vert \underline{p}(2)\rangle\vert \lambda_2-\underline{p}(2)\rangle \\
C_-\vert \underline{p}(2)\rangle\vert 1\rangle &=& \vert \underline{p}(2)\rangle\vert 1-\underline{p}(2)\rangle \\
C_-\vert \underline{p}(1)\rangle \vert 1-\underline{p}(2)\rangle &=& \vert \underline{p}(1)\rangle \vert 1-\underline{p}(2) -\underline{p}(1)\rangle \end{eqnarray}\begin{eqnarray}
C_-\vert 1-\underline{p}(2)-\underline{p}(1)\rangle\vert \lambda_3\rangle &=& \vert 1-\underline{p}(2) -\underline{p}(1)\rangle\vert  \lambda_3-1+\underline{p}(2)+\underline{p}(1)\rangle \nonumber\\
\end{eqnarray}
\end{subequations}
\item \textbf{Rotation gate $R^T(\theta,F)$:} This operation rotates the qutrit $\vert p\rangle$ in accordance with (3.42) and (3.43).
\begin{itemize}
\item \textbf{Change of variables:} We observe that the arguments of $\theta,F$ in the third row of  $\hat{\rho}, \hat{\sigma}$ differ from those in their first two rows. Furthermore, Table 3.1 shows that the transformed states corresponding to $p=0$ are just the same as the transformed states corresponding to $p=2,1$ with $k+1,l+1,m+1$ replacing $k,l,m$. To make equations compact, we introduce variables $k'',l'',m''$, and write rows of the $\hat{\rho},\hat{\sigma}$ matrices as: \footnotesize
\begin{eqnarray}\begin{aligned}
&\left[\hat{\rho}(\theta,F)\right]= \begin{bmatrix}
-sin(\theta_{k'',l'',m''})F^{u,p}_{k'',l''-1} & cos(\theta_{k'',l'',m''})F^{u,p}_{k'',l''-1}& F^{s,p}_{k'',l''}
\end{bmatrix}\\
&\left[\hat{\sigma}(\theta,F)\right]= \begin{bmatrix}
cos(\theta_{k'',l'',m''})F^{u,p}_{k''-1,l''} & sin(\theta_{k'',l'',m''})F^{u,p}_{k''-1,l''}& F^{s,p}_{k'',l''}
\end{bmatrix}
\end{aligned}\end{eqnarray} \normalsize  Here \footnotesize 
\begin{eqnarray}
\vert k'',l'',m'' \rangle &=&\left\lbrace \begin{array}{ccc} \vert k,l,m \rangle & \text{for} & \vert p \rangle = \vert 2 \rangle , \vert 1 \rangle \\ \vert k+1,l+1,m+1 \rangle & \text{for} & \vert p \rangle = \vert 0 \rangle
\end{array} \right. \\
\vert k''+l'',l''+m'',k''+m'' \rangle &=&\left\lbrace \begin{array}{ccc} \vert k+l,l+m,k+m \rangle & \text{for} & \vert p \rangle = \vert 2 \rangle , \vert 1 \rangle \\ \vert k+l+2,l+m+2,k+m+2 \rangle & \text{for} & \vert p \rangle = \vert 0 \rangle
\end{array} \right. \nonumber
\end{eqnarray} \normalsize All three combinations $k+l,l+m,k+m$ transform to $k''+l'',l''+m'',k''+m''$ the same way, and can be obtained using the following logic gates.
\footnotesize
\begin{subequations}
\begin{eqnarray}
C_+\vert \underline{\left(1-\underline{p}(2)-\underline{p}(1)\right)0}\rangle  \vert k+m\rangle  &=& \vert \underline{\left(1-\underline{p}(2)-\underline{p}(1)\right)0}\rangle \vert k+m+ \underline{\left(1-\underline{p}(2)-\underline{p}(1)\right)0}\rangle  \nonumber\\ &=&\vert k''+m''\rangle \\
 C_+\vert \underline{\left(1-\underline{p}(2)-\underline{p}(1)\right)0}\rangle  \vert l+m\rangle  &=& \vert \underline{\left(1-\underline{p}(2)-\underline{p}(1)\right)0}\rangle \vert l+m+ \underline{\left(1-\underline{p}(2)-\underline{p}(1)\right)0}\rangle  \nonumber\\ &=&\vert l''+m''\rangle \\ 
 C_+\vert \underline{\left(1-\underline{p}(2)-\underline{p}(1)\right)0}\rangle  \vert k+l\rangle  &=& \vert \underline{\left(1-\underline{p}(2)-\underline{p}(1)\right)0}\rangle \vert k+l+ \underline{\left(1-\underline{p}(2)-\underline{p}(1)\right)0}\rangle  \nonumber\\ &=&\vert k''+l''\rangle \end{eqnarray}
\end{subequations}\normalsize 
\item \textbf{Recovery of ancilla:} Just before the rotation gate is applied to $\vert p \rangle$, we recover the ancilla qubit used in the $C_{-}$ operations that changed the quantum numbers. This qubit is then reused in the later part of the circuit.
\begin{subequations}
\begin{eqnarray}
C_+ \vert \underline{p}(1) \rangle \vert  1-\underline{p}(2)-\underline{p}(1) \rangle &=& \vert \underline{p}(1) \rangle \vert 1-\underline{p}(2) \rangle \\
C_+ \vert \underline{p}(2) \rangle \vert  1-\underline{p}(2) \rangle &=& \vert \underline{p}(2) \rangle \vert 1 \rangle
\end{eqnarray}\end{subequations} 
\item \textbf{Choice of  $\hat{\sigma}$ or $\hat{\rho}$:} Before we rotate $\vert p\rangle$ using the transformation (3.43), we need to choose whether to apply $\hat{\sigma}$  or $\hat{\rho}$ on $\vert p \rangle$. As pointed out before, when expressed in the new labels, the choice depends on whether  $F^{u,p}_{k''-1,l''}$ or $F^{u,p}_{k'',l''-1}$ is non-zero. We use this condition to construct a control gate $D$ that acts on an ancilla bit $\vert 0 \rangle$, converting it to $\vert 1 \rangle$ for $T'=T+1/2$ and leaving it unchanged for $T'=T-1/2$.
\footnotesize \begin{eqnarray}
D\vert 0 \rangle = \frac{1}{N}\big(|F^{u,p}_{k''-1,l''}| +  \sigma_x |F^{u,p}_{k'',l''-1}|\big)\vert 0 \rangle \equiv \vert w \rangle,
\end{eqnarray}\normalsize
where $N=\sqrt{(F^{u,p}_{k'',l''-1})^2+(F^{u,p}_{k''-1,l''})^2}$. Then,  \begin{eqnarray}
\vert q\rangle =\left[\hat{R}(\theta,F)\right]^T\vert p\rangle &= \sum _{p=1}^3 (\delta_{T',T+1/2} \hat{\rho}_{3-p,3-q}+\delta_{T',T-1/2}\hat{\sigma}_{3-p,3-q}) \vert p\rangle \nonumber\\
&= \sum _{p=1}^3 (w\times\hat{\rho}_{3-p,3-q}+(1-w)\times\hat{\sigma}_{3-p,3-q}) \vert p\rangle . \nonumber\\ \end{eqnarray} 
Here $\hat{\rho}_{3-p,3-q},\hat{\sigma}_{3-p,3-q}$ are the matrix elements of the $\left[\hat{\rho}\right]^T,\left[\hat{\sigma}\right]^T$ matrices, and $ \vert q\rangle$ represents the quark states. (3.51) is achieved in the logic circuit by using a $C_{\left[R(\theta,F)\right]^T}=C_{\left[\rho\right]^T}\cdot \tilde{C}_{\left[\sigma\right]^T}$ gate controlled by $\vert w \rangle$.
\end{itemize}

\item \textbf{Control-B:} The action of the rotation matrix on the qutrit gives a superposition of states with their CG coefficients. Thereafter, the corresponding tensor products of $T+1/2$ or $T-1/2$ states have to be obtained according to Table 3.1 in terms of $k'',l'',m''$ as defined in (3.47).
\footnotesize
 \begin{eqnarray}
 \vert k',l',m'\rangle = \left\lbrace \begin{array}{cc} \left. \begin{array}{cc}
 \vert k''-1,l'',m''-1\rangle & \text{for} \hspace*{.2cm}q    =2 \\ \vert k''-1,l'',m''\rangle & \text{for} \hspace*{.2cm}q    =1\\ \vert k'',l'',m''\rangle & \text{for} \hspace*{.2cm}q    =0    \end{array} \right\rbrace & \text{for} \hspace*{.2cm}T-1/2 \hspace*{.1cm}\text{states}\\ \\ 
 \left. \begin{array}{cc}
 \vert k'',l''-1,m''-1\rangle & \text{for} \hspace*{.2cm}q    =2 \\ \vert k'',l''-1,m''\rangle & \text{for} \hspace*{.2cm}q    =1\\\vert k'',l'',m''\rangle & \text{for} \hspace*{.2cm}q    =0     \end{array}\right\rbrace &\text{for} \hspace*{.2cm} T+1/2 \hspace*{.1cm} \text{states}
 \end{array}\right.
 \end{eqnarray}\normalsize The changes needed in the quantum numbers $k''+l'',l''+m'',k''+m''$ to obtain $ k'+l',l'+m',k'+m'$ are then as given in Table 3.2. \\
 \begin{center}
 \begin{tabular}{|c|ccc|ccc|}
 \hline 
  $T'$&  & $T-\dfrac{1}{2}$ &  &  & $T+\dfrac{1}{2}$ &  \\ 
 \hline 
 $\vert q     \rangle$ & $\vert 2 \rangle$ & $\vert 1 \rangle$ & $\vert 0 \rangle$ & $\vert 2 \rangle$ & $\vert 1 \rangle$ & $\vert 0 \rangle$ \\ 
 \hline 
 $\vert k''+l'' \rangle$ & -1 & -1 & 0 & -1 & -1 & 0 \\ 
 \hline 
 $\vert l''+m'' \rangle$ & -1 & 0 & 0 & -2 & -1 & 0 \\ 
 \hline 
 $\vert k''+m'' \rangle$ & -2 & -1 & 0 & -1 & 0 & 0 \\ 
 \hline 
 \end{tabular}
 \captionof{table}{Changes in quantum numbers ($k''+l'',l''+m'',l''+m''$) during $U_{CG}^{-1}$}
 \end{center} 
We see that firstly, the quantum numbers remain unchanged for $q    =0$ for both $T \pm 1/2$ states. Secondly, the variable $k''+l''$ changes by the same amount for both $T \pm 1/2$ states. Thirdly, the changes in variables $k''+m''$ and $l''+m''$ get swapped for $T+1/2$ and $T-1/2$ states.\\ 
To implement these changes, we first use logic gates that carry out the required changes for $T-1/2$ states, and then make extra corrections for $T+1/2$ states. The extra corrections are addition of 1 to $k''+m''$ and subtraction of 1 from $l''+m''$, only when $q \neq 0$. They require an operation which does not make any corrections to the quantum numbers when either $q=0$ or the state is $T-1/2$. We have already constructed the qubit $\vert w \rangle$ in (3.50), which is $\vert 0 \rangle$ when the state is $T-1/2$ and $\vert 1 \rangle$ when the state is $T+1/2$. We also define the qubit $ \vert \underline{q} \rangle \equiv \vert \underline{q }(1) \rangle \oplus \vert \underline{q }(2) \rangle$ which is $\vert 0 \rangle$ for $ q =0$  and $\vert 1\rangle$ for $q =2/1$. Now let
$\vert z'\rangle =(1-\underline{q})\vert 0\rangle+\underline{q}\vert w \rangle =\vert \underline{q}w\rangle$.  We first convert the ancilla qubit recovered in (3.49) to $\vert q \rangle$, and then obtain the qubit $\vert z' \rangle$ by operating a $C^2_{NOT}$ gate, controlled by $\vert \underline{q} \rangle$ and $\vert  w \rangle$, on another ancilla qubit initialized to $\vert 0 \rangle$,
\begin{equation}
C^2_{NOT} \vert \underline{q}\rangle \vert w \rangle \vert 0\rangle=\vert \underline{q}\rangle \vert w \rangle \vert z'\rangle.
\end{equation} 

With this strategy, the following logic gates yield the desired values of $k'+l',l'+m',k'+m'$:
\begin{subequations}\begin{eqnarray}
C_- \vert z'\rangle\vert l''+m''\rangle  &=& \vert z'\rangle\vert l''+m''-z'\rangle \\ 
C_+\vert z'\rangle \vert k''+m''\rangle  &=&\vert z'\rangle \vert k''+m''+z'\rangle \\ 
C_-\vert \underline{q    }(1)\rangle \lbrace C_-\vert \underline{q    }(2)\rangle\vert k''+l''\rangle \rbrace &=& \vert\underline{q    }(1) \rangle\vert \underline{q    }(2)\rangle\vert k''+l''-\underline{q    }(1)-\underline{q    }(2)\rangle \nonumber\\ &=& \vert\underline{q    }(1) \rangle\vert \underline{q    }(2)\rangle\vert k'+l'\rangle \\
C_-\vert q    \rangle\vert k''+m''+z'\rangle &=&\vert q    \rangle\vert k''+m''+z'-q    \rangle\nonumber\\
&=&\vert q    \rangle\vert k'+m'\rangle\\
C_- \vert \underline{q    }(1)\rangle\vert l''+m''-z'\rangle &=&\vert \underline{q    }(1)\rangle\vert l''+m''-z'-\underline{q    }(1)\rangle \nonumber\\ &=&\vert \underline{q    }(1)\rangle\vert l'+m' \rangle
\end{eqnarray} \end{subequations}After the operation (3.54b), we restore the ancilla qubits to their initial states, from $\vert z' \rangle$ to $\vert 0 \rangle$ by a $C^2_{NOT}$ gate and $\vert \underline{q} \rangle$ to $\vert 0 \rangle$ by $C_-$ gates.
\end{itemize}

The final result combining (3.45)-(3.54) is \footnotesize \begin{eqnarray}
&&U_{CG}^{-1} \vert \lambda_1,\lambda_2,\lambda_3;k+l,l+m,k+m:p\rangle =\nonumber\\
&&\sum_{possibilities}\vert \lambda_1',\lambda_2',\lambda_3';k''+l''-\underline{q}(2)-\underline{q}(1),l''+m''-\underline{q}(1)-z',k''+m''+z'-q:q\rangle.\nonumber\\
\end{eqnarray}\normalsize 

\subsection{Efficient quantum circuit for $U_{CG}^{-1}$}
\begin{figure}[!]
\centering
\framebox{\includegraphics[scale=0.8,angle=90]{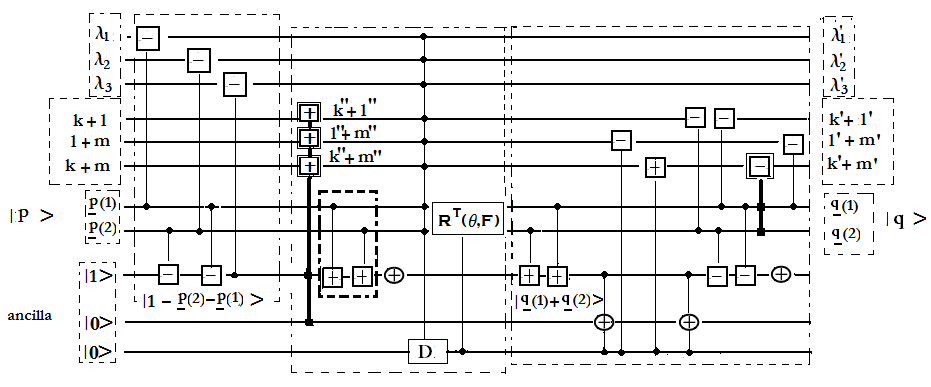}}
\caption{\label{y7}Efficient quantum circuit to implement $U_{CG}^{-1}$ operation for $SU(3)$}
\end{figure}

 The logic operations of the previous section are depicted in Figure \ref{y7}, with dotted boxes grouping the quantum numbers and enclosing the three parts of the whole transformation.
  We point out the following:
 \begin{itemize}
 \item The variables in the dotted boxes expand the variables shown on a single line in the schematic diagram of Figure \ref{y5}.
 \item Lines other than the ancilla and $\underline{p}(1), \underline{p}(2)$ are qutrit registers, with sufficient number of carry bits to implement additions and subtractions.
 \item Every qutrit is represented by padded qubits as per (3.36). The circuit uses padded qutrit $\vert p \rangle$ as well as individual qubits $\vert \underline{p}(1)\rangle,\vert \underline{p}(2)\rangle$.
 \item  Qubits are subtracted from and added to qutrits using binary subtractors ($\boxminus$) and binary adders ($\boxplus$). The gates with thick lines and double boxed operators are ternary subtractors and adders. They can be implemented using conventional qubit gates on padded qubits [Appendix B].
 \item The first ancilla qubit stores the values of $\vert 1-\underline{p}(1)-\underline{p}(2)\rangle$ and $\vert \underline{q}(1)+\underline{q}(2)\rangle$. The second ancilla qubit stores the value of $\vert z' \rangle$, and also forms, together with the transformed qubit (3.45e), the qutrit needed in (3.48). Both these qubits are recovered at the end of $U_{CG}^{-1}$.  
 \item The last ancilla qubit  stores the value of $\vert w \rangle$. It is not recovered, and has to be reset to $\vert 0 \rangle$ before the execution of the next $U_{CG}^{-1}$. 
 \end{itemize}

\subsection{Examples}
We illustrate our algorithm with some simple examples. The Schur basis is written as $ \vert P,Q;T,T_3,Y\rangle =\vert \lambda_1,\lambda_2,\lambda_3;k+l,m+l,k+m,p_1,p_2,..p_{n-1}\rangle,$ with the relations among the quantum numbers  given by (3.27). We follow the scheme of (3.55), with $p_i$ in padded qubit notation.  In the following, the specific $p_i$ undergoing transformation is represented in boldface.

In every example below, first $k'',l'',m''$ are determined according to (3.47), and then the isoscalar factors are obtained for $T+1/2$ and $T-1/2$ states using the rotation gate $R^T(\theta,F)$. Subsequently, the corresponding $SU(2)$ CG-coefficients are calculated to carry out the desired transformation on $p_i$.  

\footnotesize
\begin{enumerate}
  \item[\textbf{I.}] $ n=3,P=3,Q=0; T=1,T_3=0,Y=0; p_1=2,p_2=2 $\\ $\longleftrightarrow \Yvcentermath1 \young(uds) \longleftrightarrow
         \vert 3,0,0;2,1,3:\underline{10},\underline{10}\rangle_{sch}$ \\ 
  \textit{$1^{st}$ iteration:} $ \vert p \rangle = \vert 2 \rangle \Rightarrow (k'',l'')=(k,l).$
\vspace{-0.4cm}\begin{subequations} 
  \begin{eqnarray} 
 F^s_{k'',l''}&=& F^s_{k,l}= \sqrt{\frac{1}{3}}. \nonumber\\
   F^u_{k'',l''-1}&=&F^u_{k,l-1}=0, \hspace*{.7cm} F^u_{k''-1,l''}=F^u_{k-1,l}=\sqrt{\frac{2}{3} }.\nonumber\\
   \therefore T'&=&T-1/2, \hspace*{1cm} \vert w \rangle = \vert 0 \rangle. \nonumber\\
  cos(\theta_{k-1,l})&=&\sqrt{\dfrac{1}{2}} , \hspace*{1cm} sin(\theta_{k-1,l})=\sqrt{\dfrac{1}{2}}. \nonumber\\\nonumber\\
  \therefore \vert 3,0,0;2,1,3:\underline{10},\underline{\textbf{10}} \rangle_{sch} & \rightarrow & \sqrt{\frac{1}{3}} \left(\vert 2,0,0;1,0,1:\underline{10},\underline{\textbf{10}}\rangle +\vert 2,0,0;1,1,2:\underline{10},\underline{\textbf{01}}\rangle \right.\nonumber\\ &&+ \left. \vert 2,0,0;2,1,3:\underline{10},\underline{\textbf{00}}\rangle \right) .
  \end{eqnarray}
   \textit{$2^{nd}$ iteration:}$\vert p\rangle= \vert 2\rangle \Rightarrow(k'',l'')=(k,l)$
   \begin{eqnarray}
   \vert 2,0,0;1,0,1:\underline{\textbf{10}},\underline{10}\rangle &\rightarrow& \sqrt{\frac{1}{2}} \left(\vert 1,0,0;0,0,0:\underline{\textbf{01}},\underline{10}\rangle +\vert 1,0,0;1,0,1:\underline{\textbf{00}},\underline{10}\rangle \right) \nonumber\\\\
   \text{with}\hspace*{1cm}F^s_{k'',l''}&=& F^s_{k,l}= \sqrt{\frac{1}{2}}, \hspace*{.7cm} F^u_{k''-1,l''}=F^u_{k-1,l}=\sqrt{\frac{1}{2} },\nonumber\\
     cos(\theta_{k-1,l})&=&0,  \hspace*{1cm} sin(\theta_{k-1,l})=1 .\nonumber\\        
    \vert 2,0,0;1,1,2:\underline{\textbf{10}},\underline{01}\rangle &\rightarrow& \sqrt{\frac{1}{2}} \left(\vert 1,0,0;1,1,2:\underline{\textbf{00}},\underline{01}\rangle +\vert 1,0,0;0,0,0:\underline{\textbf{10}},\underline{01}\rangle \right) \nonumber\\\\\text{with}\hspace*{1cm}
   F^s_{k'',l''}&=& F^s_{k,l}= \sqrt{\frac{1}{2}}, \hspace*{.7cm} F^u_{k''-1,l''}=F^u_{k-1,l}=\sqrt{\frac{1}{2} },\nonumber \\
     cos(\theta_{k-1,l})&=&1,  \hspace*{1cm} sin(\theta_{k-1,l})=0 .\nonumber \end{eqnarray} 
    \begin{eqnarray}
    \vert 2,0,0;2,1,3:\underline{\textbf{10}},\underline{00}\rangle &\rightarrow& \sqrt{\frac{1}{2}} \left(\vert 1,0,0;1,1,2:\underline{\textbf{01}},\underline{00}\rangle +\vert 1,0,0;1,0,1:\underline{\textbf{10}},\underline{00}\rangle \right) \nonumber\\\\ \text{with}\hspace*{1cm}
   F^s_{k'',l''}&=& F^s_{k,l}= 0, \hspace*{.7cm} F^u_{k''-1,l''}=F^u_{k-1,l}=1,\nonumber\\
     cos(\theta_{k-1,l})&=&\sqrt{\frac{1}{2}},  \hspace*{1cm} sin(\theta_{k-1,l})=\sqrt{\frac{1}{2}} .\nonumber
    \end{eqnarray}
  Combining the two iterations, we have 
     \begin{eqnarray}
   &&\vert 3,0,0;2,1,3:\underline{10},\underline{10} \rangle_{sch}  \rightarrow  \sqrt{\frac{1}{6}} \left\lbrace \vert 1,0,0;1,1,\textbf{2}:\underline{\textbf{01}},\underline{\textbf{00}}\rangle +\vert 1,0,0;1,0,\textbf{1}:\underline{\textbf{10}},\underline{\textbf{00}}\rangle \right. \nonumber\\ &&\left. + \vert 1,0,0;1,1,\textbf{2}:\underline{\textbf{00}},\underline{\textbf{01}}\rangle +\vert 1,0,0;0,0,\textbf{0}:\underline{\textbf{10}},\underline{\textbf{01}}\rangle + \vert 1,0,0;0,0,\textbf{0}:\underline{\textbf{01}},\underline{\textbf{10}}\rangle \right. \nonumber\\ &&\left. +\vert 1,0,0;1,0,\textbf{1}:\underline{\textbf{00}},\underline{\textbf{10}}\rangle\right\rbrace \\
   &&\hspace*{3.7cm}  =\sqrt{\frac{1}{6}}(\vert uds\rangle + \vert dus\rangle +\vert usd\rangle +\vert sud\rangle +\vert sdu\rangle +\vert dsu\rangle) .\nonumber
   \end{eqnarray}
  \end{subequations}
  \item[\textbf{II.}] $n=3,P=1,Q=1,T=1,T_3=0,Y=0,p_1=2,p_2=1 $\\\\$\longleftrightarrow \Yvcentermath1 \young(ud,s)$ \hspace*{0.5cm} as  \hspace*{0.5cm}$ \Yvcentermath1\young(12,3) \longleftrightarrow  \vert 2,1,0;2,1,3:\underline{10},\underline{01}\rangle_{sch}$\\\\
   \textit{$1^{st}$ iteration:} $ \vert p \rangle = \vert 1 \rangle \Rightarrow (k'',l'')=(k,l)$
\begin{subequations} 
  \begin{eqnarray}
 F^s_{k'',l''}&=&F^s_{k,l}= \sqrt{\frac{2}{3}} ,\nonumber\\ F^u_{k''-1,l''}&=&F^u_{k-1,l}= -\sqrt{\frac{1}{3}}, \hspace*{.5cm} F^u_{k'',l''-1}= F^u_{k,l-1}= 0.\nonumber\\
  \therefore T'&=&T-1/2, \hspace*{1cm} \vert w \rangle = \vert 0 \rangle  . \nonumber\\
   cos(\theta_{k-1,l})&=&\sqrt{\frac{1}{2}},  \hspace*{1cm} sin(\theta_{k-1,l})=\sqrt{\frac{1}{2}}. \nonumber\\\nonumber\\
   \therefore \vert 2,1,0;2,1,3:\underline{10},\underline{\textbf{01}}\rangle_{sch} &\rightarrow & \sqrt{\frac{2}{3}}\vert 2,0,0;2,1,3:\underline{10},\underline{\textbf{00}}\rangle   - \sqrt{\frac{1}{6}}\left( \vert 2,0,0;1,0,1:\underline{10},\underline{\textbf{10}}\rangle \right.\nonumber\\&+& \left.\vert 2,0,0;1,1,2:\underline{10},\underline{\textbf{01}}\rangle \right).
  \end{eqnarray} 
  \textit{$2^{nd}$ iteration:} \\
   The terms on RHS of (3.57a) have already been decomposed in (3.56b)-(3.56d), giving the final result
    \vspace*{-.5cm}\begin{eqnarray}&&\vert 2,1,0;2,1,3:2,1\rangle_{sch}\rightarrow 
    \sqrt{\frac{1}{3}}\left(\vert 1,0,0;1,0,\textbf{1}:\underline{\textbf{10}},\underline{\textbf{00}}\rangle + \vert 1,0,0;1,1,\textbf{2}:\underline{\textbf{01}},\underline{\textbf{00}}\rangle \right) \nonumber\\&&\hspace*{3.7cm} -\sqrt{\frac{1}{12}}\left(\vert 1,0,0;1,1,\textbf{2}:\underline{\textbf{00}},\underline{\textbf{01}}\rangle +\vert 1,0,0;0,0,\textbf{0}:\underline{\textbf{10}},\underline{\textbf{01}}\rangle \right) \nonumber\\&&\hspace*{3.7cm}- \sqrt{\frac{1}{12}}\left(\vert 1,0,0;1,0,\textbf{1}:\underline{\textbf{00}},\underline{\textbf{10}}\rangle +\vert 1,0,0;0,0,\textbf{0}:\underline{\textbf{01}},\underline{\textbf{10}}\rangle  \right)\nonumber\\\nonumber 
    &&\hspace*{3.3cm}= \frac{1}{\sqrt{3}}\left(\vert dus\rangle + \vert uds\rangle \right)-\frac{1}{\sqrt{12}} \left( \vert usd\rangle + \vert sud\rangle +\vert dsu\rangle +\vert sdu\rangle\right) .\nonumber\\
    \end{eqnarray}
  \end{subequations}
  
  \item[\textbf{III.}] $n=3,P=1,Q=1,T=0,T_3=0,Y=0,p_1=1,p_2=2.$\\\\$\longleftrightarrow \Yvcentermath1 \young(us,d)$ \hspace*{0.5cm} as  \hspace*{0.5cm}$ \Yvcentermath1\young(13,2) \longleftrightarrow  \vert 2,1,0;2,2,2:\underline{01},\underline{10}\rangle_{sch}$\\\\
  \textit{$1^{st}$ iteration:} $ \vert p \rangle = \vert 2 \rangle \Rightarrow (k'',l'')=(k,l)$
\begin{subequations} 
  \begin{eqnarray}
 F^s_{k'',l''}&=&F^s_{k,l}= \sqrt{\frac{2}{3}} ,\nonumber\\F^u_{k'',l''-1}&=&F^u_{k,l-1}= -\sqrt{\frac{1}{3}}, \hspace*{.5cm} F^u_{k''-1,l''}= F^u_{k-1,l}= 0.\nonumber\\
  \therefore T'&=&T+1/2, \hspace*{1cm}\vert w \rangle = \vert 1 \rangle  . \nonumber\\
   cos(\theta_{k,l-1})&=&\sqrt{\frac{1}{2}},  \hspace*{1cm} sin(\theta_{k,l-1})=\sqrt{\frac{1}{2}}. \nonumber\\\nonumber
   \therefore \vert 2,1,0;2,2,2:\underline{01},\underline{\textbf{10}}\rangle_{sch} &\rightarrow & \sqrt{\frac{2}{3}}\vert 1,1,0;2,2,2:\underline{01},\underline{\textbf{00}}\rangle   + \sqrt{\frac{1}{6}}\left( \vert 1,1,0;1,0,1:\underline{01},\underline{\textbf{10}}\rangle \right.\nonumber\\&-& \left.\vert 1,1,0;1,1,2:\underline{01},\underline{\textbf{01}}\rangle \right).
  \end{eqnarray} 
   \textit{$2^{nd}$ iteration:} $ \vert p \rangle = \vert 1 \rangle \Rightarrow (k'',l'')=(k,l)$. 
      \begin{eqnarray}
      \vert 1,1,0;1,0,1:\underline{\textbf{01}},\underline{10}\rangle &\rightarrow & \sqrt{\frac{1}{2}} \left(\vert 1,0,0;1,0,1:\underline{\textbf{00}},\underline{10}\rangle - \vert 1,0,0;0,0,0:\underline{\textbf{01}},\underline{10}\rangle \right) \nonumber\\ \text{with}\hspace*{1cm}
  F^s_{k'',l''}&=&F^s_{k,l}= \sqrt{\frac{1}{2}}, \nonumber\\
   F^u_{k''-1,l''}&=&F^u_{k-1,l}= -\sqrt{\frac{1}{2}}, \\
   cos(\theta_{k-1,l})&=&0,  \hspace*{2.8cm} sin(\theta_{k-1,l})=1. \nonumber \end{eqnarray}\begin{eqnarray}
   \vert 1,1,0;1,1,2:\underline{\textbf{01}},\underline{01}\rangle &\rightarrow & \sqrt{\frac{1}{2}}\left(\vert 1,1,0;1,1,2:\underline{\textbf{00}},\underline{01}\rangle-\vert 1,0,0;0,0,0:\underline{\textbf{10}},\underline{01}\rangle \right) \nonumber\\ 
 \text{with}\hspace*{1cm}F^s_{k'',l''}&=&F^s_{k,l}= \sqrt{\frac{1}{2}}, \nonumber\\
   F^u_{k''-1,l''}&=&F^u_{k-1,l}= -\sqrt{\frac{1}{2}}, \\
   cos(\theta_{k-1,l})&=&1,  \hspace*{2.8cm} sin(\theta_{k-1,l})=0. \nonumber\\\nonumber 
    \vert 1,1,0;2,2,2:\underline{\textbf{01}},\underline{00}\rangle &\rightarrow & \sqrt{\frac{1}{2}}\left( \vert 1,0,0;1,0,1:\underline{\textbf{10}},\underline{00}\rangle -\vert 1,0,0;1,1,2:\underline{\textbf{01}},\underline{00}\rangle \right)  \nonumber\\\text{with}\hspace*{1cm}
 F^s_{k'',l''}&=&F^s_{k,l}=0,  \nonumber\\
   F^u_{k'',l''-1}&=&F^u_{k,l-1}= -1 .\\
  cos(\theta_{k,l-1})&=&\sqrt{\frac{1}{2}},  \hspace*{.8cm} sin(\theta_{k,l-1})=\sqrt{\frac{1}{2}}.\nonumber 
  \end{eqnarray}
     Combining the two iterations, we have
   \begin{eqnarray}&&\vert 2,1,0;2,2,2:1,2\rangle_{sch} \rightarrow\nonumber\\
  && \frac{1}{\sqrt{12}}\left( \vert 1,0,0;1,0,\textbf{1}:\underline{\textbf{00}},\underline{\textbf{10}}\rangle - \vert 1,0,0;0,0,\textbf{0}:\underline{\textbf{01}},\underline{\textbf{10}}\rangle +\vert 1,0,0;0,0,\textbf{0}:\underline{\textbf{10}},\underline{\textbf{01}}\rangle \right. \nonumber\\ &&\left. -\vert 1,1,0;1,1,\textbf{2}:\underline{\textbf{00}},\underline{\textbf{01}}\rangle +2\vert 1,0,0;1,0,\textbf{1}:\underline{\textbf{10}},\underline{\textbf{00}}\rangle -2\vert 1,0,0;1,1,\textbf{2}:\underline{\textbf{01}},\underline{\textbf{00}}\rangle
\right)\nonumber\\
 &&=\frac{1}{\sqrt{12}}\left( \vert dsu\rangle -\vert sdu\rangle +\vert sud\rangle -\vert usd\rangle +2\vert dus\rangle -2\vert uds\rangle\right).
   \end{eqnarray}
  \end{subequations}

  \item[\textbf{IV.}] $n=3,P=0,Q=0,T=0,T_3=0,Y=0,p_1=1,p_2=0$\\\\$ \longleftrightarrow \Yvcentermath1 \young(u,d,s)$
  $ \longleftrightarrow  \vert 1,1,1;0,0,0:\underline{01},\underline{00}\rangle_{sch}$\\\\
   \textit{$1^{st}$ iteration:} $ \vert p \rangle = \vert 0 \rangle \Rightarrow (k'',l'')=(k+1,l+1)$
\begin{subequations} 
  \begin{eqnarray}
  F^s_{k'',l''}&=&F^s_{k+1,l+1}= \sqrt{\frac{1}{3}}, \nonumber\\
   F^u_{k'',l''-1}&=&F^u_{k+1,l}= \sqrt{\frac{2}{3}}, \hspace*{.7cm} F^u_{k''-1,l''}=F^u_{k,l+1}= 0.\nonumber\\
   \therefore T'&=&T+1/2, \hspace*{1cm} \vert w \rangle = \vert 1 \rangle .\nonumber\\
   cos(\theta_{k+1,l})&=&\sqrt{\frac{1}{2}},  \hspace*{1cm} sin(\theta_{k+1,l})=\sqrt{\frac{1}{2}}. \nonumber\\
  \therefore \vert 1,1,1;0,0,0:\underline{01},\underline{\textbf{00}}\rangle_{sch} &\rightarrow & \sqrt{\frac{1}{3}}\left(- \vert 1,1,0;1,0,1:\underline{01},\underline{\textbf{10}}\rangle +\vert 1,1,0;1,1,2:\underline{01},\underline{\textbf{01}}\rangle \right. \nonumber\\ &&\left.  + \vert 1,1,0;2,2,2:\underline{01},\underline{\textbf{00}}\rangle \right) .
   \end{eqnarray}
   \textit{$2^{nd}$ iteration:} \\ The terms on RHS of (3.59a) have already been decomposed in (3.58b)-(3.58d), giving the final result
   \begin{eqnarray}&&\vert 1,1,1;0,0,0:1,0\rangle_{sch} \rightarrow\nonumber\\
  && \frac{1}{\sqrt{6}}\left( -\vert 1,0,0;1,0,\textbf{1}:\underline{\textbf{00}},\underline{\textbf{10}}\rangle + \vert 1,0,0;0,0,\textbf{0}:\underline{\textbf{01}},\underline{\textbf{10}}\rangle -\vert 1,0,0;0,0,\textbf{0}:\underline{\textbf{10}},\underline{\textbf{01}}\rangle \right. \nonumber\\ &&\left. +\vert 1,1,0;1,1,\textbf{2}:\underline{\textbf{00}},\underline{\textbf{01}}\rangle +\vert 1,0,0;1,0,\textbf{1}:\underline{\textbf{10}},\underline{\textbf{00}}\rangle -\vert 1,0,0;1,1,\textbf{2}:\underline{\textbf{01}},\underline{\textbf{00}}\rangle
\right)\nonumber\\
 &&=\frac{1}{\sqrt{6}}\left( \vert sdu\rangle -\vert dsu\rangle -\vert sud\rangle +\vert usd\rangle +\vert dus\rangle -\vert uds\rangle\right).
   \end{eqnarray}
  \end{subequations}
   \end{enumerate}\normalsize
   \section{Complexity analysis}
The structure of the $SU(3)$ algorithm is the same as that for the $SU(2)$ case.
\begin{itemize}
\item  Number of iterations (each consisting of a single $U_{CG}^{-1}$) increase linearly with the number of qutrits $n$.
\item The resources needed by the algorithm to generate an $n$-qutrit state are:
\begin{itemize}
\item \textit{\textbf{Space:}}  Each of $\lambda_1, \lambda_2,\lambda_3, k''+l'',l''+m'',k''+m''$ are represented by $\log_{_{3}} n +1$ qutrits. Furthermore, $n-1$ qutrits for $\vert p_i \rangle$'s and three ancilla qubits are required to run the algorithm. Since each qutrit is represented as a padded qubit, we need $6\lbrace 2\left( log_{_{3}} n+1\right)\rbrace + 2(n-1)+3$ qubits, which scales as $\mathcal{O}\left[ log_{_{3}}\left( N\left(log_{_{3}} N\right)^6\right)\right]$ for large $N$.
\item \textit{\textbf{Time:}} Each iteration has a specific sequence of logic gates. They are tabulated below with the resources needed [Appendix B] to implement them.
\end{itemize}\footnotesize
\begin{center}
\begin{tabular}{|c|c|c|c|c|}
\hline 
Operation Type & $\substack{No.\hspace*{.1cm} of\hspace*{.1cm} operations \\in\hspace*{.2cm} U_{CG}^{-1}}$&$C_{NOT}$ & $C^2_{NOT}$ & $NOT$   \\ 
\hline 
Qutrit $\pm$ Qubit &8 &$3\left(log_{_{3}} n+1\right)$ & $3\left(log_{_{3}} n+1\right)$ & $2\left(log_{_{3}} n+1\right)$   \\ 
\hline 
Qutrit Register$\pm$ Qutrit &  4 &$3\left(log_{_{3}} n+2\right)$ & $3\left(log_{_{3}} n+2\right)$ & $2\left(log_{_{3}} n+2\right)$  \\ 
\hline 
Qubit operations & &8 &2  &2    \\ 
\hline 
\end{tabular} \captionof{table}{Number of different types of logic gates needed for different operations in $U_{CG}^{-1}$. The third, fourth and fifth columns indicate the number of gates required for a single operation.}\end{center} 
\normalsize\begin{itemize}
\item Therefore for $(n-1)$ iterations, the total number of logic gates needed are \footnotesize
$$(n-1)\lbrace \left( 56+36log_{_{3}} n\right)C_{NOT} + \left( 50+36log_{_{3}} n\right)C^2_{NOT} + \left( 34+24 log_{_{3}} n\right) NOT\rbrace$$ \normalsize
 which scales as $\mathcal{O}\left[log_{_{3}}N \cdot log_{_3}\left( log_{_{3}} N\right)\right]$ for large $N$.
 \item \textbf{\textit{Controlled rotations:}} The algorithm needs $(n-1)$ controlled qubit rotations for $D$, and $(n-1)$ controlled qutrit rotations for $R(\theta,F)$. The resources needed to implement them depend on the available hardware, and are $\mathcal{O}[log_{_{3}} N]$.
\end{itemize}
\item Thus, as per the arguments given in  Section 2.7, the algorithm belongs to the class BQP \citep{elementary-gates}, and can be used to generate any $n$-qutrit eigenstate of any $SU(3)$ representation.
  \end{itemize}

\chapter{$SU(d)$}
%\ifpdf
 %   \graphicspath{{sud/sudFigs/PNG/}{sud/sudFigs/PDF/}{sud/sudFigs/}}
%\else
 %   \graphicspath{{sud/sudFigs/EPS/}{sud/sudFigs/}}
%\fi
\normalsize
In previous Chapters, we have explicitly constructed efficient quantum circuits to obtain eigenstates of $SU(2)$ and $SU(3)$ representations in computational basis states. Here we discuss the possibility of extending this framework by mathematical induction to construct eigenstates of arbitrary representations of $SU(d)$ group, and then discuss the computational complexity of the algorithm. 

In case of $SU(d)$, the computational basis states are formed using qudits, \textit{i.e.,} $d$-dimensional vectors in $\mathbb{C}^d$ Hilbert space with the basis  $\{ 0,1,...,d-1 \}$. Each qudit can be represented by a register of $[\![ log_{_{2}}d+1 ]\!]$ qubits, where $[\![...]\!]$  denotes the integer part of the expression. The Schur basis state description of an eigenstate (formed by $n$ qudits) of $SU(d)$ representation is $\vert  [\lambda], q_\lambda, p_\lambda \rangle_{sch}$, where $[\lambda]$ is a collection of $d$ registers $\big( \lambda_1,\lambda_2, ...,\lambda_d\big)$ and $p_\lambda$ is a register of $n-1$ qudits \citep{schur-2}. $[\lambda]$ describes an irreducible representation of $SU(d)$ by a Young diagram of $d$ rows, with $k^{th}$ row having $\lambda_k$ boxes. $p_\lambda$ identifies the unique path by which the irreducible representation is  constructed, starting from the fundamental representation and adding one box at a time. $[\lambda]$ with $q_\lambda$ describes an eigenstate completely in terms of all its quantum numbers. Inductive use of the result $SU(d) \supset SU(d-1) \times U(1)$ from the general description of special unitary groups \citep{georgii}, enables us to identify $q_\lambda$ with $\frac{d(d-1)}{2}$ quantum numbers that uniquely specify an eigenstate corresponding to the irreducible representation $[\lambda]$. The largest value that a register $\lambda_k$ can have is $n$, and so all registers corresponding to $[\lambda]$ and $q_\lambda$ contain at most $[\![log_{_{d}}n+1 ]\!]$ qudits.

For implementing the algorithm to generate eigenstates of $SU(d)$ representation, we need to calculate the $SU(d)$ CG-coefficients. For this, we use the result that any $U_{CG}$ for $SU(d)$ can be factored as a product of $U_{CG}$ for $SU(d-1)$  and the corresponding reduced Wigner coefficient matrix, as described in \citep{schur-2}. Recursive use of this decomposition $(d-2)$ times then breaks down the $SU(d)$ CG-coefficients to $SU(2)$ CG-coefficients and various reduced Wigner coefficients. Once the $U_{CG}$ for $SU(d)$ is calculated by this method, one needs to run $n-1$ iterations of $U_{CG}$ to obtain the desired eigenstate in computational basis, as depicted in Figure 1.2. 

As a particular case, the reduced Wigner coefficients \citep{wigner} in case of $SU(3)$ are the isoscalar factors, and the $\left[  \hat{R}(\theta,\phi)\right]$ corresponding to (3.46) can be written as \footnotesize
\begin{subequations}
\begin{eqnarray}
&&\left[\hat{R}(\theta,F)\right]=\left[\hat{\mathcal{R}}(F)\right] \cdot \left[\hat{\mathcal{R}}(\theta)\right]\hspace*{2cm}\text{where,}\\
&&\left[\hat{\mathcal{R}}(F)\right]= \begin{bmatrix}
F^{u,2}_{k''-1,l''} &F^{u,2}_{k'',l''-1} & F^{s,2}_{k'',l''}\\
F^{u,1}_{k''-1,l''} &F^{u,1}_{k'',l''-1} & F^{s,1}_{k'',l''}\\
F^{u,0}_{k''-1,l''} &F^{u,0}_{k'',l''-1} & F^{s,0}_{k'',l''}
\end{bmatrix}\\&&\left[\hat{\mathcal{R}}(\theta)\right]=\begin{bmatrix}
cos(\theta_{k'',l'',m''})&sin(\theta_{k'',l'',m''})&0\\
-sin(\theta_{k'',l'',m''})& cos(\theta_{k'',l'',m''})&0\\0&0&1
\end{bmatrix}
\end{eqnarray}\end{subequations}\normalsize
\\Here
$\left[\hat{\mathcal{R}}(\theta)\right]$ contains only the $SU(2)$ CG-coefficients, and $\left[\hat{\mathcal{R}}(F)\right]$ contains only the isoscalar factors. Thus the operation of rotation corresponding to $U_{CG}^{-1}$ for $SU(3)$ is decomposed into a rotation corresponding to  $U_{CG}^{-1}$ for $SU(2)$ and another rotation according to isoscalar factors, as shown schematically in Figure 4.1.
\begin{figure}[!]
\centering
\framebox{\includegraphics[scale=.7]{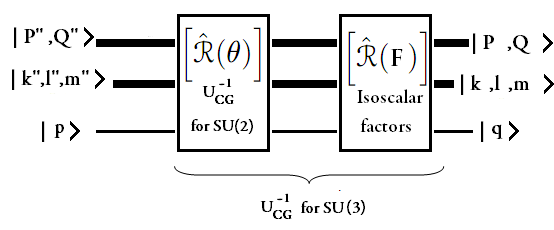}}
\caption{Schematic representation of $U_{CG}^{-1}$ for $SU(3)$ in terms of $U_{CG}^{-1}$ for $SU(2)$ and a rotation matrix containing isoscalar factors}
\end{figure}

This procedure can be extended to arbitrary group representations of $SU(d)$ \citep{schur-2}. It requires total $(d-1)$ rotation gates in $U_{CG}^{-1}$ for $SU(d)$, with one $U_{CG}^{-1}$ for $SU(2)$ and $(d-2)$ reduced Wigner coefficient matrices $\left[\hat{\mathcal{R}}_k(F)\right]$ corresponding to the groups $SU(k)$ with $k>2$.

\section{Computational complexity}
From the structure of the algorithm, we can infer the following requirements.\\ 

\textit{\textbf{Space:}}  We need $d$ registers for $[\lambda]$ and $\dfrac{d(d-1)}{2}$ registers for $q_\lambda$, each of size $[\![log_{_{d}}n+1 ]\!]$ qudits. For $p_\lambda$ we need a register of $(n-1)$ qudits. That adds up to
\begin{eqnarray}
&&\left(\left( d+\dfrac{d(d-1)}{2}\right) [\![log_{_{d}}n+1 ]\!]+n-1 \right) \text{qudits}\nonumber\\
=&&\left( \dfrac{d(d+1)}{2}[\![log_{_{d}}n+1 ]\!]+n-1 \right)[\![ log_{_{2}}d+1 ]\!] \text{qubits}.
\end{eqnarray} The algorithm also needs ancilla bits, whose number will be at most of the order of the number of qubits calculated above. Therefore the space requirement scales as  $\mathcal{O}\left[log_{_{d}}\left(N\left( log_{_{d}} N \right)^{d(d+1)/2}\right)\right]$. 

\textit{\textbf{Time:}} Just as in case of $SU(2)$ and $SU(3)$, we perform the $U_{CG}^{-1}$ operation $n-1$ times, with each $U_{CG}^{-1}$ having a total of $(d-1)$ rotation gates. The computational complexity of the rotation gates is a function of $d$. Specifically, the number of elementary logic gates needed to evaluate the CG and reduced Wigner coefficients is a polynomial in $d$ and $log_{_{d}}n$, say   $\mathcal{P}(d,log_{_{d}}n)$.  Altogether, $n-1$ iterations of $U_{CG}^{-1}$ need temporal resources scaling as $\mathcal{O}\left[ log_{_{d}}N \cdot \mathcal{P}(d,log_{_{d}}\left( log_{_{d}}N \right))\right]$. 

Therefore, the algorithm belongs to the class BQP \citep{elementary-gates} for any $d$, and can be used to generate any $n$-qudit eigenstate of any $SU(d)$ representation. 

\appendixpage
\renewcommand{\thechapter}{A}
\makeatletter
\renewcommand{\@chapapp}{Appendix}
\makeatother
\chapter{Young Diagrams and Young Tableaux}
%\begin{center}
%\textbf{\Large{Appendix A}:\\Young Diagrams and Young Tableaux}
%\end{center}\normalsize
 \section{Young diagrams}
  Young diagrams provide a pictorial notation that labels the irreducible representations of the permutation group $S_n$  \citep{georgii,hooklength}. Permutation group is the group of all  possible permutations of an $n$-object system. For example, consider 5 objects with a particular permutation
 $$\left( \begin{array}{ccccc}
  1&2&3&4&5 \\ 5&3&2&4&1
  \end{array}\right).$$ This permutation can be represented in its cycle notation as  (15)(23)(4), which stands for the movements: \begin{align}
  &1\longrightarrow 5\longrightarrow 1 \nonumber\\ &2\longrightarrow 3\longrightarrow 2 \nonumber\\ &4\longrightarrow 4 \nonumber
  \end{align}

Every permutation operation can be expressed in the cycle notation. The convention is that the cycle containing the largest number of elements is written first and  the smaller ones later. The above example has 3 cycles described by the notation (2,2,1). In general, each cycle has length $c_k$, and the sum of all cycle lengths is the total number of objects permuted.  Since, the number of cycles cannot be more than the number of objects, the index $k$ goes from 1 to n.  Thus we have
\begin{eqnarray}
 c_1 \geq c_2 \geq...\geq c_n\geq0, \hspace*{1cm} \sum_{k=1}^nc_k =n.
\end{eqnarray}
Every irreducible representation of the permutation group corresponds to a specific cycle structure. When each of these cycles of length $c_k$, is depicted as a number of consecutive square boxes in the $k^{th}$ column, we obtain a \textit{valid Young diagram}.

It can be shown that any arbitrary permutation of $n$ objects is equivalent to at most $n$ pairwise exchanges of objects called transpositions  \citep{georgii}. The parity of a permutation is defined as $(-1)^p$ with $p$ being the number of transpositions. Permutation with parity 1 (-1) is called even (odd) permutation. Thus a permutation can be labelled by symmetric or antisymmetric combinations of objects under exchange. These exchanges translate to rows and columns respectively in the language of Young diagrams, \textit{i.e.} for a specific state, exchange of boxes in a row is symmetric and that in a column is antisymmetric. In particular, objects in a cycle are in an antisymmetric state, while all cycles are mutually symmetric.

A dual description of a Young diagram can be given in terms of its rows, by defining partitions $\lambda_k$ as
\begin{equation}
\lambda_k =\sum_{i=n}^k [c_i], \end{equation} 
where $[c_i]$ is the number of cycles with cycle length $c_i$. From (A.1), it follows that the set of partitions $(\lambda_1,\lambda_2,...,\lambda_n)$ satisfies 
\begin{eqnarray}
\lambda_1\geq \lambda_2 \geq...\geq \lambda_n\geq0, \hspace*{1cm}
\sum_{i=1}^n\lambda_i = \sum_{i=1}^n i[c_i]=n.
\end{eqnarray}  
The example described earlier has partition set (3,2). The objects in a partition are symmetric to each other under exchange, while objects in different partitions are mutually antisymmetric under exchange. Throughout this thesis, we use the partition description of the Young diagrams. 

 \section{Young tableaux}  
 \subsection{Representations of $S_n$}

A Young tableau for $S_n$ is a Young diagram of $n$ boxes, filled with numbers from 1 to $n$ exactly once, such that the numbers  from left to right in a row and top to bottom in a column are strictly increasing. Such tableaux are known as \textit{legal Young tableaux.} They describe all possible states belonging to the corresponding representation of $S_n$.

For example, consider a Young diagram for $S_6$ with $\lambda_1=3, \lambda_2=2, \lambda_3=1$.  Some legal Young tableaux of this young diagram are as follows.
$$ \young(123,45,6)\hspace*{.6cm} \young(124,35,6)\hspace*{.6cm} \young(146,23,5)\hspace*{.6cm} \young(123,46,5)\hspace*{.6cm} \young(125,36,4)\hspace*{.6cm} \young(146,25,3)\hspace*{.6cm}\young(126,34,5)$$
The number of ways of writing legal Young tableaux for a given diagram is the dimension of the corresponding irreducible representation of $S_n$. For instance, for the group $S_3$, representations (3,0,0) and (1,1,1) are one dimensional and representation (2,1,0) is two-dimensional.
\begin{eqnarray*}
\Yvcentermath1 \young(\ \ \ )\rightarrow \young(123) \hspace*{.4cm}, \hspace*{1cm}  \Yvcentermath1 \young(\ ,\ ,\ )\rightarrow\young(1,2,3) \hspace*{.4cm},\hspace*{1cm}  \Yvcentermath1 \young(\ \ ,\ )\rightarrow \young(12,3) \hspace*{.5cm} \young(13,2) \hspace*{.4cm}.
\end{eqnarray*}

For an arbitrary irreducible representation $(S_n)_\lambda$, its dimensionality can be obtained by using the definition of \textit{hook length}  \citep{hooklength}. Every box in the Young diagram is assigned a hook length $h$, such that $h= r+b+1$, where $r$ and $b$ denote the number of boxes to the right  of it in its row and below it in its column respectively. The dimensionality $D_\lambda$ of the irreducible representation is then the order of the group (number of all possible permutations) divided by the product of all the hook lengths. 
\begin{equation}
D_\lambda = \frac{n!}{\prod_i h_i}
\end{equation}
For the earlier mentioned example of the representation (3,2,1) of $S_6$, its dimension is 
\begin{eqnarray}
\Yvcentermath1 \young(\ \ \ ,\ \ ,\ ) \xrightarrow{ \hspace*{.2cm} h_i} \young(531,31,1) \hspace*{1cm} D_{(3,2,1)} = \frac{6!}{5.3.3.1.1.1} = 16.
\end{eqnarray}\newline
\subsection{Representations of $SU(d)$}
As a result of Schur-Weyl duality \citep{schur-weyl}, Young diagrams and Young tableaux are also useful for labelling the irreducible representations of the special unitary group $SU(d)$ \citep{biedenharn}. The Young diagrams and Young tableaux describe the symmetry of a collection of $n$  particles, and we let each particle corresponding to a box belong to the fundamental representation of the group $SU(d)$. From the representation theory of unitary groups, we know that the $SU(d)$ group has only $d$ distinct species of particles that label the states \citep{georgii}.  Consequently, the irreducible representations of  $SU(d)$ group are represented by Young diagrams, with each particle represented by a box and at most $d$ rows.

As discussed in the previous section, in a \textit{legal Young diagram} the particles are symmetric under exchange in a row and are antisymmetric under exchange in a column. This property differs somewhat between $S_n$ and $SU(d)$, due to the specific choice of orthonormal bases. For example, let us consider the case of $S_3$ described earlier. The Young diagrams with the partitions (3,0,0) and (1,1,1) have dimension one, corresponding to the 3 particles being in completely symmetric and antisymmetric states respectively. On the other hand, the partition (2,1,0) involves one symmetrisation  and one antisymmetrisation.  Since the order of particles can be freely chosen,  it is a two dimensional representation:
$$ \Yvcentermath1 \young(\ \ ,\ )\rightarrow \young(12,3) \hspace*{.5cm} \young(13,2) $$
In the first diagram, the particles in $1^{st}$ and $2^{nd}$ position are symmetric under exchange. After the system of 1 and 2 is formed, the $3^{rd}$ particle is not antisymmetric to the $1^{st}$ particle. In fact it is not antisymmetric to any of the particles. Such representations are said to have \textit{mixed symmetry.} Similarly, the second Young diagram represents a mixed symmetry state that is antisymmetric among particles in $1^{st}$ and $2^{nd}$ positions.

A \textit{valid Young tableau} for $SU(d)$ is a Young diagram filled with $d$ integers (conventionally from 1 to $d$, each representing a particle) such that the numbers are weakly increasing across the row from left to right and strictly increasing down the column. This respects the constraint on the number of rows of a Young diagram and a valid Young tableau to be at most $d$. Therefore,  we have the relations
\begin{eqnarray}
\lambda_1 \geq \lambda_2 \geq .... \lambda_d \geq 0, \hspace*{1cm} \sum_{i=1}^d \lambda_i =n.
\end{eqnarray}
For example, the Young diagram for the (4,3,1) representation of $SU(3)$ can have the following valid Young tableaux.   
$$  \Yvcentermath1 \young(\ \ \ \ ,\ \ \ ,\ )\rightarrow \young(1111,222,3)\hspace*{.5cm} \young(1122,223,3)\hspace*{.5cm} \young(1123,233,3)\hspace*{.5cm} \young(1222,233,3)$$

Each Young tableau of a given Young diagram represents a possible eigenstate of the irreducible representation of the $SU(d)$ group. Therefore, the number of possible valid Young tableaux of a Young diagram  represents the dimensionality of the $SU(d)$ representation. 

The Schur-Weyl duality associates any Young diagram with appropriate irreducible representations of both $S_n$ and $SU(d)$. But the dimensionalities of the mapped $S_n$ and $SU(d)$ representations differ in general. Also, because of mixed symmetries in $SU(d)$ representations, the symmetry properties of Young tableaux for $S_n$ do not fully appear in the Young tableaux for $SU(d)$.
 % ------------------------------------------------------------------------

%%% Local Variables: 
%%% mode: latex
%%% TeX-master: "../thesis"
%%% End: 

\renewcommand{\thechapter}{B} %command is to change number of Chapter to 'B'

\makeatletter
\renewcommand{\@chapapp}{Appendix}%command to change the word 'chapter' to 'appendix'
\makeatother

\chapter{Quantum Circuits for Binary and Ternary Full Adders and Subtractors}
%\begin{center}
%\textbf{\Large{Appendix B}:\\Quantum Circuits of Binary and ternary Full Adders/subtractors }
%\end{center}\normalsize

%\ifpdf
 %   \graphicspath{{appendix2/appendix2Figs/PNG/}{appendix2/appendix2Figs/PDF/}{appendix2/appendix2Figs/}}
%\else
 %   \graphicspath{{appendix2/appendix2Figs/EPS/}{appendix2/appendix2Figs/}}
%\fi

Quantum circuits are presented in Chapters 2 and 3 (Figure 2.5, Figure 3.6) encoding the algorithm that generates eigenstates of $SU(2)$ and $SU(3)$ representations. In these circuits, except for the controlled rotation gates, all other gates are simple arithmetic gates (adders and subtractors). These arithmetic gates have their addend and subtrahend as single qubits or qutrits, and we present here simple binary and ternary reversible logic circuits for them, following established conventions \citep{elementary-gates}.

\section{$SU(2)$}
The basic gates used to construct binary adder/subtractor circuits are the $NOT$ and the $C_{NOT}$ gates operating linearly on qubits. The $NOT$ gate  simply inverts the qubit, and the $C_{NOT}$ gate inverts the target qubit \textit{iff} the control qubit is in the state $\vert 1 \rangle $.
\begin{eqnarray}
 \begin{aligned}
  \text{\underline{$NOT$}:} \hspace*{2cm}&\vert 0 \rangle \rightarrow \vert 1 \rangle \\
 &\vert 1 \rangle \rightarrow \vert 0 \rangle
 \end{aligned}
\end{eqnarray}
  \begin{eqnarray}
 \begin{aligned}
\hspace*{-2cm}\text{\underline{$C_{NOT}$}:}\hspace*{2cm}&\vert 0\rangle\vert 0\rangle \rightarrow \vert 0\rangle\vert 0\rangle \\ &\vert 0\rangle\vert 1\rangle \rightarrow \vert 0\rangle\vert 1\rangle \\&\vert 1\rangle\vert 0\rangle \rightarrow \vert 1\rangle\vert 1\rangle \\ &\vert 1\rangle\vert 1\rangle \rightarrow \vert 1\rangle\vert 0\rangle
\end{aligned}
\end{eqnarray} 
A useful extension of these operations is the Toffoli or the $C^2_{NOT}$ gate which inverts the target qubit \textit{iff} both the control qubits are in the state $\vert 1 \rangle$. 
 
 The corresponding logic circuit notation is 
\begin{figure}[H]
\centering
\includegraphics[scale=.8]{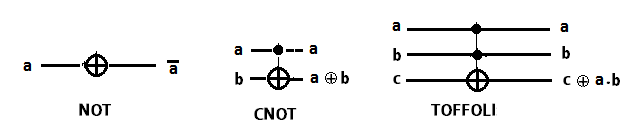}
\caption{Logic circuit representation of $NOT$, $C_{NOT}$ and $C^2_{NOT}$ gates}
\end{figure}

\subsection{Quantum binary full adder}

A simple  binary full adder, adding two qubits $a$ and $b$ to produce a sum qubit $s=a \oplus b$ and a carry qubit $c=a.b$,  has the following truth table and logic circuit.\\\\
\begin{minipage}{\textwidth}
   \begin{minipage}[b]{0.40\textwidth}
    \centering
    \begin{tabular}{|c|c|c|c|}\hline
     $a$&$b$ & $s=a \oplus b$ & $c=a.b$\\ \hline
      0&0&0&0 \\
        0&1&1&0 \\
        1&0&1&0\\
        1&1&0&1 \\
        \hline 
      \end{tabular}
      \captionof{table}{Truth table for binary full adder }
    \end{minipage}\hspace*{1cm}
    \begin{minipage}[b]{0.49\textwidth}
  \centering \framebox{\includegraphics[scale=.8]{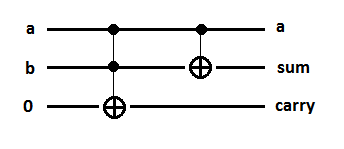}}
    \captionof{figure}{Simple binary full adder circuit adding two qubits}
  \end{minipage}
  \hfill
  \end{minipage}\\\\
  
Now adding a single qubit $a$ to a register of $n$-qubits ($b_nb_{n-1}...b_{2}b_1$) involves $n$ binary full adders, performing the following arithmetic with $c_m$ being the $m^{th}$ carry qubit.
\begin{eqnarray}
\begin{array}{cccccc}
& b_n&b_{n-1}&...&b_{2}&b_1\\
+ & & & & &  a \\ \hline\\  \hline
\end{array}
= \begin{array}{cccccc}
& b_n&b_{n-1}&...&b_{2}&b_1\\
\oplus & c_{n-1}&c_{n-2}&...&c_1&a \\  \hline  c_n & b_n'&b_{n-1}'&...&b_{2}'&b_1' \\   \hline
\end{array}
\end{eqnarray} 
The final carry qubit $c_n$ vanishes when there is no overflow. (In our applications this is always true with appropriate choice of $n$.) The sum is  then represented by the register of qubits ($b_n'b_{n-1}'...b_2'b_1'$). Thus a quantum binary full adder which adds a qubit to an $n$-qubit register needs $(n-1)$ carry qubits, $(n)$ $C_{NOT}$ gates and $(n-1)$ $C^2_{NOT}$ gates.

For example, the logic circuit for adding a qubit to a 3-qubit register is:
\begin{figure}[h!]
  \centering \framebox{\includegraphics[scale=.7]{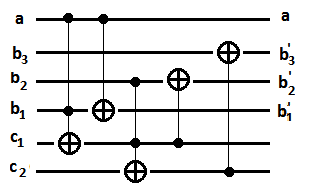}}
    \captionof{figure}{Logic circuit for adding a qubit to a $3-$qubit register} 
  \end{figure}

\subsection{Quantum binary full subtractor}

A similar implementation can be carried out for constructing a binary full subtractor that subtracts a single qubit from an $n$-qubit register. The truth table and logic circuit of a simple full subtractor, subtracting qubit $a$ from qubit $b$, to produce remainder $r=b \oplus a$ and borrow qubit $c=a.\bar{b}$  is as follows.\\
 
\begin{minipage}{\textwidth}
   \hspace*{-.5cm}\begin{minipage}[b]{0.40\textwidth}
    \centering
    \begin{tabular}{|c|c|c|c|}\hline
      $b$&$a$ & $r=b \oplus a$ & $c=a.\bar{b}=a.r$\\ \hline
      0&0&0&0 \\
        0&1&1&1 \\
        1&0&1&0\\
        1&1&0&0 \\
        \hline 
      \end{tabular}
      \captionof{table}{Truth table for binary full subtractor}
    \end{minipage}\hspace*{.5cm}
    \begin{minipage}[b]{0.49\textwidth}
  \centering \framebox{\includegraphics[scale=.8]{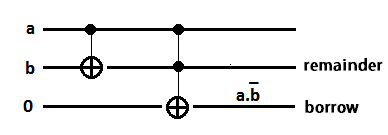}}
    \captionof{figure}{Simple binary full subtractor\\ circuit subtracting qubit $a$ from qubit $b$}
  \end{minipage}
  \hfill
  \end{minipage}\\
 
Subtracting a single qubit $a$ from an $n$-qubit register ($b_n b_{n-1}...b_{2} b_1$) needs $n$ binary full subtractors, performing the following arithmetic with $c_m$ being the $m^{th}$ borrow qubit.
\begin{eqnarray}
\begin{array}{cccccc}
& b_n&b_{n-1}&...&b_{2}&b_1\\
- & & & & &  a \\ \hline\\  \hline
\end{array}
= \begin{array}{cccccc}
& b_n&b_{n-1}&...&b_{2}&b_1\\
\ominus & c_{n-1}&c_{n-2}&...&c_1&a \\  \hline  c_n & b_n'&b_{n-1}'&...&b_{2}'&b_1' \\   \hline
\end{array}
\end{eqnarray} 
 Again, in our applications the final borrow qubit $c_n$ vanishes. Similar to the quantum  binary full adder, a quantum binary full subtractor that subtracts a qubit from an $n$-qubit register needs $(n-1)$ borrow qubits, $(n)$ $C_{NOT}$ gates and $(n-1)$ $C^2_{NOT}$ gates.

For example, the binary full subtractor logic circuit for subtracting a qubit from a 3-qubit register is:
    \begin{figure}[H]
  \centering \framebox{\includegraphics[scale=.7]{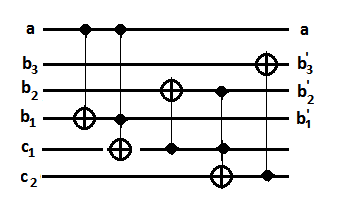}}
    \captionof{figure}{Logic circuit for subtracting a qubit from a $3-$qubit register}
  \end{figure}
  
\section{$SU(3)$}

In this case, we are interested in constructing ternary full adders/subtractors which have their addend/subtrahend as single qutrits. Instead of using the computational basis for qutrits as $\left\lbrace \vert 0\rangle, \vert 1\rangle, \vert 2\rangle \right\rbrace$,  we use padded qubits:
\begin{equation}
\vert 0\rangle_{su(3)} =\vert \underline{00}\rangle_{su(2)} \hspace*{1cm}  \vert 1\rangle_{su(3)} =\vert \underline{01}\rangle_{su(2)} \hspace*{1cm}\vert 2\rangle_{su(3)} =\vert \underline{10}\rangle_{su(2)}
\end{equation}
These padded qubits allow us to perform the arithmetic using the $NOT$, $C_{NOT}$ and $C^2_{NOT}$ gates  defined earlier. Note that for any padded qubit $\vert \underline{b_1b_2}\rangle$, $b_1.b_2=0$ always.

\subsection{Quantum ternary full adder} 

To construct a quantum ternary full adder circuit which adds a qutrit to an $n$-qutrit register, we first consider adding a single qubit to an $n$-qutrit register. The first step is to add the qubit $a$ to a qutrit $\underline{b_1b_2}$, with the sum $\underline{b_1'b_2'}$ and carry qubit $c_1$. The truth table and the logic circuit are as follows. \\\\
\begin{minipage}{\textwidth}
  \centering
    \begin{tabular}{|c|c|c|c|c|c|}\hline
    $b_1$ & $b_2$ & $a$& $b_1'=b_1 \oplus ((b_1 \oplus b_2). a)$ & $b_2'=b_2 \oplus (\bar{b}_1.a)$&$c_1=b_1.a$\\ \hline
      0&0&0&0&0&0 \\
        0&1&0&0&1&0 \\
        1&0&0&1&0&0\\
        0&0&1&0&1&0 \\
        0&1&1&1&0&0\\
        1&0&1&0&0&1\\
        \hline 
      \end{tabular}
      \captionof{table}{Truth table of a simple ternary adder adding a qubit to a qutrit}
      \end{minipage}
      \begin{figure}[H]
      \centering
      \framebox{\includegraphics[scale=.83]{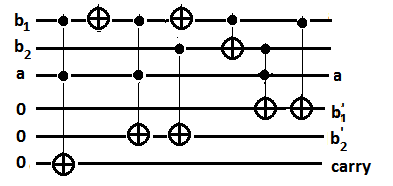}}
      \caption{A simple ternary full adder circuit adding a qutrit and a qubit}
      \end{figure}
      
      \begin{comment}
      The above circuit helps us to build a circuit which adds a qubit ($a$) to a $n$-qutrit register ($\underline{b_1^nb_2^n}\underline{b_1^{n-1}b_2^{n-1}}...\underline{b_1^{2}b_2^{2}}\underline{b_1^1b_2^1}$) in the following way. 
      \footnotesize
      \begin{eqnarray}
      \begin{array}{cccccc}
      &\underline{b_1^1b_2^1}&\underline{b_1^2b_2^2}&...&\underline{b_1^{n-1}b_2^{n-1}}&\underline{b_1^nb_2^n}\\
      +& & & & &a\\ \hline \\ \hline
      \end{array} = \begin{array}{cccccc}
      &\underline{b_1^1b_2^1}&\underline{b_1^2b_2^2}&...&\underline{b_1^{n-1}b_2^{n-1}}&\underline{b_1^nb_2^n}\\
      +& c_{n-1}&c_{n-2}& ...&c_1 &a\\ \hline  c_n &\underline{b_1^1b_2^1}'&\underline{b_1^2b_2^2}'&...&\underline{b_1^{n-1}b_2^{n-1}}'&\underline{b_1^nb_2^n}'\\  \hline
      \end{array}
      \end{eqnarray} \normalsize
      Here $c_m$ are the $m^{th}$ carry qubit. A single qutrit needs 3 extra qubits and 9 gates, a circuit adding a qubit to a $n$-qutrit register needs $3n$ extra qubits (including $n$ carry qubits) and $9n$ gates. 
      \end{comment}
      
Now we can construct a ternary full adder, which adds qutrit $\underline{a_1a_2}$ to qutrit  $\underline{b_1b_2}$, as follows:
\begin{eqnarray}
\begin{array}{cc}
 \begin{array}{ccc}
&(b_1&b_2)\\+&(a_1&a_2)\\ \hline\\\hline
\end{array} = \begin{array}{ccc}
&b_1&b_2\\ \oplus &c_1'&a_2 \\\hline  c_1''&b_1''& b_2' \\ 
  \oplus c_1'''&a_1&    \\\hline c_1&b_1'&b_2'\\\hline \end{array} &  \text{ $\underline{c_1''b_1''}$ is qutrit  and $a_1$ is a qubit.}
\end{array}
\end{eqnarray} 
In the above equation, the LHS denotes addition of two qutrits, whereas the RHS describes all qubit additions separately. Here $c_1$ is the final carry qubit corresponding to the sum and it can only be 0 or 1. To add two qutrits we thus use two ternary full additions of a qubit to a qutrit, which needs twice the resources used in Figure B.6.  

In the subsequent steps of adding a qutrit to an $n$-qutrit register, all the carry qubits are either 0 or 1, and so these steps involve only additions of a qubit to a qutrit. Thus the total  calculation requires one addition of a qutrit $\underline{b_1^1 b_2^1}$ with a qutrit $\underline{a_1a_2}$ and $(n-1)$ additions of a qubit (carry qubit $c_{k-1}$) to a qutrit ($\underline{b_1^kb_2^k}$ with $k>1$). It is therefore equivalent to  $(n+1)$ additions of a qubit to a qutrit. \footnotesize
\begin{eqnarray}
\begin{array}{cccccc}
&\underline{b_1^nb_2^n}&\underline{b_1^{n-1}b_2^{n-1}}&...&\underline{b_1^{2}b_2^{2}}&\underline{b_1^1b_2^1}\\
      +& & & & &\underline{a_1a_2}\\ \hline \\ \hline
      \end{array} = \begin{array}{cccccc}
     & \underline{b_1^n b_2^n} &\underline{b_1^{n-1}b_2^{n-1}}&...&\underline{b_1^{2}b_2^{2}}&\underline{b_1^1b_2^1}\\
     \oplus & c_{n-1}&c_{n-2}& ...&c_1 &\underline{a_1a_2}\\ \hline  c_n &\underline{b_1^{n'}b_2^{n'}}&\underline{b_1^{n-1'}b_2^{n-1'}}&...&\underline{b_1^{2'}b_2^{2'}}&\underline{b_1^{1'}b_2^{1'}}\\  \hline 
      \end{array}
\end{eqnarray}\normalsize
In our application the borrow qubit $c_n$ vanishes with a suitable choice of $n$.

A single qutrit-qubit addition needs 3 extra qubits and (2) $NOT$, (3) $C_{NOT}$ and (3) $C^2_{NOT}$ logic gates. Therefore to implement the circuit for adding a single qutrit to an $n$-qutrit register, we need $3(n+1)$ extra qubits and $8(n+1)$ logic gates. 

\subsection{Quantum ternary full subtractor}

The quantum ternary full subtractor, subtracting a qutrit from an $n$-qutrit register, is constructed in a similar way as the full adder constructed  in the preceding section. We first construct the quantum logic circuit for subtracting a qubit $a$ from a single qutrit $\underline{b_1b_2}$. It has the following truth table and logic circuit. \\\\
\begin{minipage}{\textwidth}
\centering
    \begin{tabular}{|c|c|c|c|c|c|}\hline
     $b_1$ & $b_2$ & $a$& $b_1'=b_1 \oplus(\bar{b}_2.a)$ & $b_2'=b_2 \oplus((b_1 \oplus b_2).a)$&$c_1= (b_1 \oplus \bar{b}_2).a$\\ \hline
      0&0&0&0&0&0 \\
        0&1&0&0&1&0 \\
        1&0&0&1&0&0\\
        0&0&1&1&0&1 \\
        0&1&1&0&0&0\\
        1&0&1&0&1&0\\
        \hline 
      \end{tabular}
      \captionof{table}{Truth table of  ternary subtractor subtracting a qubit from a qutrit}
      \end{minipage}
      \begin{figure}[H]
      \centering
      \framebox{\includegraphics[scale=.8]{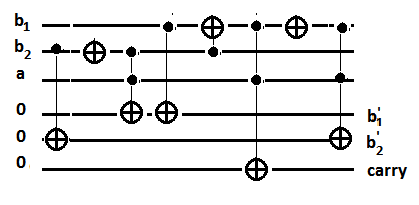}}
      \caption{A simple ternary full subtractor circuit subtracting a qubit from a qutrit}
      \end{figure}

Next, we construct the ternary full subtractor subtracting a qutrit $\underline{a_1a_2}$ from another qutrit $\underline{b_1b_2}$, using  2 ternary full subtractors that subtract a qubit from a qutrit, as follows: 
\begin{eqnarray}
\begin{array}{cc}
 \begin{array}{ccc}
&(b_1&b_2)\\-&(a_1&a_2)\\ \hline\\\hline
\end{array} = \begin{array}{ccc}
&b_1&b_2\\ \ominus &c_1'&a_2 \\\hline  c_1''&b_1''& b_2' \\ 
  \ominus c_1'''&a_1&    \\\hline c_1&b_1'&b_2'\\\hline \end{array} &  \text{ $\underline{c_1''b_1''}$ is qutrit  and $a_1$ is a qubit.}
\end{array}
\end{eqnarray} 
Here $c_1$ is the final borrow qubit of the difference and it can only be 0 or 1. This calculation needs twice the resources of the circuit in Figure B.7. \\

Now we can extend this procedure to subtract a qutrit from an $n$-qutrit register. It first involves subtraction of a qutrit $\underline{a_1a_2}$ from a qutrit $\underline{b_1^1b_2^1}$, and then $(n-1)$ subtractions of a qubit (borrow qubit $c_{k-1}$) from a qutrit ($\underline{b_1^kb_2^k}$ with $k >1$). In total that is equivalent to $(n+1)$ subtractions of a qubit from a qutrit.
     \footnotesize
\begin{eqnarray}
\begin{array}{cccccc}
&\underline{b_1^nb_2^n}&\underline{b_1^{n-1}b_2^{n-1}}&...&\underline{b_1^{2}b_2^{2}}&\underline{b_1^1b_2^1}\\
      -& & & & &\underline{a_1a_2}\\ \hline \\ \hline
      \end{array} = \begin{array}{cccccc}
      &\underline{b_1^nb_2^n}&\underline{b_1^{n-1}b_2^{n-1}}&...&\underline{b_1^{2}b_2^{2}}&\underline{b_1^1b_2^1}\\\ominus & c_{n-1}&c_{n-2}& ...&c_1 &\underline{a_1a_2}\\ \hline  c_n &\underline{b_1^{n'}b_2^{n'}}&\underline{b_1^{n-1'}b_2^{n-1'}}&...&\underline{b_1^{2'}b_2^{2'}}&\underline{b_1^{1'}b_2^{1'}}\\  \hline 
      \end{array}
\end{eqnarray}\normalsize
In our applications, the borrow qubit $c_n$ vanishes. 

Subtraction of a qubit from a qutrit needs 3 extra qubits and (2) $NOT$, (3) $C_{NOT}$ and (3) $C^2_{NOT}$ logic gates. So overall, the ternary full subtractor, subtracting a qutrit  from an $n$-qutrit register, therefore needs  $3(n+1)$ extra qubits and $8(n+1)$ logic gates.
% ------------------------------------------------------------------------
\begin{comment}
Therefore when a qubit ($a$) is subtracted from a $n$-qutrit register ($\underline{b_1^1b_2^1}...\underline{b_1^{n-1}b_2^{n-1}}\underline{b_1^nb_2^n}$), it needs $3n$ extra qubits and $11n$ gates. The circuit implements the following logic.
 \footnotesize
      \begin{eqnarray}
      \begin{array}{cccccc}
      &\underline{b_1^1b_2^1}&\underline{b_1^2b_2^2}&...&\underline{b_1^{n-1}b_2^{n-1}}&\underline{b_1^nb_2^n}\\
      -& & & & &a\\ \hline \\ \hline
      \end{array} = \begin{array}{cccccc}
      &\underline{b_1^1b_2^1}&\underline{b_1^2b_2^2}&...&\underline{b_1^{n-1}b_2^{n-1}}&\underline{b_1^nb_2^n}\\
      -& c_{n-1}&c_{n-2}& ...&c_1 &a\\ \hline  c_n &\underline{b_1^1b_2^1}'&\underline{b_1^2b_2^2}'&...&\underline{b_1^{n-1}b_2^{n-1}}'&\underline{b_1^nb_2^n}'\\  \hline
      \end{array}
      \end{eqnarray} \normalsize
      
\end{comment}

%%% Local Variables: 
%%% mode: latex
%%% TeX-master: "../thesis"
%%% End: 

\renewcommand{\thechapter}{C} %command is to change number of Chapter to 'B'

\makeatletter
\renewcommand{\@chapapp}{Appendix}%command to change the word 'chapter' to 'appendix'
\makeatother

\chapter{Isoscalar Factors for $(P_1,Q_1) \otimes (1,0)$ }
Here we derive the isoscalar factors for the tensor product $(P_1,Q_1) \otimes(1,0)$ in $SU(3)$ \citep{williams}. We specify the states using the $k,l,m$ notation of (3.24) and (3.27). We start with $\vert P_1,Q_1,k_1,l_1,m_1\rangle$ as a state belonging to the irreducible representation $(P_1,Q_1)$, and after the tensor product obtain $\vert P,Q,k,l,m \rangle$ as a resultant state of the irreducible representation $(P,Q)$. The possible values of $(P,Q)$ are:
\begin{equation*}
(P,Q)=\left\lbrace\begin{array}{c}
(P_1+1,Q) \\ 
(P_1-1,Q_1+1) \\ 
(P_1,Q_1-1)
\end{array} \right..
\end{equation*}
Isospin of the states that are formed by an addition of a $u/d$ quark or an $s$ quark to these irreducible representations take the values $T=T_1\pm 1/2$ or $T=T_1$ respectively. The changes in $T_3$ and $Y$ values are additive only. So we need to consider the possibilities $(k,l)=\lbrace (k_1+1,l_1),(k_1,l_1+1)\rbrace$ when adding $u/d$ quark and $(k,l)=\lbrace(k_1,l_1), (k_1-1,l_1-1)\rbrace$ when adding $s$ quark. 

Since isoscalar factors do not depend on $T_3$ (or equivalently $m$), we simplify our analysis by considering tensor products of states $\vert k_1,l_1,m_1\rangle$ with $u$ or $s$ quarks, such that $k_1=m_1$ and $k=m$. This restriction only to $T=T_3$ states reduces the possibilities to $(k,l)=\lbrace (k_1+1,l_1),(k_1,l_1+1)\rbrace$ and $(k,l)=\lbrace(k_1,l_1),(k_1-1,l_1-1)\rbrace$ when adding a $u$ and $s$ quark respectively. Then the required isoscalar factors for the tensor product $(P_1,Q_1)\otimes(1,0) $ are $F(k_1,l_1:1,0;k,l)$ and $F(k_1,l_1:0,0;k,l)$, when adding $u$ and $s$ quarks respectively. 

Our derivation consists of three steps. First, the recursion relations with operations of $\hat{U}_-$ and $\hat{V}_+$, starting with the HWS (\textit{i.e.,} $k=m=P+Q,l=0$), are considered. They lead to the isoscalar factors mentioned in (3.34d). Then, the  states with $s$-quarks in the first and second row of the reduced Young tableau are considered (\textit{i.e.,} $k<P+Q,l=0$). Their recursion relations with operation of $\hat{V}_-$ lead to the isoscalar factors mentioned in (3.34c). Finally arbitrary states with $d$- quarks in the second row of reduced Young tableau are considered (\textit{i.e.,} $k \leq P+Q,l>0$). Recursive application of  $\hat{U}_+$ on these states results in (3.34a) and (3.34b) . 

The action of the ladder operators $\hat{V}_-,\hat{V}_+,\hat{U}_-$ and $\hat{U}_+$ on an arbitrary state is \citep{williams}
\begin{eqnarray}
\begin{aligned}
&\hat{V}_{+}\vert P,Q;k,l,m \rangle = v_{1+}\vert  P,Q;k+1,l,m+1 \rangle + v_{2+}\vert  P,Q;k,l+1,m+1 \rangle \\ 
&\hat{V}_{-}\vert  P,Q;k,l,m \rangle = v_{1-}\vert  P,Q;k-1,l,m-1 \rangle + v_{2-}\vert  P,Q;k,l-1,m-1 \rangle \\ 
&\hat{U}_{+}\vert  P,Q;k,l,m \rangle = u_{1+}\vert  P,Q;k+1,l,m \rangle - u_{2+}\vert  P,Q;k,l+1,m \rangle \\
&\hat{U}_{-}\vert  P,Q;k,l,m \rangle = u_{1-}\vert  P,Q;k-1,l,m \rangle - u_{2-}\vert  P,Q;k,l-1,m \rangle 
\end{aligned}
\end{eqnarray}
All the coefficients in the above equations implicitly  have $(P,Q,k,l,m)$ as their arguments.
The values of these coefficients are given by the subequations of (3.28). Since we will always consider $k = m$, we have
\footnotesize
\begin{eqnarray*} 
\begin{aligned}
&v_{1+} = \sqrt{\frac{(k+2)(k-Q+1)(P+Q-k)}{(k-l+2)}},  \hspace*{1cm}
 v_{1-} = \sqrt{\frac{(k+1)(k-Q)(P+Q-k+1)}{(k-l+1)}}, \\\\
&u_{1+} = \sqrt{\frac{(k+2)(k-Q+1)(P+Q-k)}{(k-l+2)(k-l+1)}},  \hspace*{1cm}
v_{2-} = \sqrt{\frac{l(Q-l+1)(P+Q-l+2)}{(k-l+2)(k-l+1)}},\\ \\ \end{aligned}
\end{eqnarray*}\begin{eqnarray} 
\begin{aligned}
&u_{2+} = \sqrt{\frac{(l+1)(Q-l)(P+Q-l+1)}{(k-l+1)}}, \hspace*{1cm}
u_{2-} = \sqrt{\frac{l(Q-l+1)(P+Q-l+2)}{(k-l+2)}},\\\\
&u_{1-} = v_{2+} = 0,
\end{aligned}
\end{eqnarray}
\normalsize with the arguments ($P,Q,k,l$).

\textbf{\textit{Note:}} In the main text of Chapter 3, the state corresponding to $\vert P_1,Q_1,k_1,l_1,m_1\rangle$ is labelled as $\vert P',Q',k',l',m' \rangle$. There the isoscalar factors are defined as \begin{eqnarray*}
F(k',l':1,0;k,l)\equiv F^{u,p}_{k',l'}, \hspace*{1cm} F(k',l':0,0;k,l)\equiv F^{s,p}_{k',l'}.
\end{eqnarray*} 

\section{Recursion relations starting with HWS}
Consider the HWS with $k=m=P+Q,l=0$. The first recursion relation is obtained by operating with $\hat{U}_-=\hat{U}_{1-}+\hat{U}_{2-}$, which annihilates the state.
\footnotesize
\begin{equation}
\begin{split}
0=\langle  k_{1},l_{1},m_{1} \otimes k_{2},l_{2},m_{2} | \hat{U}_{-} | P+Q,0,P+Q \rangle  =  &\langle  \hat{U}_{+}(k_{1},l_{1},m_{1})\otimes k_{2},l_{2},m_{2} | P+Q,0,P+Q \rangle \\ +&  \langle  k_{1},l_{1},m_{1} \otimes \hat{U}_{+}(k_{2},l_{2},m_{2}) | P+Q,0,P+Q \rangle 
\end{split}
\end{equation}
\normalsize
Let the second state in the tensor product be the $u$-quark \textit{i.e.,} $k_{2} =m_2= 1,l_{2} = 0$,  $P_{2} = 1,Q_{2} = 0$. With $k_{1} = m_{1}$, we have\footnotesize
\begin{eqnarray} 
\begin{aligned}
0=&u_{1+}[P_1,Q_1,k_1,l_1] \langle  k_{1}+1,l_{1},k_{1} \otimes 1,0,1 | P+Q,0,P+Q \rangle\\-& u_{2+}[P_1,Q_1,k_1,l_1] \langle k_{1},l_{1}+1,k_{1} \otimes 1,0,1 | P+Q,0,P+Q \rangle
\end{aligned}  
\end{eqnarray}\normalsize
We suppress the arguments $P_1,Q_1$ in the coefficients of the ladder operators, and write the above equation as 
\footnotesize
\begin{equation} 
\begin{split}
\frac{u_{1+}[k_1,l_1]}{u_{2+}[k_1,l_1]}  \left( \begin{array}{ccc}
\frac{k_{1}+1-l_{1}}{2} & \frac{1}{2} & \frac{P+Q}{2} \\ 
\frac{k_{1}-1-l_{1}}{2} & \frac{1}{2} & \frac{P+Q}{2} \\ 
\end{array} \right) F(k_{1}+1,l_{1} : 1,0;P+Q,0)=&\left( \begin{array}{ccc}
\frac{k_{1}-1-l_{1}}{2} & \frac{1}{2} & \frac{P+Q}{2} \\ 
\frac{k_{1}-1-l_{1}}{2} & \frac{1}{2} & \frac{P+Q}{2} \\ 
\end{array} \right)\\& \times F(k_{1},l_{1}+1 : 1,0;P+Q,0)
\end{split}
\end{equation}\normalsize
Evaluating the $SU(2)$ CG-coefficients using the angular momentum algebra, with the condition $ k_{1}-l_{1} = P+Q $ that ensures the addition rule for $T_3$, we obtain \footnotesize
\begin{equation}
\frac{u_{1+}[k_{1},l_{1}]}{u_{2+}[k_{1},l_{1}]} \left( \frac{-1}{\sqrt{k_{1}-l_{1}+2}} \right) F(k_{1}+1,l_{1}:1,0;P+Q,0) = F(k_{1},l_{1}+1 : 1,0;P+Q,0 ). 
\end{equation}\normalsize

Next, consider the action of the operator $\hat{V}_{+}$ on HWS, which also annihilates the state.\footnotesize
\begin{eqnarray} \begin{aligned}
0=\langle k_{1},l_{1},m_{1} \otimes k_{2},l_{2},m_{2} | \hat{V}_{+} | P+Q,0,P+Q  \rangle     =   \langle  \hat{V}_{-}(k_{1},l_{1},m_{1}) \otimes k_{2},l_{2},m_{2} | P+Q,0,P+Q \rangle  \\+   
\langle  k_{1},l_{1},m_{1} \otimes \hat{V}_{-}(k_{2},l_{2},m_{2}) | P+Q,0,P+Q\rangle  \end{aligned}
\end{eqnarray}\normalsize
Again considering the $u$-quark as the second state, and $k_1=m_1$, we have \footnotesize
\begin{eqnarray}\begin{aligned}
0= v_{1-}[k_{1},l_{1}] \langle k_{1}-1,l_{1},k_{1}-1 \otimes 1,0,1|P+Q,0,P+Q  \rangle   
 \\   + v_{2-}[k_{1},l_{1}]\langle k_{1},l_{1}-1,k_{1}-1 \otimes 1,0,1|P+Q,0,P+Q  \rangle\\ + \langle k_{1},l_{1},k_{1} \otimes 0,0,0|P+Q,0,P+Q  \rangle ,  
 \end{aligned}\end{eqnarray}\normalsize
where we have used $v_{1-}[1,0,1,0]=1$ from (C.2). This can be rewritten as \footnotesize
\begin{eqnarray}\begin{aligned}
-\left( \begin{array}{ccc}
\frac{k_{1}-l_{1}}{2} & 0 & \frac{P+Q}{2} \\
\frac{k_{1}-l_{1}}{2} & 0 & \frac{P+Q}{2} \\
\end{array} \right)  F(k_{1}l_{1} : 0,0;P+Q,0 ) = v_{1-}[k_{1}l_{1}]  \left( \begin{array}{ccc}
\frac{k_{1}-l_{1}-1}{2} & \frac{1}{2} & \frac{P+Q}{2} \\
\frac{k_{1}-l_{1}-1}{2} & \frac{1}{2} & \frac{P+Q}{2} \\
\end{array} \right)\\
\times F(k_{1}-1,l_{1} : 1,0;P+Q,0 ) + v_{2-}[k_{1}l_{1}] \left( \begin{array}{ccc}
\frac{k_{1}-l_{1}+1}{2} & \frac{1}{2} & \frac{P+Q}{2} \\
\frac{k_{1}-l_{1}-1}{2} & \frac{1}{2} & \frac{P+Q}{2} \\
\end{array} \right) F(k_{1},l_{1}-1;1,0;P+Q,0 ) \end{aligned}
\end{eqnarray} \normalsize
Again using $SU(2)$ CG-coefficients with $ k_{1}-l_{1} = P+Q$, and inserting (C.6), we get
\footnotesize
\begin{equation}
\begin{split}
F(k_{1},l_{1} : 0,0;P+Q,0 ) = &\left\lbrace v_{1-}[k_{1},l_{1}] \frac{u_{1+}[k_{1}-1,l_{1}-1]}{u_{2+}[k_{1}-1,l_{1}-1]} \frac{1}{\sqrt{k_{1}-l_{1}+2}} + v_{2-}[k_{1},l_{1}] \frac{1}{\sqrt{k_{1}-l_{1}+2}} \right\rbrace \\ 
& \hspace*{3cm}\times F(k_{1},l_{1}-1 : 1,0;P+Q,0 ).
\end{split}
\end{equation}\normalsize
This can be rewritten as
\footnotesize
\begin{eqnarray}
F(k_{1},l_{1} : 0,0;P+Q,0 ) = &\frac{1}{u_{2+}[k_{1}-1,l_{1}-1]\sqrt{k_{1}-l_{1}+2}}\left\lbrace \sum_{i=1}^2 v_{i-}[k_{1},l_{1}] u_{i+}[k_{1}-1,l_{1}-1]  \right\rbrace \nonumber\\ 
& \hspace*{3cm} \times F(k_{1},l_{1}-1 : 1,0;P+Q,0 ).
\end{eqnarray}\normalsize

Now we use the orthonormality condition  $$ \sum_{k_{1}l_{1}k_{2}l_{2}} |F(k_{1},l_{1}:k_{2},l_{2};P+Q,0 )|^{2} = 1 ,$$ in the form
\footnotesize
\begin{equation}
|F(k_{1}-1,l_{1} : 1,0;P+Q,0 )|^{2} +|F(k_{1},l_{1}-1:1,0;P+Q,0 )|^{2} + |F(k_{1},l_{1}:0,0;P+Q,0 )|^{2} = 1.
\end{equation}\normalsize
Then from (C.6) and (C.11), we have\footnotesize
\begin{equation}
\begin{split}
&|\frac{u_{1+}[k_{1}-1,l_{1}-1]}{u_{2+}[k_{1}-1,l_{1}-1]\sqrt{k_{1}-l_{1}+2}} F(k_{1},l_{1}-1:1,0:P+Q,0 )|^{2} + |F(k_{1},l_{1}-1:1,0;P+Q,0 )|^{2} 
\\& + | F(k_{1},l_{1}-1;1,0:P+Q,0 ) \frac{1}{u_{2+}[k_{1}-1,l_{1}-1]}\left\lbrace \sum_{i=1}^2 \frac{v_{i-}[k_{1},l_{1}] u_{i+}[k_{1}-1,l_{1}-1]}{\sqrt{k_{1}-l_{1}+2}}  \right\rbrace |^{2} = 1
\end{split}
\end{equation}\normalsize The arguments of $u_{i \pm}$ are $(P_1,Q_1,k_1-1,l_1-1)$ and those of $v_{i \pm}$ are $(P_1,Q_1,k_1,l_1)$. Suppressing the fixed arguments, we obtain
\begin{equation}
\left\lbrace \frac{\left( u_{1+} \right)^{2}}{(u_{2+})^2(k_{1}-l_{1}+2)} + 1 + \frac{\left( v_{1-}u_{1+}+v_{2-}u_{2+}\right)^{2}}{(u_{2+})^2 (k_{1}-l_{1}+2)} \right\rbrace |F(k_{1},l_{1}-1:1,0;P+Q,0 )|^{2} = 1.
\end{equation}

We follow the De Swart phase convention \citep{deswart} $F(I_{1max},Y_{1max}:I_2,Y_2 \vert I_{max},Y_{max}) >0$, which  allows us to set $F(P_1+Q_1,0:1,0;P_1+Q_1+1,0)=1$ and makes all the square roots positive. Then the above relation with (C.6) and (C.10) yields : \footnotesize
\begin{eqnarray}
\begin{aligned} 
&F(k_{1},l_{1}-1:1,0;P+Q,0 ) = \delta[k_1,l_1-1]+(1-\delta[k_1,l_1-1])\sqrt{\frac{(u_{2+})^{2}(k_{1}-l_{1}+2)}{G[P_1,Q_1,k_{1},l_{1}]}}, \\ 
&F(k_{1}-1,l_{1};1,0;P+Q,0 ) =\delta[k_1-1,l_1]-(1-\delta[k_1-1,l_1])\sqrt{\frac{(u_{1+})^{2}}{G[P_1,Q_1,k_{1},l_{1}]}}, \\ 
&F(k_{1},l_{1}:0,0:P+Q,0 ) = \delta_{k_1-l_1,P+Q}\sqrt{\frac{(v_{1-}u_{1+}+v_{2-}u_{2+})^2}{G[P_1,Q_1,k_{1},l_{1}]}}\\
& G[P_1,Q_1,k_{1}l_{1}] = (u_{2+})^{2}(k_{1}-l_{1}+2) + (u_{1+})^{2} + (v_{1-}u_{1+}+v_{2-}u_{2+})^2,
\end{aligned}
\end{eqnarray}\normalsize
with $\delta[k_1,l_1] \equiv \delta_{k_1-l_1,P_1+Q_1}\delta_{P+Q,P_1+Q_1+1}$. The $\delta[k_1,l_1]$ terms ensure the condition $F(P_1+Q_1,0:1,0;P_1+Q_1+1,0)=1$. Otherwise the ladder operators determine the isoscalar factors.  Note that the non-negativity of $k_i,l_i$ holds good in the arguments of $F$ but not in the arguments of $u_{i \pm},v_{i \pm}$, \textit{i.e.}, $F(k_1,l_1:k_2,l_2;P+Q,0)=0$ for any $k_i<0$  or $l_i<0$ but $u_{i \pm}[k_1,l_1]$ may not vanish. 

\section{Recursion relations for $ k=P+Q-s ,l = 0$ } 
Next consider the states with $ k=P+Q-s ,l = 0, \hspace*{.3cm}\forall \hspace*{.1cm}s \in [1,P+Q]$. For these states, we derive recursion relations using the operator $\hat{V}_-$,
\begin{equation}
\begin{split}
\langle k_{1},l_{1},m_{1} \otimes k_{2},l_{2},m_{2}|\hat{V}_{-}|k,0,m \rangle   = &\langle \hat{V}_{+}(k_{1},l_{1},m_{1}) \otimes k_{2},l_{2},m_{2}|k,0,m \rangle  \\& + \langle k_{1},l_{1},m_{1} \otimes \hat{V}_{+}(k_{2},l_{2},m_{2})|k,0,m \rangle . 
\end{split}
\end{equation}
Again we only consider states with $ k = m $ and $k_{1} = m_{1}$, to derive the isoscalar factors. 
\subsubsection*{Case I:  $u$-quark}  
When the second state in the tensor product is a $u$-quark, $ k_{2} = m_{2} = 1 , l_{2} = 0$.\\
\textbf{Step 1:} Let us first find the isoscalar factor for $ k = P+Q-1 $. From (C.16) we get,\footnotesize
\begin{equation}
v_{1-}[k,l] \langle k_{1},l_{1},k_{1} \otimes 1,0,1|P+Q-1,0,P+Q-1 \rangle= v_{1+}[k_{1},l_{1}] \langle k_{1}+1, l_{1}, k_{1}+1  \otimes 1,0,1|P+Q,0,P+Q \rangle .   
\end{equation}\normalsize
Separating the $SU(2)$ CG-coefficients, this can be rewritten as
\begin{equation}
\begin{split}
&v_{1-}[P+Q,0]  \left( \begin{array}{ccc}
\frac{k_{1}-l_{1}}{2} & \frac{1}{2} & \frac{P+Q-1}{2} \\
\frac{k_{1}-l_{1}}{2} & \frac{1}{2} & \frac{P+Q-1}{2} \\
\end{array} \right)  F(k_{1},l_{1}:1,0;P+Q-1,0) = \\ &v_{1+}[k_{1}l_{1}]
F(k_{1}+1,l_{1}:1,0;P+Q,0 ) \left( \begin{array}{ccc}
\frac{k_{1}-l_{1}+1}{2} & \frac{1}{2} & \frac{P+Q}{2} \\
\frac{k_{1}-l_{1}+1}{2} & \frac{1}{2} & \frac{P+Q}{2} \\
\end{array} \right).  
\end{split}
\end{equation}
Both the $SU(2)$ CG-coefficients  are 1 in (C.18). Therefore we have,
\begin{equation}
F(k_{1},l_{1}:1,0;P+Q - 1,0) = \frac{v_{1+}[k_{1}l_{1}]}{v_{1-}[P+Q,0]} F(k_{1}+1,l_{1}:1,0;P+Q,0 ).
\end{equation}
\textbf{Step 2:} We repeat the above process to obtain the isoscalar factor for $k=P+Q-2$, obtaining
\footnotesize
\begin{equation}
F(k_{1},l_{1}:1,0;P+Q-2,0) = \frac{v_{1+}[k_{1}l_{1}]}{v_{1-}[P+Q,0]} \frac{v_{1+}[k_{1}+1,l_{1}]}{v_{1-}[P+Q-1,0]} F(k_{1}+2,l_{1};1,0:P+Q,0 ).
\end{equation}\normalsize
Continuing by induction, for an arbitrary $ s  = P+Q-k $ with $s\in[1,P+Q]$, we have the relation:\footnotesize
\begin{eqnarray}
 F(k_{1},l_{1}:1,0;k,0)= \frac{v_{1+}[k_{1},l_{1}]....v_{1+}[k_{1}+s-1,l_{1}]}{v_{1-}[P+Q,0]....v_{1-}[P+Q-s+1,0]} F(k_{1}+s,l_{1}:1,0:P+Q,0 ) \nonumber\\
\end{eqnarray}\normalsize
Let $d[k_1,l_1,s] \equiv v_{1+}[k_{1},l_{1}]....v_{1+}[k_{1}+s-1,l_{1}]$,  and \\$B[P,Q,s] \equiv \left( v_{1-}[P+Q,0]v_{1-}[P+Q-1,0]...v_{1-}[P+Q-s+1,0] \right)^{-1}$. \\
Using $v_{1-}[k_1,l_1]$ from (C.2), we have
\footnotesize
\begin{eqnarray}
\begin{aligned}
B[P,Q,s] =& \left\lbrace (P+Q-Q)(P+Q-P-Q+1) \right\rbrace^{-\frac{1}{2}} \\ &\times \left\lbrace (P+Q-1-Q)(P+Q-P-Q+1+1) \right\rbrace^{-\frac{1}{2}} \\ &\hspace*{3cm}\vdots\\&  \times \left\lbrace  (P+Q-s+1-Q)(P+Q-P-Q+s-1+1)\right\rbrace^{-\frac{1}{2}} \\
 =& (P(1)\times (P-1)(2) \times ...\times (P-s+1)(s))^{-\frac{1}{2}}  \\ 
=& \left(\frac{(P-s)!}{(P)! (s)!}\right)^{\frac{1}{2}} .
\end{aligned}
\end{eqnarray}\normalsize
Similarly using $v_{1+}[k_1,l_1]$ from (C.2), we have
\footnotesize
\begin{eqnarray*}
\begin{aligned}
 d[P_1,Q_1,k_{1},l_{1},s] =& \left\lbrace \frac{(k_{1}+2)(k_{1}-Q_1+1)(P_1+Q_1-k_{1})}{(k_{1}-l_{1}+2)}\right\rbrace^{\frac{1}{2}}\\ \times &\left\lbrace \frac{(k_{1}+3)(k_{1}-Q_1+2)(P_1+Q_1-k_{1}-1)}{(k_{1}-l_{1}+3)}\right\rbrace^{\frac{1}{2}} \end{aligned}
\end{eqnarray*}  \begin{eqnarray}
\begin{aligned} 
&\hspace*{3cm}\vdots\\ \times & \left\lbrace \frac{(k_{1}+s+1)(k_{1}+s-Q_1)(P_1+Q_1-k_{1}-s+1)}{(k_{1}+s+1-l_{1})} \right\rbrace^{\frac{1}{2}} \\
=& \left\lbrace \frac{ \left\lbrace(k_{1}+2)...(k_{1}+s+1)\right\rbrace \left\lbrace(k_{1}-Q_1+1)...(k_{1}+s-Q_1)\right\rbrace}{(k_{1}-l_{1}+2)(k_{1}-l_{1}+3)...(k_{1}+s+1-l_{1})} \right\rbrace^{\frac{1}{2}} \\
&\times  \left\lbrace (P_1+Q_1-k_{1})...(P_1+Q_1-k_{1}-s+1) \right\rbrace^{\frac{1}{2}} \\ 
=& \left\lbrace \frac{(k_{1}+s+1)!(k_{1}+s-Q_1)!(P_1+Q_1-k_{1})!(k_{1}-l_{1}+1)!}{(k_{1}+1)!(k_{1}-Q_1)!(P_1+Q_1-k_{1}-s)!(k_{1}-l_{1}+s+1)!} \right\rbrace^{\frac{1}{2}} .
\end{aligned}
\end{eqnarray}\normalsize
Combining (C.21),(C.22),(C.23) with (C.15), we obtain isoscalar factors of states with $k=P+Q-s,l=0$ as
\begin{equation}
F(k_{1},l_{1}:1,0;k,0) = B[P,Q,s] \hspace*{.1cm} d[P_1,Q_1,k_{1},l_{1},s] \hspace*{.1cm}F(k_{1}+s,l_{1}:1,0;P+Q,0 ).
\end{equation}

\subsubsection*{Case II: $s$-quark} 
When the second state in the tensor product is an $s$-quark,  $k_{2} =m_{2}=l_2=0$.\\\\
\textbf{Step 1:} Let us first find the isoscalar factor for $ k = P+Q-1 $. From (C.16) we get\footnotesize
\begin{equation}
\begin{split}
&v_{1-}[P+Q,0] \langle k_{1},l_{1},k_{1} \otimes 0,0,0|P+Q-1,0,P+Q-1 \rangle =  \langle k_{1},l_{1},k_{1} \otimes 1,0,1|P+Q,0,P+Q  \rangle \\&\hspace*{5cm}+ v_{1+}[k_{1},l_{1}]  \langle k_{1}+1,l_{1},k_{1}+1 \otimes 0,0,0|P+Q,0 ,P+Q \rangle,
\end{split}
\end{equation}\normalsize
since $v_{1+}[0,0]=1$. Separating the $SU(2)$ CG-coefficients, we have 
\footnotesize
\begin{equation}
\begin{split}
 &\left( \begin{array}{ccc}
\frac{k_{1}-l_{1}}{2} & 0 & \frac{P+Q-1}{2}\\
\frac{k_{1}-l_{1}}{2} & 0 & \frac{P+Q-1}{2} \\
\end{array} \right) F(k_{1}l_{1}:0,0;P+Q-1,0) = \frac{1}{v_{1-}[P+Q,0]}
\left\lbrace   \left( \begin{array}{ccc}
\frac{k_{1}+1-l_{1}}{2} & 0 & \frac{P+Q}{2} \\
\frac{k_{1}+1-l_{1}}{2} & 0 & \frac{P+Q}{2} \\
\end{array} \right) \right.\\ & \times\left. v_{1+}[k_{1}l_{1}] F(k_{1}+1l_{1}:0,0;P+Q,0 )  
+ F(k_{1},l_{1},1,0;P+Q,0 ))\left( \begin{array}{ccc}
\frac{k_{1}-l_{1}}{2} & \frac{1}{2} & \frac{P+Q}{2} \\
\frac{k_{1}-l_{1}}{2} & \frac{1}{2} & \frac{P+Q}{2} \\
\end{array} \right) \right\rbrace .
\end{split}
\end{equation}\normalsize
All the $SU(2)$ CG-coefficients are 1 here, and the equation simplifies to \footnotesize
\begin{equation}
\begin{split}
F(k_{1},l_{1}:0,0;P+Q-1,0) = &\frac{1}{v_{1-}[P+Q,0]}\left\lbrace v_{1+}[k_{1},l_{1}] F(k_{1}+1,l_{1}:0,0;P+Q,0 ) \right.\\ & \left. + F(k_{1},l_{1};1,0;P+Q,0 ) \right\rbrace .
\end{split}
\end{equation}\normalsize
The isoscalar factors on the RHS can be obtained from (C.15), determining the value of  $F(k_{1},l_{1}:0,0;P+Q-1,0)$.\\\\
\textbf{Step 2:} Now repeat the above process for $k=P+Q-2$. \footnotesize
\begin{equation}
\begin{split}
&v_{1-}[P+Q-1,0] \langle  k_{1}l_{1}k_{1} \otimes 0,0,0|P+Q-2,0,P+Q-2 \rangle   = v_{1+}[k_{1}l_{1}] \\ & \times \langle k_{1}+1,l_{1},k_{1}+1 \otimes 0,0,0|P+Q-1,0,P+Q-1 \rangle  
+ \langle k_{1}l_{1}k_{1} \otimes 1,0,1|P+Q-1,0,P+Q-1 \rangle  
\end{split}
\end{equation}\normalsize
With all the $SU(2)$ coefficients  equal to 1, and using (C.27)and (C.19), we get\footnotesize
\begin{equation}
\begin{split}
&F(k_{1},l_{1}:,0;P+Q-2,0) = \frac{1}{v_{1-}[P+Q-1,0]v_{1-}[P+Q,0]} 
\left\lbrace v_{1+}[k_{1}+1,l_{1}]\right. \\ &\times\left.v_{1+}[k_{1}l_{1}]    F(k_{1}+2,l_{1}:0,0;P+Q,0 ) + 2 \times v_{1+}[k_{1}l_{1}] F(k_{1}+1,l_{1}:1,0;P+Q,0 ) \right\rbrace.
\end{split}
\end{equation}\normalsize
Continuing by induction, till an arbitrary $s=P+Q-k$, the result is
\begin{equation}
\begin{split}
F(k_{1},l_{1}:0,0;k,0) = & B[P,Q,s] \times 
\left\lbrace d[P_1,Q_1,k_{1},l_{1},s]F(k_{1}+s,l_{1}:0,0:P+Q,0 ) \right.\\& \left. + s \times d[P_1,Q_1,k_{1},l_{1},s-1] F(k_{1}+s-1,l_{1}:1,0;P+Q,0 ) \right\rbrace, 
\end{split}
\end{equation}
where $B$ and $d$ are given by  (C.22) and (C.23). The RHS can be evaluated explicitly using the isoscalar factors in (C.15).
\section{Recursion relations for $k=P+Q-s , l>0$}
Finally, consider the states with $k=P+Q-s,l>0, \hspace*{0.2cm}\forall s\in[0,P+Q]$. For these states, we derive recursion relations using the operator $\hat{U}_+$,\\\vspace*{-0.8cm}
\begin{equation}
\begin{split}
\langle k_{1},l_{1},m_{1}k_{2},l_{2},m_{2}|\hat{U}_{+}|k,l,m \rangle   =& \langle \hat{U}_{-}(k_{1},l_{1},m_{1}) \otimes k_{2},l_{2},m_{2}|k,l,m \rangle \\&  + \langle  k_{1},l_{1},m_{1} \otimes \hat{U}_{-}(k_{2},l_{2},m_{2})|k,l,m \rangle .
\end{split} 
\end{equation}
\indent In this case, none of the terms vanish in general. 
For isoscalar factor with addition of a $u$-quark and $m_1=k_1-1$, we can have upto 4 terms in the recurrence relation and the calculation is cumbersome. Instead, we consider $m_1=k_1$, which gives isoscalar factor for most of the combinations, and the rest can be extracted by exploiting the orthonormality condition \vspace*{-0.4cm}
$$ \sum_{k_{1}l_{1}k_{2}l_{2}}|F (k_{1},l_{1},k_{2},l_{2}:k,l)|^{2} = 1.$$
Note that finding the isoscalar factor with addition of an $s$-quark is sufficient to determine the magnitude of the isoscalar factor with addition of a $u$-quark. Therefore, let us first find the isoscalar factor for the addition of an $s$-quark.

\subsubsection*{Case I: $s$-quark} 
In calculation of the isoscalar factor for the addition of an $s$-quark, we use the states with  $k_1=m_1$. Considering (C.31) for the state $\vert k,l-1,k\rangle$, we get 
\footnotesize
\begin{equation}
\begin{split}
&-u_{2-}[k_{1},l_{1}] \langle k_{1},l_{1}-1,k_1 \otimes 0,0,0,|k,l-1,k \rangle= -u_{2+}[k,l-1]\langle k_{1},l_{1},k_{1} \otimes 0,0,0|k,l,k \rangle \\& \hspace*{5cm} + u_{1+}[k,l-1]\langle k_{1},l_{1},k_{1} \otimes 0,0,0|k+1,l-1,k \rangle. 
\end{split}
\end{equation}\normalsize
Separating the $SU(2)$ CG-coefficients, it can be rewritten as
\footnotesize \begin{equation}
\begin{split}
&- u_{2-}[k_{1},l_{1}]  \left( \begin{array}{ccc}
\frac{k_{1}-l_{1}+1}{2} & 0 & \frac{k-l+1}{2} \\
\frac{k_{1}-l_{1}+1}{2} & 0 & \frac{k-l+1}{2}
\end{array}\right) F(k_{1},l_{1}-1:0,0;k,l-1)=-u_{2+}[k,l-1]\\&  \left( \begin{array}{ccc}
\frac{k_{1}-l_{1}}{2} & 0 & \frac{k-l}{2} \\
\frac{k_{1}-l_{1}}{2} & 0 & \frac{k-l}{2} \\
\end{array} \right) F(k_{1}l_{1}:0,0;k,l)+ u_{1+}[k,l-1]\left( \begin{array}{ccc}
\frac{k_{1}-l_{1}}{2} & 0 & \frac{k-l}{2} + 1 \\
\frac{k_{1}-l_{1}}{2} & 0 &  \frac{k-l}{2} \\
\end{array} \right) \\ & \hspace*{7cm} \times F(k_1,l_1:0,0;k+1,l-1).
\end{split} 
\end{equation}\normalsize
 In this relation, the $SU(2)$ CG-coefficient for the second term on RHS vanishes and the other two happen to be 1. Furthermore, the constraint 
 $ \frac{k_{1}-l_{1}}{2} = \frac{k-l}{2}$ leads to
 \vspace*{-0.4cm}\begin{equation}
 u_{2+}[k,l-1] F(k_{1},l_{1}:0,0;k,l) = u_{2-}[k_{1},l_{1}]F(k_{1},l_{1}-1:0,0:k,l-1).
 \end{equation}\\
 \textbf{step 1:} Let us start with $l=1$. Then
\vspace*{-0.4cm} \begin{equation}
 u_{2+}[k,0]F(k_{1},l_{1}:0,0;k,1) = u_{2-}[k_{1}l_{1}] F(k_{1},l_{1}-1:0,0;k,0).
 \end{equation}
\textbf{step 2:} Next, for $l=2$ in (C.34), we have using (C.35),
\vspace*{-0.4cm} \begin{equation}
 u_{2+}[k,1]F(k_{1},l_{1}:0,0;k,2) = \frac{u_{2-}[k_{1}l_{1}]u_{2-}[k_{1},l_{1}-1]}{u_{2+}[k,0]} F(k_{1},l_{1}-2:0,0;k,0).
 \end{equation}
 Iterating the process, we obtain for an arbitrary $l$,
\vspace*{-0.4cm} \begin{equation}
 F(k_{1},l_{1}:0,0:k,l) = \frac{u_{2-}[k_{1},l_{1}] ... u_{2-}[k,l_{1}-l+1]}{u_{2+}[k,0]....u_{2+}[k,l-1]} F(k_{1},l_{1}-l:0,0:k,0).
 \end{equation}
 The denominator on RHS is, using (C.2), 
\vspace*{-0.4cm}\footnotesize \begin{eqnarray} 
\begin{aligned}
A[P,Q,k,l] \equiv & \left(u_{2+}[k,0]u_{2+}[k,1].... u_{2+}[k,l-1]\right)^{-1} \\ 
=& (-1)^{l} \left\lbrace \frac{(1)(Q)(P+Q+1)}{(k+1)} \times \frac{(2)(Q-1)(P+Q)}{(k)} ......\right.\\ & \hspace*{3cm}\left.\times \frac{(l)(Q-l+1)(P+Q-l+2)}{(k-l+2)} \right\rbrace^{-\frac{1}{2}} \\ 
=& (-1)^{l} \left\lbrace \frac{(Q-l)!(P+Q-l+1)!(k+1)!}{(l)!(Q)!(P+Q+1)!(k+1-l)!} \right\rbrace^{\frac{1}{2}} 
\end{aligned}
\end{eqnarray}\normalsize
 Similarly, the numerator on RHS contains \footnotesize
 \begin{eqnarray}
\begin{aligned}
 c[P_1,Q_1,k_1,l_1,l] \equiv &u_{2-}[k_{1},l_{1}]u_{2-}[k_{1},l_{1}-1]...u_{2-}[k_{1},l_{1}-l+1] \\
 =& (-1)^{l} \left\lbrace \frac{l_{1}(Q_1-l_{1}+1)(P_1+Q_1-l_{1}+2)}{k_{1}-l_{1}+2} \right.\\ &  \times \frac{(l_{1}-1)(Q_1-l_{1}+2)(P_1+Q_1-l_{1}+3)}{(k_{1}-l_{1}+3)}  .... \\ &\times \left. \frac{(l_{1}-l+1)(Q_1-l_{1}+l)(P_1+Q_1-l_{1}+l+1)}{(k_{1}-l_{1}+l+1)} \right\rbrace^{\frac{1}{2}}  \end{aligned}
\end{eqnarray}  \begin{eqnarray}
\begin{aligned}
=& (-1)^{l}\left\lbrace \frac{(l)!(Q_1-l_{1}+l)!(P_1+Q_1-l_{1}+l+1)!(k_{1}-l_{1}+1)!}{(l_{1}-l)!(Q_1-l_{1})!(P_1+Q_1-l_{1}+1)!(k_{1}-l_{1}+l+1)!} \right\rbrace^{\frac{1}{2}} .
\end{aligned}
\end{eqnarray}\normalsize
Thus we can express 
 \begin{equation}
 F(k_{1},l_{1}:0,0:k,l) = c[P_1,Q_1,k_{1},l_{1},l]\hspace*{.2cm} A[P,Q,k,l]\hspace*{.2cm}F(k_{1},l_{1}-l:0,0:k,0),
 \end{equation}
  where the RHS can be evaluated using (C.39),(C.38) and (C.30). 
\subsubsection*{Case II: $u$-quark}  
We again consider $m_1=k_1$, and use (C.31) for the state $\vert k,l,k \rangle$. We get
\footnotesize
\begin{equation}
\begin{split}
&-u_{2-}[k_{1},l_{1}] \langle k_{1},l_{1}-1,k_1 \otimes 1,0,1,|k,l,k \rangle= -u_{2+}[k,l]\langle k_{1},l_{1},k_{1} \otimes 1,0,1|k,l+1,k \rangle \\& \hspace*{5cm} + u_{1+}[k,l]\langle k_{1},l_{1},k_{1} \otimes 1,0,1|k+1,l,k \rangle. 
\end{split}
\end{equation}\normalsize Separating the $U(2)$ CG-coefficients, this can be written as \footnotesize\begin{equation}
\begin{split}
&- u_{2-}[k_{1},l_{1}]  \left( \begin{array}{ccc}
\frac{k_{1}-l_{1}+1}{2} & \frac{1}{2} & \frac{k-l}{2} \\
\frac{k_{1}-l_{1}+1}{2} & \frac{1}{2} & \frac{k-l}{2}
\end{array}\right) F(k_{1},l_{1}-1:1,0;k,l)=-u_{2+}[k,l]\\&\times  \left( \begin{array}{ccc}
\frac{k_{1}-l_{1}}{2} & \frac{1}{2} & \frac{k-l-1}{2} \\
\frac{k_{1}-l_{1}}{2} & \frac{1}{2} & \frac{k-l-1}{2} \\
\end{array} \right) F(k_{1}l_{1}:1,0;k,l+1)+ u_{1+}[k,l]\left( \begin{array}{ccc}
\frac{k_{1}-l_{1}}{2} & \frac{1}{2} & \frac{k-l+1}{2}  \\
\frac{k_{1}-l_{1}}{2} & \frac{1}{2} &  \frac{k-l-1}{2} \\
\end{array} \right) \\ & \hspace*{8.6cm} \times F(k_1,l_1:1,0;k+1,l).
\end{split} 
\end{equation}\normalsize
Again the $SU(2)$ CG-coefficient for the second term on RHS vanishes, and the other two happen to be 1. Therefore, we have
\begin{eqnarray}
u_{2-}[k_{1},l_{1}] F(k_{1},l_{1}-1:1,0;k,l)=u_{2+}[k,l]F(k_{1}l_{1}:1,0;k,l+1).
\end{eqnarray}
 \textbf{step 1:} Let us start with $l=0$. Then
 \begin{equation}
 u_{2-}[k_1,l_1]F(k_{1},l_{1}-1:1,0;k,0) = u_{2+}[k,0] F(k_{1},l_{1}:1,0;k,1).
 \end{equation}
\textbf{ step 2:} Next, for $l=1$ in (C.43), we have using (C.44),
 \begin{equation}
 u_{2+}[k,1]F(k_{1},l_{1}:1,0;k,2) = \frac{u_{2-}[k_{1},l_{1}]u_{2-}[k_{1},l_{1}-1]}{u_{2+}[k,0]} F(k_{1},l_{1}-2:1,0;k,0).
 \end{equation}
 Iterating the process, we obtain for an arbitrary $l$,
 \begin{equation}
 F(k_{1},l_{1}:1,0:k,l) = \frac{u_{2-}[k_{1},l_{1}] ... u_{2-}[k,l_{1}-l+1]}{u_{2+}[k,0]....u_{2+}[k,l-1]} F(k_{1},l_{1}-l:1,0:k,0).
 \end{equation}
 Thus we obtain
 \begin{equation}
 F(k_{1},l_{1}:1,0:k,l) = c[P_1,Q_1,k_{1},l_{1},l]\hspace*{.2cm} A[P,Q,k,l]\hspace*{.2cm}F(k_{1},l_{1}-l:1,0:k,0),
 \end{equation} where the RHS can be evaluated using (C.39),(C.38) and (C.24). 
  
\noindent\textbf{step 3:} As pointed out before, the choice $k_1=m_1$ has a limitation, and does not yield results for all the required isoscalar factors. The recurrence relation in (C.47) works for $l\leq l_1$. But we also need the isoscalar factor for the case $l=l_1+1$. This situation occurs when the single quark is added to the second row of the Young diagram, transforming $(P_1,Q_1)\otimes(1,0)$ to $(P_1-1,Q_1+1)$, \textit{i.e.} for $p=1$. The increase in the value of $l$ is associated with the decrease in isospin from $T_1=T+1/2$ to $T$, when the box corresponds to isospin 1/2. We can nevertheless use the orthonormality condition to relate the situation to the case where the added box is $s$-quark,  \begin{equation}
  |F(k_1,l_1:1,0;k,l)|=\sqrt{1-|F(k_1,l_1+1:0,0,k,l)|^2},
  \end{equation}
 and evaluate the RHS using (C.40). To determine the sign of the isoscalar factor, we use the orthogonality condition of the CG-coefficients. Even a single orthogonality constraint is sufficient for our purpose. We consider the state $\vert P,Q,k,l,m\rangle$ of the irreducible representation $(P,Q)=(P_1-1,Q_1+1)$, which is orthogonal to the highest weight state of the same irreducible representation, \textit{i.e.}, $\vert P,Q,P+Q,0,P+Q\rangle$, when both are formed by  $(P_1,Q_1)\otimes(1,0)$. In general, we have \footnotesize
  \begin{eqnarray}
  &&\hspace*{-0.8cm}\vert P_1-1,Q_1+1,k,l,m\rangle = \alpha \vert P_1,Q_1,k,l-1,m-1 \rangle \vert u\rangle + \beta \vert P_1,Q_1,k,l-1,m \rangle \vert d\rangle \nonumber\\ && \hspace*{4cm}+\gamma \vert P_1,Q_1,k,l,m \rangle \vert s\rangle ,\\
 &&\hspace*{-0.8cm}\vert P_1-1,Q_1+1,P_1+Q_1,0,P_1+Q_1\rangle = \alpha' \vert P_1,Q_1,P_1+Q_1-1,0,P_1+Q_1-1 \rangle \vert u\rangle  \nonumber\\&&+ \beta' \vert P_1,Q_1,P_1+Q_1-1,0,P_1+Q_1 \rangle \vert d\rangle  +\gamma' \vert P_1,Q_1,P_1+Q_1,0,P_1+Q_1 \rangle \vert s\rangle ,
  \end{eqnarray}\normalsize 
  with $\alpha's, \beta's$ being the $SU(3)$ CG-coeffficients. Clearly $\beta'=0$ since it does not satisfy the condition $k \geq m$. Therefore we have $ \alpha \alpha'+\gamma \gamma'=0$, with  the Wigner-Eckart decomposition giving \begin{eqnarray}\begin{aligned}
  &\alpha =\frac{-1}{\sqrt{k-l+2}}F(k,l-1:1,0;k,l), \hspace*{1cm} \gamma =F(k,l:0,0;k,l),\\
  &\alpha' = F(P_1+Q_1-1,0:1,0;P_1+Q_1), \hspace*{1cm} \gamma' = F(P_1+Q_1,0:0,0;P_1+Q_1).
  \end{aligned}
\end{eqnarray}   
From (C.15) and (C.2) we have 
\begin{eqnarray}
&&\alpha'= -\sqrt{\dfrac{1}{P_1+1}}, \hspace*{2cm} \gamma'= \sqrt{\dfrac{P_1}{1+P_1}}.
\end{eqnarray}
Then to calculate the value of $\gamma$, we use the recursion relations (C.40) and (C.30), substitute the coefficients from (C.22),(C.23),(C.38) and (C.39), and obtain \footnotesize
\begin{eqnarray}
\begin{aligned}
\gamma =& F(k,l:0,0;k,l)=c[P_1,Q_1,k,l,l]\hspace*{0.1cm} A[P_1-1,Q_1+1,k,l]\hspace*{0.1cm}F(k,0:0,0;k,0) \\=& c[P_1,Q_1,k,l,l]\hspace*{0.1cm} A[P_1-1,Q_1+1,k,l]\hspace*{0.1cm}B[P_1-1,Q_1+1,s]\\&\times \left\lbrace d[P_1,Q_1,k,0,s]\hspace*{0.1cm}F(P_1+Q_1,0:0,0;P_1+Q_1,0) \right.\\&\left. +s\times d[P_1,Q_1,k,0,s-1] F(P_1+Q_1-1,0:1,0;P_1+Q_1,0) \right\rbrace\\
=& \sqrt{\dfrac{(1-l+Q_1)(P_1-1-s)!}{(Q_1+1)(P_1-1)!(s)!}}\left\lbrace d[P_1,Q_1,k,0,s]\hspace*{0.1cm}\gamma' +s\times d[P_1,Q_1,k,0,s-1]\hspace*{0.1cm}\alpha' \right\rbrace\\
=&\sqrt{\dfrac{(1-l+Q_1)(P_1-1-s)!}{(Q_1+1)(P_1-1)!(s)!}}\left\lbrace \sqrt{\dfrac{(P_1)!(s)!}{(P_1-s)!}\dfrac{P_1}{P_1+1}} -s \sqrt{\frac{(P_1-1)!(s)!}{(P_1-s)!}\dfrac{1}{P_1+1}} \right\rbrace\\
=&\sqrt{\dfrac{(1-l+Q_1)(P_1-s)}{(Q_1+1)(P_1+1)}},
\end{aligned}
\end{eqnarray} \normalsize where $s=P_1+Q_1-k$.
Using this value in (C.51),\footnotesize
\begin{eqnarray}\begin{aligned}
\alpha =-\dfrac{\gamma\gamma'}{\alpha'}\Rightarrow
F(k,l-1:1,0;k,l)=-\sqrt{k-l+2}\left(\sqrt{\dfrac{(1-l+Q_1)(P_1-s)P_1}{(Q_1+1)(P_1+1)}} \right).
\end{aligned}\end{eqnarray}\normalsize
Thus the isoscalar factor has a negative sign when $l=l_1+1$. Combining (C.54) with (C.47) and (C.48), we have
\begin{eqnarray}\begin{aligned}
F(k_1,l_1:1,0;k,l)&=\delta_{l,l_1+1}(-\sqrt{1-|F(k_1,l_1+1:0,0;k,l)|^2})\\&+c[P_1,Q_1,k_{1},l_{1},l]\hspace*{.2cm} A[P,Q,k,l]\hspace*{.2cm}F(k_{1},l_{1}-l:1,0:k,0).
\end{aligned}\end{eqnarray}
which gives isoscalar factors for all the cases of interest.

%\backmatter % book mode only
%\appendix

%\addappheadtotoc
%\include{Appendix1/appendix1}
%\include{Appendix2/appendix2}

%new
%\bibliographystyle{plainnat}
%endnew
%\bibliographystyle{Classes/CUEDbiblio}
%\bibliographystyle{Classes/jmb}
%\bibliographystyle{plainnat} %this works with package natbib
%\bibliographystyle{Classes/jmb} % bibliography style
\renewcommand{\bibname}{References} % changes default name Bibliography to References

%------------------important for thesis
\bibliography{references} % References file

\end{document}